\documentclass[aps,prl,twocolumn,showkeys,amsmath,amssymb,longbibliography,floatfix,superscriptaddress]{revtex4-1}
\usepackage[T1]{fontenc}
\usepackage{graphicx,tabularx}
\usepackage{dcolumn}
\usepackage{amsmath,amssymb,amsfonts,bm,dsfont,units}
\usepackage{hyperref}
\hypersetup{colorlinks=true,citecolor=blue,linkcolor=blue,urlcolor=blue,pdfstartview=FitH}
\usepackage{appendix} 
\usepackage{braket}
\usepackage{float,latexsym}
\usepackage{multirow}
\usepackage{diagbox}
\usepackage[normalem]{ulem} % for \sout
\usepackage[caption=false]{subfig} % for subfloat
\usepackage{xcolor}
\newcommand{\vect}[1]{\boldsymbol{#1}}
\usepackage{array}
\usepackage{graphicx} 
\usepackage{hhline,booktabs}
\usepackage{makecell}
\usepackage{wasysym}

\newcommand{\cdotnone}{}

\newcolumntype{P}[1]{>{\centering\arraybackslash}p{#1}}

\usepackage{natbib}
\newcommand{\diag}{\text{diag}}
 
\begin{document}

\title{Theory of Next-Generation Even-Denominator States}
\author{Misha Yutushui}
\affiliation{Department of Condensed Matter Physics, Weizmann Institute of Science, Rehovot 76100, Israel} 
\affiliation{Institute for Theoretical Physics, University of Cologne, 50937 Cologne, Germany}
\author{David F. Mross}
\affiliation{Department of Condensed Matter Physics, Weizmann Institute of Science, Rehovot 76100, Israel} 

\begin{abstract}
Even-denominator quantum Hall states are leading candidates for realizing non-Abelian topological orders, with the $\nu=\frac{5}{2}$ plateau in GaAs the first and most-studied example. Recent experiments in GaAs and bilayer graphene (BLG) have observed many `next-generation' even-denominator states at filling factors such as $\nu=\frac{3}{4}$, $\frac{3}{8}$, and $\frac{3}{10}$. We develop the theory of these states, including analyses of their bulk quasiparticles, of methods for distinguishing between pairing channels in edge transport measurements, and of their trial wavefunctions. As part of this study, we derive general relations of how flux attachment affects many universal properties of states. In particular, we prove that the topological stability of interface modes is invariant under flux attachment. We compare next-generation paired states to Bonderson-Slingerland states at the same filling factors, and demonstrate that their quasiparticles carry identical charges and obey the same exchange statistics. The next-generation and Bonderson-Slingerland states still describe distinct phases, and we find that the former are energetically favored in the lowest Landau level, while the latter are favored in the first excited level. 

\end{abstract}

\date{\today}
\maketitle
Fractional quantum Hall (FQH) states are paradigmatic examples of topologically ordered phases of matter~\cite{Wen_topological_1990}. Most of the observed states can be understood as an integer quantum Hall effect of composite fermions (CFs), emergent quasiparticles formed from electrons bound to an even number of flux quanta~\cite{Jain_composite_2007,Halperin_FQH_2020}. The first observed state that defied such an interpretation was observed at the filling $\nu=\frac{5}{2}$ in GaAs~\cite{Willett_observation_1987}. Even-denominator plateaus cannot arise from integer quantum Hall states of CFs, but occur when they form paired superfluids~\cite{Haldane_fqh_1983,Moore_nonabelions_1991,Greiter_half_filled_1991,Read_paired_2000}.

Over the past years, many additional even-denominator FQH states were found in GaAs wide quantum wells~\cite{Suen_Observation_1992,Suen_Correlated_1992,Suen_Origin_1994,Shabani_Correlated_2009,shabani_evidence_2009,Shabani_Phase_2013,Hasdemir_tilted_2015,Dorozhkin_Unconventional_2023}, hole-doped GaAs~\cite{Shabani_Phase_2013,Liu_hole_2014},  monolayer~\cite{Zibrov_Even_Denominator_2018,Kim_Even_Denominator_f_wave_2019},  bilayer~\cite{Ki_bilyaer_graphene_2014,Kim_bilayer_graphene_2015,Li_bilayer_graphene_2017,Zibrov_Tunable_bilayer_graphene_2017,Huang_Valley_bilayer_graphene_2022,Kumar_Quarter_2024,Assouline_Energy_Gap_bilayer_graphene_2024,Haug_Interaction_2025,Singh_fractional_2025,Kumar_Orbitally_2025}, and trilayer graphene~\cite{chen_tunable_2023,chanda_Even_TLG_2025}, and other heterostructures like ZnO~\cite{Falson_Zno_2015,Falson_Zno_2018,Falson_Phase_2019} and WSe$_2$~\cite{Shi_even_wse2_2020}. Even-denominator states may realize topological states that permit non-Abelian excitations~\cite{Moore_nonabelions_1991,Read_paired_2000}, which could have applications in quantum information~\cite{Stern_non_Abelian_2010}.

Despite considerable efforts, it remains a formidable challenge to determine if any of these states indeed realizes non-Abelian quasiparticles. Distinguishing between different topological orders at the same filling factor requires sensitivity to neutral excitations, which do not contribute to conventional charge responses. The most direct observable that can achieve such a distinction is the thermal Hall conductance, which is affected by all excitations independent of their charge. In the $\nu=\frac{5}{2}$ case, thermal Hall conductance experiments~\cite{Banerjee_Observation_2018,Dutta_Isolated_2022,Paul_Thermal_2024} identified a particular non-Abelian topological order known as PH-Pfaffian~\cite{Son_is_2015}.

The difficulty of thermal conductance measurements, their disagreement with numerics~\cite{Mishmash_numerical_2018,Yutushui_Large_scale_2020,Wojs_3body_2005,wang_particle_hole_2009,Rezayi_breaking_2011,zaletel_infinite_2015,Tylan_Phase_2015,Rezayi_Landau_2017,Antoni_Paired_2018,Zhao_CF_pairing_LLM_2023}, questions about thermal equilibration~\cite{Simon_equilibration_2018,Feldman_equilibration_2019,Simon_equilibration_2020,Asasi_equilibration_2020,Lotric_Majorana_2025}, and the role of disorder \cite{Mross_theory_2018,Wang_topological_2018,Lian_theory_2018,zhu2020,fulga2020} leave a need for alternative probes. One possibility is upstream-noise measurements~\cite{Spanslatt_Noise_2019,Park_noise_2020,Manna_Full_Classification_2022,Yutushui_Identifying_2023,Manna_Experimentally_2023}, which were carried out at $\nu=\frac{5}{2}$ in Ref.~\onlinecite{Dutta_Isolated_2022} and corroborated the PH-Pfaffian topological order. Additional identification schemes of half-filled FQH states include the scaling of tunneling currents~\cite{Ken_sixteenfold_2019} and measurements of the coherent conductance in certain mesoscopic structures~\cite{Yutushui_Identifying_2022,Yutushui_Universal_2025}.

At other even-denominator fillings such as $\nu=\frac{3}{8}$ and $\frac{3}{10}$, similar questions about competing topological orders arise. Plateaus at these fillings cannot be explained as paired superfluids of weakly interacting CFs. Instead, these states can be viewed as strongly interacting CFs forming half-filled states. This work generalizes the theoretical understanding of the half-integer quantum Hall states and their experimental signatures to such `next-generation' (NG) even-denominator states. 

Two distinct families of paired FQH states could account for the plateaus at NG even-denominator fillings. Firstly, half-filled FQH states of CFs that can be described as paired superfluids of a second CF generation~\cite{Lee_Stripe_2001,Scarola_Possible_2002,Wojs_Fractional_2004,Mukherjee_Possible_2012,Mukherjee_incompressible_2015}. Secondly, the Bonderson-Slingerland (BS) states occur at the same fillings~\cite{Bonderson_hierarchy_2008} and do not permit a simple flux-attachment interpretation. The two families coincide at half-filling but further enrich the possibilities at general even denominators.

We construct trial wave functions for NG paired states and BS states at $\nu=\frac{3}{8}$ and $\frac{3}{10}$ and study their energetic competition in different Landau levels. We find that the NG states are favored in the lowest Landau level and the BS states in the first excited Landau level. In higher levels, Wigner crystals are favorable over both families of paired states.

The paper is organized as follows: In Section~\ref{sec.CF}, we briefly summarize relevant aspects of CF theory and derive a duality between two different CF approaches. In Section~\ref{sec.BS}, we introduce BS states as alternative candidate states. In Section~\ref{sec.edge_theory}, we discuss the topological stability of Abelian and paired states, and prove its invariance under flux attachment. In Section~\ref{sec.experimental}, we discuss experimental signatures distinguishing pairing channels of even-denominator states. In Section~\ref{sec.numerics}, we present our numerical techniques and findings. Finally, in the Appendices, we prove the equivalence of different CF descriptions for the NG states, discuss the relationship between BS and NG states, analyze their edge theories to derive experimental signatures, and provide details of our numerical calculations.

\section{Composite fermion theory}\label{sec.CF}
In this section, we briefly introduce the pertinent aspects of CF theory used in this study; a comprehensive treatment is available in Ref.~\onlinecite{Jain_composite_2007}. We then use this framework to derive some key properties of the NG FQH states. 

\subsubsection{Composite-fermion charge and filling factor}
Composite fermions arise by attaching $2p$ quantized phase vortices, commonly referred to as `fluxes', to each electron. This transformation retains the fermionic statistics, but results in a CF filling factor that differs from the electron filling $\nu =\frac{\rho \phi_0}{B}$, with the electron density $\rho$, flux quantum $\phi_0=\frac{e}{h}$, and magnetic field $B$. There are two equivalent perspectives for obtaining the relation between $B$ and the effective magnetic field experienced by CFs. In the first approach, one notes that the attached fluxes amount to an \textit{average} magnetic flux $2p \rho \phi_0$, which partially cancels the physical magnetic field $B$ and reduces it to $B^* = B - 2 p \rho \phi_0$.  In the second approach, one observes that in a system with Hall response $\sigma_{xy} = \nu\frac{e^2}{h}$, a flux $\phi_0$ is associated with an electric charge $Q_{\phi_0}= \sigma_{xy}\phi_0 = \nu e$. Attaching $2p$ flux quanta to an electron with charge $e$ thus results in a composite particle with charge $Q_\text{CF}= e - 2p Q_{\phi_0} = (1-2 p \nu)e$. The effective magnetic field experienced by these particles, $B^* = \frac{Q_\text{CF}}{e} B$, is identical to the one computed above.

The CF density equals the electron density, which leads to the CF filling factor $\nu^* \equiv \frac{\rho \phi_0}{B^*}$. (Notice that for $\nu>\frac{1}{2p}$, $B^*$, and $\nu^*$ are negative.) Solving for $\nu$, one finds the relationship
\begin{align}
    \nu=\frac{\nu^*}{2p\nu^*+1}. \label{eqn.nuvfnu}
\end{align}

The main advantage of this mapping is that many strongly correlated states of electrons can be captured by simple states of CFs.

\subsubsection{Quasiparticle content and $K$-matrix}
To obtain the quasiparticle content of a FQH state described by CFs, it is helpful to follow the second perspective. Above, we computed the effective charge $Q_\text{CF}$ of a CF. The exchange statistics of the corresponding bulk excitation are modified from the fermionic phase $\pi$ by an accumulated phase acquired as this charge encircles the flux $2p\phi_0$, i.e., a single exchange results in a phase $\theta_\text{CF} = \pi\left(1 -2 p\frac{Q_\text{CF}}{e}\right)$.

In general, the quasiparticle content of an Abelian topological order is efficiently encoded in a symmetric, integer-valued, and non-singular $K$-matrix and an integer-valued charge vector $\vect{t}$. The filling factor is $\nu = \vect t^T K^{-1 }\vect{t}$, and the number of independent quasiparticles (not relatable by a finite number of electrons) is given by the determinant of the $K$-matrix. The quasiparticles are parametrized by integer-valued vectors $\vect{m}$ such that their charge and topological spin (self-statistics) are $Q_{\vect{m}} = e\,\vect{t}^TK^{-1}\vect{m}$ and $\theta_{\vect{m}}=\pi\,\vect{m}^T K^{-1} \vect{m}$, respectively. The mutual exchange phase of two such quasiparticles is $\theta_{12}=\pi\,\vect{m}_1^T K^{-1} \vect{m}_2$.

The effective topological field theory describing the bulk of the topological order specified by $K,\vect t$ is
\begin{align}\label{eq.L_chern_simon}
    {\cal L} =  -\frac{K_{IJ}}{2\pi}\vec {a}_{I}\cdot (\vec \nabla \times \vec a_J)  + e\frac{t_I}{2\pi} \vec A \cdot  ( \vec\nabla \times \vec {a}_{I})+ \ldots,
\end{align}
where $\vec A = (A_\tau,A_x,A_y)$ is the electro-magnetic vector potential, $\vec a = (a_\tau,a_x,a_y)$ are emergent gauge fields and $\vec \nabla = (\partial_\tau,\partial_x,\partial_y)$ . The ellipsis denotes irrelevant, subleading contributions allowed by symmetry and gauge invariance, e.g., kinetic terms $\propto(\vec\nabla \times \vec {a}_{I})(\vec\nabla \times \vec {a}_{I'})$ for the emergent gauge fields and gapped matter fields. Such terms are generically present with non-universal coefficients, but do not affect topological properties.

In the $K$-matrix formalism, the attachment of $2p$ fluxes relates the CF $K^*$-matrix to its electronic counterpart as 
\begin{align}\label{eq.flux_attachment}
    K = K^* + 2p \vect{t}\vect{t}^T,
\end{align}
while the charge vector remains the same. The determinants of $K$ and $K^*$ are related by
\begin{align} 
    \det K = (1+ 2 p \nu^*)\det K^*~,\label{eq.flux_attachment.det}
\end{align}
where $\nu^*=\vect{t}^T[K^*]^{-1}\vect{t}$. Notice that for $|1+2p\nu^*|>1$, the topological order of $\nu$ has a larger number of distinct quasiparticle excitations compared to $\nu^*$. To obtain the filling factor after flux attachment, we compute $\nu  = \vect t^T K^{-1 }\vect{t}$ using
\begin{align}
   K^{-1} = (K^*)^{-1 } - \frac{2 p}{{1+2p \nu^*}}(K^*)^{-1 }\vect{t}\vect{t}^T(K^*)^{-1 }~,\label{eq.flux_attachment.inverse}
\end{align}
which reproduces Eq.~\eqref{eqn.nuvfnu}.

Integer quantum Hall states of CFs at filling  $\nu^*=\pm q$ with $q>0$ are described by a $q$-dimensional identity matrix $K^*=\pm \mathbb{I}_q$
and the charge vector $\vect t=(1,1,\ldots)$. The resulting electron phase occurs at the Jain-state filling factor $\nu=\frac{q}{2pq\pm1}$, and the quasiparticle charges and statistics match those obtained directly from the CF analysis.

The topological order does not uniquely specify a $K$-matrix and a charge vector. The matrix $K'=W^TKW$ and charge vector $\vect t'=W^T\vect{t}$ for any $W\in \text{SL}(\text{dim}(K),\mathbb{Z})$ encode the same universal properties. In particular, a quasiparticle parametrized by $\vect{m}$ in the basis $(K,\vect{t})$ becomes $W^T \vect{m}$ in the primed basis.

\subsubsection{Thermal Hall conductance}
A measurable topological invariant that distinguishes different topological orders at the same filling factor is the thermal Hall conductance $\kappa_{xy}$. For Abelian states, it is the signature of the $K$-matrix, i.e.,
\begin{align}\label{eq.kappa}
    \kappa_{xy}=\kappa_0 \sum_i \text{sgn}(\lambda_i),
\end{align}
where $\lambda_i$ are the eigenvalues of $K$. Notice that the signature is invariant under $W\in \text{SL}(\text{dim}(K),\mathbb{Z})$ transformation, as expected for a topological quantity. 

The transformation of $\kappa_{xy}$ under flux attachment can be inferred from Eq.~\eqref{eq.flux_attachment.det}, which extends mathematically to non-integer values of $p$. When $2 p \nu^*+1 >0 $, the prefactor in front of $\det K^*$ is positive, implying that no eigenvalue crosses zero. Consequently, $\kappa_{xy}$ remains unchanged. However, if $2 p \nu^*+1 <0$, one eigenvalue must cross zero. Since $\nu$ and $\nu^*$ have opposite signs in this case [cf.~Eq.~\eqref{eqn.nuvfnu}], this change can be attributed to the charge mode. We thus arrive at the general relation
\begin{align}
    \kappa_{xy} + \kappa_0\text{sgn}(\nu) = \kappa_{xy}^* +\kappa_0 \text{sgn}(\nu^*)~.\label{eqn.kappa.under.flux.attachment}
\end{align}
This relation is most apparent in a basis where charge and neutral modes are decoupled. The flux attachment does not affect neutral modes and can at most change the chirality of the charge mode contributing $\kappa_0$. When the effective magnetic field vanishes ($B^*=0$), the $K^*$-matrix possesses one zero eigenvalue that does not contribute to $\kappa_{xy}$, and $\nu^*$ is ill-defined. Still, Eq.~\eqref{eqn.kappa.under.flux.attachment} holds, provided we take  $\text{sgn}(\nu^*)=0$ in this case.

\subsubsection{The shift quantum number}
In finite systems, the number of electrons $N_e$ generically deviates from $\nu$ times the number of fluxes $N_\phi$ by a constant of order unity that depends on the genus of the system and the topological order. We write
\begin{align}\label{eq.shift}
    N_\phi=\nu^{-1}N_e - S~,
\end{align}
where $S$ is called the shift quantum number~\cite{Wen_shift_1992}. For example, the lowest Landau level on a sphere has $N_\phi+1$ single-particle states, and each higher level has two more states than the previous one. Filling the first $n$ levels to create a $\nu=n$ state thus requires $N_e = n (N_\phi+1) + n(n-1)$ electrons, which implies $S=n$. When CFs fill $n$ levels, then $S^*=n$, and the shift of the electrons is
\begin{align}\label{eq.cfshift}
    S =  S^* + 2p.
\end{align} 
More generally, for a topological order parameterized by a $K$-matrix and a charge vector $\vect{t}$, an additional invariant is defined, the half-odd-integer (integer) shift vector $\vect{s}$, which reflects the effective angular momentum of a fermionic (bosonic) mode~\cite{Wen_shift_1992}. The shift quantum number on a sphere is determined by the spin and charge vectors as
\begin{align}
S=\frac{2}{\nu}\vect{t}^T\cdotnone K^{-1}\cdotnone \vect{s}.    
\end{align}
For a $\nu=n$ integer quantum Hall state, the shift vector is given by $\vect s=(\frac{1}{2},\frac{3}{2},\ldots)$ when $n$ is positive and $\vect s=(-\frac{1}{2},-\frac{3}{2},\ldots)$ when it is negative. The shift vector transforms as the charge vector, $\vect{s}'=W^T\vect{s}$.  Unlike the charge vector, $\vect s$ changes under flux attachment, which reflects a contribution of the attached fluxes to the angular momentum of the composite particle. After attaching $2p$ fluxes, the shift vector transforms according to 
\begin{align}
    \vect{s} = \vect{s}^* + p\vect{t}.
\end{align}
(For a half-odd-integer $p$, an odd number of fluxes changes the half-odd-integer shift vector of fermions to an integer shift vector of bosons.) Using Eq.~\eqref{eq.flux_attachment.inverse}, one verifies that this choice of $\vect{s}$ reproduces Eq.~\eqref{eq.cfshift}. The shift quantum number provides a necessary condition for two states at the same filling to have the same universality class in the absence of disorder. 

\subsubsection{Particle-hole conjugation}
The particle-hole conjugation of a FQH state at  $\nu$ produces a state at $1-\nu$. In the lowest Landau level, the number of holes is $N_h=N_\phi+1-N_e$. When the electron number is given by Eq.~\eqref{eq.shift}, the number of holes satisfies $N_\phi=\bar{\nu}^{-1}N_h - \bar{S}$ with
\begin{align}\label{eq.phshift}
(\bar{\nu},\bar{S}) = \left(1-\nu,\frac{1-S\nu}{1-\nu}\right).
\end{align}
In particular, the particle-hole conjugate of the $n$th Jain state has the same filling and shift as the Jain state with $\nu^*=-(n+1)$. 

In the $K$-matrix formalism, particle-hole conjugation results in $\overline{K} = \diag(1,-K)$, $\overline{t} = (1,\vect{t})$, and $\overline{s} = (\frac{1}{2},\vect{s})$. Comparison of the $K$-matrices confirms that the particle-hole conjugate of the $n$th Jain state exhibits the same topological order as the $-(n+1)$th Jain state. Later, we show that this relation generalizes to fractional fillings of the $n$th CF Landau level.

\subsubsection{Even-denominator states}\label{sec.cf.even}
At the even-denominator fillings $\nu=\frac{1}{2p}$, CFs with $2p$ fluxes are impervious to the magnetic field, i.e., $B^*=0$. Consequently, they cannot form an integer quantum Hall state, but instead realize gapless metallic states or paired superfluids. Fermionic superfluids in two dimensions are characterized by an integer topological index $\ell$~\cite{Read_paired_2000}. In rotationally invariant systems, odd values of $\ell$ correspond to the Cooper-pair angular momentum, i.e., $\Delta(k_x,k_y) \sim (k_x + i k_y)^{\ell}$. At a spatial boundary, this index manifests itself in chiral Majorana modes. The number of these modes is $|\ell|$, and their chirality is downstream (upstream) for positive (negative) $\ell$. At half-filling $(p=1)$, the most relevant cases are the Moore-Read Pfaffian ($\ell=1$)~\cite{Moore_nonabelions_1991}, the anti-Pfaffian ($\ell=-3$)~\cite{Lee_particle_hole_2007,Levin_particle_hole_2007}, the PH-Pfaffian ($\ell=-1$)~\cite{Son_is_2015}, and the $f$-wave ($\ell=3$)~\cite{Wen_Non-Abelian_1991} states.

In addition to neutral fermionic quasiparticles, paired superfluids permit half-flux ($\frac{\phi_0}{2}$) vortices as excitations. Due to the Hall response, they carry the electric charge $Q=\frac{\nu}{2}e$. For odd $\ell$, these vortices contain Majorana zero modes, resulting in non-Abelian statistics. Allowing for such excitations requires a slight modification of the $K$-matrix formalism~\cite{Ken_sixteenfold_2019}. For $\ell=2r+1$, the $K$-matrix is constructed as a $r+1$ dimensional diagonal matrix $K=\diag(2p,\, s\, \mathbb{I}_r)$ and $\vect{t}=(1,0,\ldots,0)$. Each of the $r$ components with $s=\text{sgn}(r)$ describes a neutral complex fermion mode (two Majorana fermions) contributing $s\kappa_0$ to thermal Hall conductance. The remaining unpaired Majorana is described by an Ising topological field theory~\cite{moore_classical_1989,Moore_Lectures_1990} that consists of a trivial field $\mathbb{I}$, a fermion $\gamma$, and the non-Abelian twist field $\sigma$. Bulk quasiparticles are parametrized by combining one of these three fields with an Abelian excitation parameterized by a vector $\vect{m}$. For $\mathbb{I}$ and $\gamma$, the vector is integer-valued as for Abelian states, while for the twist field $\sigma$, the vector is half-odd-integer-valued. 

The $K$-matrices of these even-denominator states correspond to a singular $K^*=\diag(0,\, s,\ldots,s)$ via Eq.~\eqref{eq.flux_attachment}. An eigenvalue of zero implies that one of the emergent gauge fields does not acquire a topological mass. Instead, its fluctuations are described by Maxwell terms, which were omitted in Eq.~\eqref{eq.L_chern_simon} but must be restored in this case. Due to the second term in Eq.~\eqref{eq.L_chern_simon}, such a strongly fluctuating $\vec a$ implies a superconducting (Meissner) response for the physical vector potential $\vec A$, as expected for a paired superfluid.

\subsection{Next-generation states}

Integer quantum Hall states and superfluids of CFs, along with their particle-hole conjugates, account for most of the observed FQH states. However, an increasing number of states that violate this pattern have been reported in recent years. An early example of such a state was reported at filling factors $\nu=\frac{4}{11}$ in Refs.~\onlinecite{Pan_Fractional_2003,Pan_Fractional_2015,Samkharadze_Observation_411_2015}. No combination of integers $\nu^*$ and $p$ in Eq.~\eqref{eqn.nuvfnu} can realize this filling or its particle-hole-conjugate $\bar\nu=\frac{7}{11}$. Instead, $\nu^* = \frac{4}{3}$ is the simplest (smallest denominator) filling that can reproduce $\nu=\frac{4}{11}$. To create this state, the conventional `first generation' CFs fully occupy one Landau level and partially occupy a second one at $\nu_0^*=\frac{1}{3}$. Those in the partially filled level can be further converted into `second-generation' CFs, forming a $\nu^{**}_0=1$ integer quantum Hall state.

In a general second-generation state, CFs completely fill $n$ levels and partially occupy another level at filling $\nu_0^*$. (To avoid ambiguity, we always take $n,\nu_0^*$ with equal signs.)  The electronic filling factor of this state is
\begin{align}\label{eq.nu_pair}
    \nu_\text{NG}(n,p)=\frac{n+\nu_0^*}{2p(n+\nu_0^*)+1}.
\end{align}  
and we construct its $K$-matrix starting from $K_0^*$ describing the partially filled level at filling $\nu_0^*$. First, we obtain $K^*$ by appending $n$ filled levels described by $K_{n} = \text{sgn}(n) \mathbb{I}_n$. Then, we perform flux attachment according to Eq.~\eqref{eq.flux_attachment}, i.e.,
\begin{align}\label{eq.K_next_gen}
    K = \text{diag}(K_0^*,K_{n}) + 2p\vect{t}\vect{t}^T,
\end{align}
with $\vect{t}=(\vect{t}_0,1,\ldots,1)$. When the first-generation state $\nu_0^*$ occurs at even-denominator filling, the NG state $\nu_\text{NG}(n,p)$ is also an even-denominator state. Specifically, for $\nu^*=\pm (n +\frac{1}{2})$, we obtain 
\begin{align}\label{eq.filling.ng}
    \nu_\text{NG}  = \frac{1}{2}\frac{2n+1}{p(2n+1)\pm 1}.
\end{align}

Unconventional even-denominator states at filling factors $\nu=\frac{3}{8}$ and $\frac{3}{10}$ \footnote{States at $\nu=\frac{1}{4}$ can be described by conventional four-flux CFs, and $\nu=\frac{3}{4}$ as their hole conjugates~\cite{Huang_non_Abelian_2024,Yutushui_phase_2025}. At the end of Sec.~\ref{sec.equivalence}, we prove the equivalence of topological orders obtained by different CF prescriptions.} have been observed in the hole-like bands of the GaAs~\cite{Wang_even_3_4_2022,Wang_Next_generation_2023}. These states can be constructed from the two-flux CFs ($p=1$) at $\nu^*=\frac{3}{2}$ and four-flux CFs ($p=2$) states at $\nu^*=-\frac{3}{2}$, respectively. In both cases, there is a half-filled LL of CFs, $\nu_0^*=\pm \frac{1}{2}$, in addition to a filled level, $n=\pm 1$. We label the second-generation states by the pairing channel in the partially filled CF Landau level. For states where $\nu$ and $\nu^*$ have opposite signs, this convention implies that the chirality of the Majoranas with respect to the electronic charge mode is given by $-\ell$ instead of $\ell$. The neutral Majorana fermions are unaffected by flux attachment, but the chiralities of the charge modes at $\nu$ and $\nu^*$ are opposite; see the discussion leading to Eq.~\eqref{eqn.kappa.under.flux.attachment}. For example, the $\nu=\frac{3}{10}$ anti-Pfaffian state ($\ell=-3$) has three \textit{downstream} Majorana modes.

\begin{figure}
    \centering
    \includegraphics[width=0.95\linewidth]{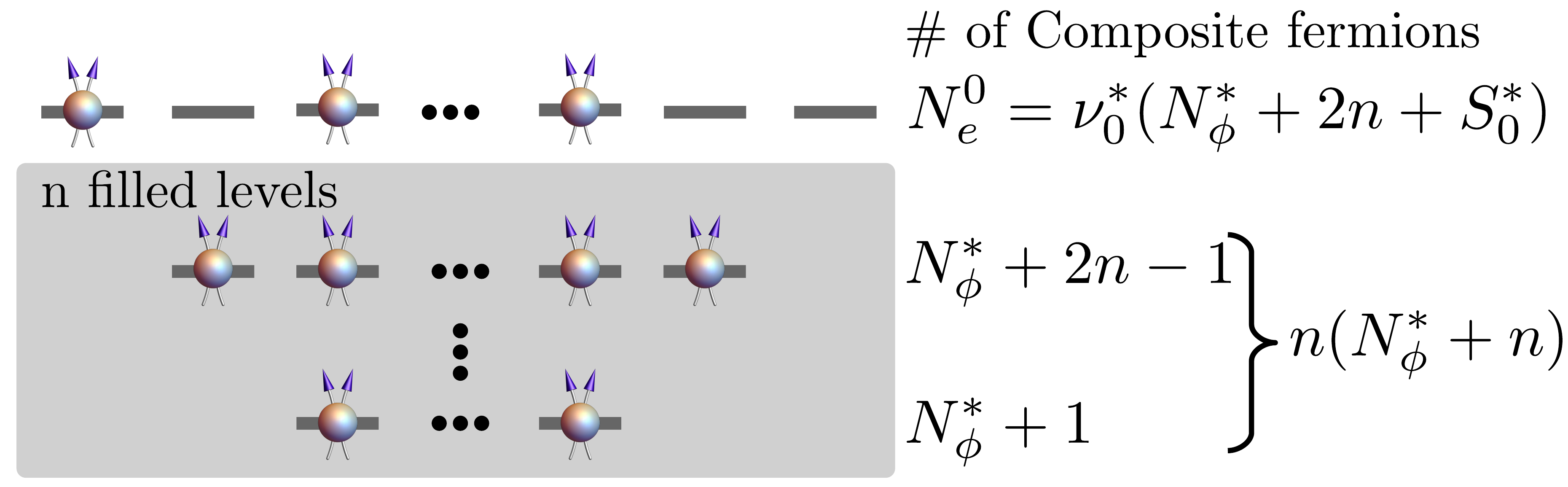}
    \caption{ Composite fermions at the effective filling factor $\nu^* = n + \nu_0^*$ form a state with $n$ fully filled CF Landau levels and a partially filled level at filling $\nu_0^*$. The latter contains $N_e^0 = \nu_0^*(N_\phi^* + 2n + S_0^*)$ CFs and realizes a FQH state characterized by the finite-size shift $S_0^*$. The remaining $n(N_\phi^* + n)$ CFs fully occupy the lower levels.
    }
    \label{fig.shift}
\end{figure}

\subsubsection{The shift of the next-generation states}
A simple way to determine the shift of NG states is by `counting', as illustrated in Fig.~\ref{fig.shift}. The partially occupied CF Landau level contains $|N_\phi^* +2n|+1$ orbitals. To form a FQH state with filling $\nu_0^*$ and shift $S_0^*$, one thus needs 
\begin{align}
  N_{e,0}=  \nu_0^*(N_\phi^* +2n+S_0^*)
\end{align} 
fermions. Another $N_{e,\text{filled}}=n(N_\phi^*+n)$ fermions are required to fill the $|n|$ lowest levels. The total number of particles $N_e=  N_{e,0}+N_{e,\text{filled}}$ satisfies

\begin{align}
    N_\phi^*=\frac{1}{\nu^*}N_e - \left(\frac{(n+S^*_0)\nu_0^*}{n+\nu_0^*}  + n\right).
\end{align}
This equation has the same form as Eq.~\eqref{eq.shift}, identifying the term in parentheses as the CF shift $S^*$. Finally, the electronic shift according to Eq.~\eqref{eq.cfshift} is given by 
\begin{align}
\label{eqn.nextgenshift}
    S=\left(\frac{(n+S^*_0)\nu_0^*}{n+\nu_0^*}  +n\right) +2p.
\end{align} 
For $n=0$, the shift reduces to the one given by Eq.~\eqref{eq.cfshift}.

Alternatively, we construct the shift vector of the NG state from the shift vector $\vect{s}^*_0$ of the $\nu^*_0$ state as
\begin{align}
    \vect{s}^* = (\vect{s}^*_0+n,\frac{1}{2},\frac{3}{2},\ldots,n-\frac{1}{2}).
\end{align}
The state $\nu_0^*$ forms in $n$th Landau level, hence its angular momentum is increased, which is accounted for by the offset of $\vect{s}^*_0$ by $n$. The equation Eq.~\eqref{eqn.nextgenshift} is reproduced by using this shift vector with Eq.~\eqref{eq.K_next_gen} and Eq.~\eqref{eq.flux_attachment.inverse}.

\begin{figure}
    \centering
    \includegraphics[width=0.95\linewidth]{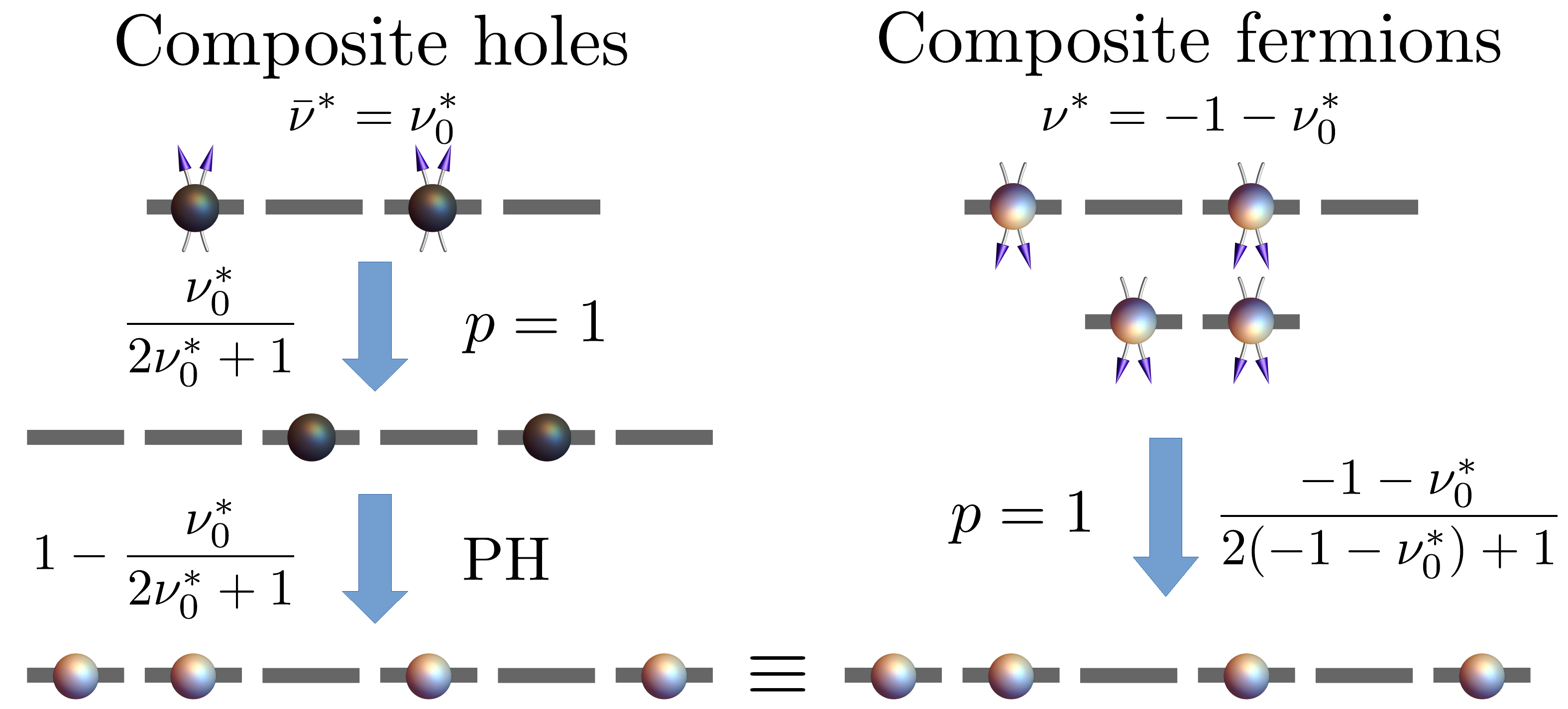}
    \caption{Equivalence between particle-hole conjugation of first-generation states and NG states. The particle-hole conjugate of a CF state at effective filling $\bar{\nu}^*=\nu_0^*$ can be equivalently described as a NG CF state at $\nu^* = -1 - \nu_0^*$, corresponding to one filled CF Landau level and a partially filled level at filling $\nu_0^*$ in the negative effective field ($N^*_\phi<0$). Both descriptions yield identical electronic filling factors, shifts, and $K$-matrices, and therefore represent the same topological phase.
    }
    \label{fig.equivalence}
\end{figure}

\subsubsection{Equivalence between next-generation states}\label{sec.equivalence} 
The same topological phase can be obtained via different CF prescriptions, as illustrated in Fig.~\ref{fig.equivalence}. In particular, NG states with one filled Landau level are describable as particle-hole conjugates of first-generation states.  For $p=1$ composite-holes at filling $\bar \nu^*=\nu_0^*$ correspond to holes at $\bar\nu = \frac{\nu_0^*}{2\nu_0^*+1}$ or electrons at
\begin{align}\label{eq.equiv}
    \nu = 1-\frac{\nu_0^*}{2\nu_0^*+1} = \frac{(-1-\nu_0^*)}{2(-1-\nu_0^*) +1}.
\end{align}
The r.h.s. of this equation describes CFs at filling $\nu^* = -(1+\nu_0^*)$; cf.~Eq.~\eqref{eqn.nuvfnu}. Notice that this analysis does not make any assumption on $\nu_0^*$; in particular, it remains unchanged for $\nu_0^*\to n+\nu_0^*$, corresponding to the NG states. Using Eqs.~\eqref{eq.phshift} and \eqref{eqn.nextgenshift} one readily confirms that the shifts of the $\bar\nu^*=\nu^*_0$ composite-hole and the $\nu^*=-1-\nu^*_0$ CF states also coincide. Finally, in Appendix~\ref{app.equiv}, we show that the two approaches also yield equivalent $K$-matrices.

This equivalence is well known for integer $\nu_0^*$, e.g., the $\frac{2}{3}$ Jain state can be equivalently obtained by CFs at $\nu^*=2$ or as the particle-hole conjugate of the $\frac{1}{3}$ Laughlin state. Similarly, the $\nu=\frac{3}{4}$ can be viewed as the hole-conjugate of $\nu=\frac{1}{4}$ or as a second-generation state with $\nu^*=-\frac{3}{2}$~\cite{Huang_non_Abelian_2024,Yutushui_phase_2025}. 

\section{Bonderson-Slingerland states}\label{sec.BS}
The NG even-denominator states with one filled CF Landau level occur at the same filling factor as the \textit{bosonic} Jain states with three filled CF Landau levels, i.e., $\nu=\frac{m}{(2q-1)m + 1}$ with $m=3$
. For electrons at such fillings,  Bonderson and Slingerland~\cite{Bonderson_hierarchy_2008,Bonderson_Competing_2012} introduced a family of states that can be understood most transparently within a `parton' construction where the electron operator is split as $c=bf$ into a boson $b$ and a fermion $f$. The bosons carry the entire electron charge and form a FQH state at filling $\nu$, while the fermions $f$ are neutral and form a paired superfluid. 

The $K$-matrix of the bosonic Jain state is obtained by attaching an \textit{odd} number $2q-1$ of fluxes to a $\nu^*=m$ integer quantum Hall state with  $K^*=\text{sgn}(m) \mathbb{I}_{|m|}$ and $\vect t=(1,1,\ldots)$, i.e., 
\begin{align}
    K = K^* + (2q-1)\vect{t}\vect{t}^T.
\end{align}
To obtain the anyon content of the BS states, the Abelian quasiparticles described by this $K$ need to be coupled to the fields $\mathbb{I},\gamma,\sigma$ of an Ising conformal field theory as prescribed above for the paired CF states.  For $m=1$, this construction reduces to the paired states at $\nu=\frac{1}{2q}$. Different pairing channels of the fermions $f$ induce distinct topological orders with the shift $S_\text{BS} = m+2q-1+\ell$ for odd $\ell$. 

For $m=\mp3$ with $q=3$, these states are competing with the NG states at $\nu=\frac{3}{8}$ and $\frac{3}{10}$. Remarkably, we find that the BS states with $\ell_\text{BS}$ Majoranas represent the same intrinsic topological order as the NG state with $\ell_\text{NG} = \pm \ell_\text{BS} -6$; see Appendix~\ref{app.Hartree}. Despite featuring anyons with the same electric charge and statistics, the NG and BS states represent different phases in rotationally invariant systems where the anyon spins are additional conserved quantum numbers. This distinction is directly apparent in the shift, which only coincides for one pairing channel, $\ell_\text{NG}=\mp\ell_\text{BS}=-3$. Different shifts for states with equal $\sigma_{xy}$ and $\kappa_{xy}$ also distinguish spin-polarized and unpolarized $\nu=\frac{2}{3}$ or $\nu=\frac{2}{5}$ states~\cite{Wen_shift_1992}. However, the case of BS and NG states is, to the best of our knowledge, the first example where both states are single-component, i.e., fully polarized.

\section{Edge theory and T-stability}\label{sec.edge_theory}

The edge of an Abelia FQH state with the topological order $(K,\vect{t})$ hosts $c=\text{dim}(K)$ boson modes $\phi_I$ governed by the Lagrangian density
\begin{align}\label{eq.LK}
     {\cal L}_{K}  = \frac{1}{4\pi}\sum^c_{I,J} \partial_x\phi_I(K_{IJ}\partial_t-V_{IJ}\partial_x)\phi_J\,,
\end{align}
where $V_{IJ}$ is a positive definite non-universal matrix describing the velocity of the edge modes and their density-density interactions. The number of boson modes, $c$, is the central charge of the edge theory.  A quasiparticle parametrized by an integer-valued vector $\vect{M}$,  in an operator formalism, is created by the vertex operator $\chi_{\vect{M}}\propto e^{i\vect{M}^T\vect{\phi}}$.

The edges of paired FQH states additionally host a number $|\ell|$ of Majorana fermions described by
\begin{align}
    {\cal L}_{\ell} = i\sum_{k=1}^{|\ell|} \gamma_k(\partial_t-\text{sgn}(\ell)v_k\partial_x)\gamma_k,
    \label{eqn.Majoranas}
\end{align}
where the sign of $\ell$ determines the chirality of the Majorana fermions. Each Majorana is associated with a central charge $\frac{1}{2}$, and the contribution of all Majoranas to the thermal Hall conductance is $\frac{\ell}{2}\kappa_0$.

To analyze cases with Majorana modes in a unified framework as used for Abelian edges, we bosonize pairs of Majoranas. For $\ell=\pm (2r+1)$ or $\pm 2r$ Majoranas, we can combine $r$ pairs of Majoranas into complex fermions  $e^{i\varphi_I}\propto \gamma_{2I-1} + i\gamma_{2I}$ with the Lagrangian density given by Eq.~\eqref{eq.LK} with $K_r=\text{sgn}(\ell) \mathbb{I}_r$. For even $\ell$, the edge is thus described by Eq.~\eqref{eq.LK} with $\tilde{K}=\diag(K,K_r)$ and $\tilde{\vect{t}}=(\vect{t},\vect{0})$. For odd $\ell$, a single Majorana fermion remains in addition to the bosonic modes, i.e., the edge theory is ${\cal L}_{\tilde{K}} + {\cal L}_{\text{sgn}(\ell)}$.

\subsection{Interface physics}
 The interface between two Abelia FQH states with $(K_\text{A},\vect{t}_\text{A})$ and $(K_\text{B},\vect{t}_\text{B})$ is governed by Eq.~\eqref{eq.LK} with
\begin{align}
    K=\begin{pmatrix}
        K_\text{A}&0\\0&-K_\text{B}
    \end{pmatrix},\qquad \vect{t}=(\vect{t}_\text{A},\;\vect{t}_\text{B}).
\end{align}
Electrons can tunnel between the two edges via 
\begin{align}\label{eq.tun}
    {\cal L}_\text{tun} \propto  e^{i\vect{L}^T\cdotnone K\cdotnone \vect{\phi}}+\text{H.c.} ,
\end{align}
where $\vect{L}$ is an integer-valued vector with $\vect{L}^T \cdotnone \vect {t}=0$. When relevant, these tunneling processes may drive the system to a fixed point similar to the one discovered by Kane-Fisher-Polchinski~\cite{Kane_Impurity_1995}. At such disorder-dominated fixed points, interactions between the total charge density at the edge, $\rho_{c} \propto \vect t^T\partial_x \vect \phi$, and the densities of neutral excitations are irrelevant~\cite{Kane_Impurity_1995,Moore_Classification_1997,Moore_Critical_2002}.  For certain $K,\vect t$, the processes described by Eq.~\eqref{eq.tun} may lead to localization, which reduces the number of long-wavelength modes. Interfaces where such a localization can occur are referred to as topologically unstable. 

\subsection{Topological stability}
For Abelian quantum Hall states, the criterion for topological stability was found by Haldane in Ref.~\onlinecite{Haldane_Stability_1995}. The interface is topologically unstable if there exists a local tunneling process $e^{i\vect{M}^T\cdotnone\vect{\phi}}$ with an integer-valued vector $\vect{M}$ satisfying 
\begin{equation}
\begin{split}
   &\text{(i) Charge neutrality }\vect{t}^T\cdotnone K^{-1}\cdotnone\vect{M}=0~,\\
      &\text{(ii) Trivial self-statistics }\vect{M}^T\cdotnone K^{-1}\cdotnone\vect{M}=0~.
    \end{split}
    \label{eqn.haldantstability}
\end{equation}

An integer-valued vector $\vect{M}$ that satisfies (i) and (ii) is called a null vector. The stability criterion can be reformulated as a condition for electron tunneling $e^{i\vect{L}\cdotnone K\cdotnone \vect{\phi}}$.  For unstable interfaces, there is an integer-valued vector $\vect L$ satisfying
\begin{equation}
\begin{split}
   &\text{(I) Charge neutrality }\vect{t}^T\cdotnone \vect{L}=0~,\\
      &\text{(II) Trivial self-statistics }\vect{L}^T\cdotnone K \cdotnone \vect{L}=0~.
    \end{split}
\end{equation}
To confirm the equivalence of the two conditions, we note that $\vect M = K\cdotnone \vect{L}$ is a null vector if $\vect L$ obeys (I),(II). Conversely, for any null-vector $\vect M$, the integer vector $\vect{L}=\text{adj}(K)\vect{M}\equiv \text{det}(K) K^{-1}\cdotnone\vect{M}$ satisfies (I),(II).

The analysis of topological stability is greatly simplified by the observation that flux attachment on both sides of the interfaces does not change topological stability. For instance, since the interfaces of $\nu^*_A=3$ and $\nu^*_B = 2$ is unstable, the interface of $\nu_A=\frac{3}{7}$ and $\nu_B=\frac{2}{5}$ is also unstable. This property can be understood by noting that the flux is attached proportionally to the charge. Since the electron tunneling between $\nu_A^*$ and $\nu_B^*$ is charge-neutral (condition i), no net flux is attached to the process. Hence, the anyon tunneling between $\nu_A$ and $\nu_B$ remains trivial self-statistic (condition ii).

\subsection{Topological stability after flux attachment}\label{sec.T_stab_flux}

We now explicitly prove that flux attachment does not affect topological stability. Our starting point is the CF interface described by $K^*= \diag(K_\text{A}^*,-K^*_\text{B})$ and $\vect{t} =(\vect{t}_\text{A},\vect{t}_\text{B})$.  We {\it separately} \footnote{Notice that off-diagonal terms of the $K$-matrix are zero, reflecting the fact that CFs of state A do not feel fluxes on CFs of state B and vice versa.} attach $p$ flux quanta to each CF in A and B, i.e.,
\begin{align}\label{eq.KAB}
    K = \begin{pmatrix}
        K^*_\text{A} + 2p\vect{t}_\text{A}\vect{t}^T_\text{A} &0^T \\0 &
        -K^*_\text{B} - 2p\vect{t}_\text{B}\vect{t}^T_\text{B}
    \end{pmatrix}.
\end{align}
 Under the assumption of topological instability of the CF interfaces, there exists a vector $\vect{L}=(\vect{L}_\text{A},\vect{L}_\text{B})$ that obeys (I) and (II) for $K^*$. We now show that the same $\vect{L}$ also satisfies (I) and (II) for $K$ in Eq.~\eqref{eq.KAB}. Since the flux attachment does not change the charge vector, the charge neutrality condition, $\vect{t}^T\cdotnone\vect{L}=0$, is trivially obeyed before and after transformation. The condition (II) for $K$ becomes
\begin{align}
    \vect{L}^T\cdotnone K \cdotnone\vect{L} =\vect{L}^T\cdotnone K^*\cdotnone \vect{L}  + 2p(\vect{t}^T_\text{A}\cdotnone \vect{L}_A)^2 - 2p (\vect{t}^T_\text{B} \cdotnone\vect{L}_B)^2=0.
\end{align}
The first term is zero by the assumption that $\vect{L}$ obeys (II) for $K^*$. The last two terms cancel each other since the tunneling process preserves charge $\vect{t}^T_\text{A}\cdotnone\vect{L}_A=-\vect{t}^T_\text{B}\cdotnone\vect{L}_B$ as follows from condition (I). 

We have thus shown that the topological stability of the interfaces remains unaffected by attaching the same number of fluxes to both states. The argument also holds for zero-dimensional $K_B$, i.e., the vacuum interface of $(K_A,\vect t_A)$. Consequently, if the vacuum interface of a state is unstable, it will remain unstable after flux attachment. 

\subsection{Topological stability with Majoranas} 
We proceed with the topological stability of states with Majoranas. The analysis of the $\ell=1$ case given by ${\cal L}_{K} + {\cal L}_{1}$ is sufficient, since pairs of Majoranas can be absorbed into the $K$-matrix~\footnote{The $\ell=-1$ case is equivalent to the present case with the extended $K$-matrix $\tilde{K}= \text{daig}(K,-1)$ and charge vector $(\vect{t},0)$.}. Here, we are only interested in the processes localizing the Majorana mode, and thus we assume there are no null vectors for $K$. To analyze topological stability similarly to the Abelian case, we add a pair of topologically trivial Majoranas, $\gamma_{R}$ and $\gamma_{L}$. We absorb the co-propagating $\gamma$ and $\gamma_R$ pair of Majoranas into the $K$-matrix, so that the edge theory is given by ${\cal L}_{\tilde{K}} + {\cal L}_{-1}$ with $\tilde{K}=\text{diag}(1,K)$ and $\tilde{\vect{t}}=(0,\vect{t})$.

If the interface described by $(\tilde{K},\tilde{\vect{t}})$ is topologically stable, then the only instability is due to the localization of $\gamma_{R}$ and $\gamma_{L}$. We thus conclude that the original theory ${\cal L}_{K} + {\cal L}_{1}$ is topologically stable. Conversely, if $(\tilde{K},\tilde{\vect{t}})$ is topologically unstable and $(K,\vect{t})$ is topological stable, the null vector is $\tilde{\vect{M}} = (g,\vect{M})$ with $g\neq0$. Thus, the original theory is topologically unstable since ${\cal L}_\text{tun}\propto e^{i\tilde{\vect{M}}^T\cdotnone\tilde{\vect{\phi}}}$ with $ e^{i\tilde{\phi}_1}$ replaced by $\gamma$ is also charge neutral and has trivial self-statistics.

We analyzed the topological stability in terms of $K$-matrices, and the derivation of the previous section applies here. Therefore, the flux attachment does not affect the topological stability of the edge with Majorana modes. Similarly, for non-Abelian states, the flux attachment does not change topological stability since the non-Abelian part is neutral~\footnote{The topological stability for non-Abelian edges has been analyzed for several examples~\cite{Lee_particle_hole_2007,Levin_particle_hole_2007, Bishara_PH_Read_Rezayi_2008,Yutushui_Identifying_2023}, but we are not aware of a generalization of Eq.~\eqref{eqn.haldantstability} for the non-Abelian case.}.

\begin{figure}[t]
 \centering
\includegraphics[width=0.9\linewidth]{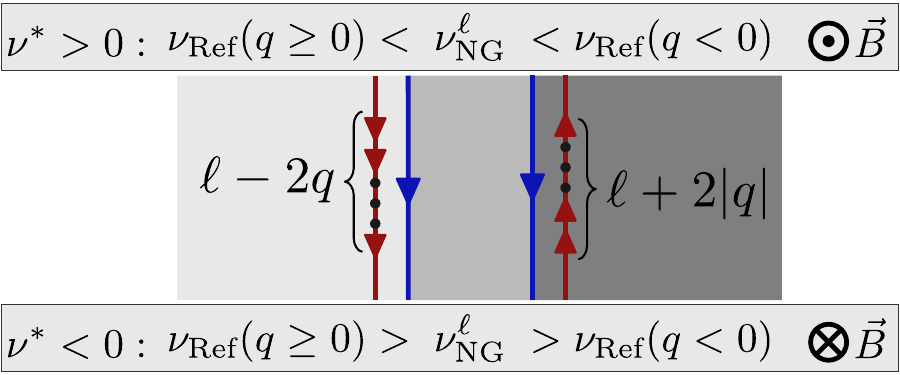}
 \caption{Edge structure at interfaces between NG paired state with pairing channel $\ell$ at filling $\nu^{\ell}_{\text{NG}}$and reference states at $\nu_{\text{Ref}}(q)$.  For the latter, we choose NG Jain states corresponding to $\nu^*_0=\frac{q}{2q+1}$ with positive or negative $q$. Both NG states have the same number $n$ of filled levels and the same number $2p$ of fluxes attached. The direction of the magnetic field is chosen to preserve the direction of the edge structure for any sign of $\nu^*$ as specified in the top ($\nu^*>0$) and bottom ($\nu^*<0$) panels.}
 \label{fig.interface_NG_nu_q} 
 \end{figure}  
 
\section{Experimental signatures}\label{sec.experimental}
In this section, we use the $K$-matrix formalism and the invariance of topological stability under flux attachment to derive the experimental signatures distinguishing different pairing channels of the NG states. We generalize several edge transport probes originally suggested for the $\nu=\frac{5}{2}$ plateau to NG states, i.e., thermal conductance, scaling of tunneling currents, upstream noise, and coherent charge conductance at mesoscopic interfaces. Since, in the absence of rotational symmetry, BS and NG states with odd $\ell_\text{NG}=\mp\ell_\text{BS} -6$ represent the same phases  (see Appendix~\ref{app.BS} and Appendix~\ref{app.Abelian} for Abelian states), our analysis is based on the NG states only.

The neutral sector of the NG states can be non-chiral, complicating their analysis compared to the half-filled case. To address this problem, we identify suitable reference states that reduce their edge structure. We show that the interfaces of any NG state at $\nu_\text{NG}(n,p)$ and the NG Jain state at $\nu_\text{Ref}(q)$ with the same $n$ and $p$ exhibit a single charge mode and $\ell-2q$ Majorana modes, see Fig.~\ref{fig.interface_NG_nu_q}. This result follows from Sec.~\ref{sec.T_stab_flux} and greatly simplifies the analysis of experimental signatures.

\begin{table}[htbp]
  \centering
  \renewcommand{\arraystretch}{1.5} % Increase the space between rows by 50%
  \caption{Quantum numbers and quasiparticle scaling dimensions at the most important first and second-generation even-denominator states with $n$ filled CF Landau levels and $2p$ attached fluxes. For the non-chiral $\nu=\frac{3}{4}$, $\frac{3}{10}$, and $\frac{5}{8}$ states, the scaling dimensions assume disordered edges, analyzed in Appendix~\ref{app.KFP_FP}.} 
   \begin{tabular}{c| c c c c c c c }
    \hline  \hline
         &  \multicolumn{7}{c}{Even-denominator states}\\ \hline
     $(n,2p)$ & $(0,0)$ &  $(0,2)$ & $(1,2)$ & $(-1,2)$ & $(-1,4)$ & $(2,2)$ & $(-2,2)$
    \\ \hline
     $\sigma_{xy}$ & $\frac{1}{2}$ &  $\frac{1}{4}$ & $\frac{3}{8}$ & $\frac{3}{4}$ & $\frac{3}{10}$ & $\frac{5}{12}$ & $\frac{5}{8}$ 
    \\ $S$ & $2+\ell$ & $4+\ell$ & $4+\frac{\ell}{3}$ & $-\frac{\ell}{3}$   & $2-\frac{\ell}{3}$  & $\frac{24+ \ell}{5}$  & $-\frac{4+ \ell}{5}$  \\
    $\kappa_{xy}/\kappa_0$   & $1+\frac{\ell}{2}$ & $1+\frac{\ell}{2}$ & $2+\frac{\ell}{2}$ & $-\frac{\ell}{2}$   & $-\frac{\ell}{2}$  & $3+\frac{\ell}{2}$  & $-1-\frac{\ell}{2}$ 
    \\ 
     $\Delta_e$   & $\frac{3}{2}$ & $\frac{5}{2}$& $\frac{3}{2}$  & $\frac{5}{6}$ & $\frac{11}{6}$& $\frac{3}{2}$ & $\frac{11}{10}$
    \\
    $\Delta_{\phi}$    & $\frac{1}{4}$ & $\frac{1}{8}$ & $\frac{3}{16}$ &$\frac{5}{24}$ &  $\frac{11}{60}$&  $\frac{5}{24}$ &  $\frac{17}{80}$
    \\
     $\Delta_{\phi/2}$   & $\frac{|\ell|+1}{16}$   & $\frac{2|\ell|+1}{32}$  & $\frac{4|\ell|+3}{64}$ & $\frac{6|\ell|+5}{96}$    & $\frac{15|\ell|+11}{240}$ & $\frac{6|\ell|+5}{96}$ &   $\frac{20|\ell|+17}{320}$
    \\ \hline\hline
  \end{tabular}
  \label{tab.thermal}
\end{table}

\subsection{Thermal Hall conductance}\label{sec.experimental.thermal}

The bosonic edge modes in  Eq.~\eqref{eq.LK} carry a thermal Hall conductance given by Eq.~\eqref{eq.kappa}, which we evaluate for NG states with the help of Eq.~\eqref{eqn.kappa.under.flux.attachment}. 
Chiral Majorana modes described by Eq.~\eqref{eqn.Majoranas} contribute an additional $\kappa_{xy}^\text{Majorana} = \ell \kappa_0/2$ to the thermal Hall conductance. In Tab.~\ref{tab.thermal}, we list the resulting thermal Hall conductances for the most relevant first and second-generation even-denominator states.

\subsection{Scaling dimensions of tunneling operators}\label{sec.experimental.scaling}
The scaling dimensions of quasiparticle operators determine how the tunneling current across quantum point contacts scales with temperature or voltage~\cite{Chang_Chiral_LL_2003,Lin_Measurements_2012,Baer_experimental_2014,Fu_Competing_2016,Radu_quasi_particle_2008}. If tunneling is dominated by quasiparticles with the scaling dimension $\Delta$, the tunneling conductance scales as $G\propto T^{4\Delta-2}$. 

For half-filled states, Ref.~\onlinecite{Ken_sixteenfold_2019} found that, depending on the pairing channel $\ell$, the most relevant excitations are either the fundamental or flux quasiparticles with the charges $q_{\phi/2}=\frac{e}{4}$ and $q_{\phi}=\frac{e}{2}$, respectively. In the context of paired states, these quasiparticles correspond to insertions of half-vortices or vortices into the superfluid. For the NG states, we adopt a naming convention according to the superfluid interpretation. The fundamental quasiparticle corresponds to a half-vortex and has charge $q_{\phi/2}=\frac{e}{4(p(2n+1)+1)}$. The charge of the `flux'  quasiparticle is given by the denominator of the filling factor, i.e., $q_{\phi}= \frac{e}{2(p(2n+1)+1)}$.  This quasiparticle should not be confused with the Laughlin quasiparticle, whose charge is $q_\text{Laughlin}=\nu e$, and that corresponds to an actual insertion of one magnetic flux quantum. (They coincide only when the numerator of $\nu$ is unity, i.e., for the first-generation particle-like states .)

In Tab.~\ref{tab.thermal}, we list the scaling dimensions of the fundamental and the flux quasiparticles for different pairing channels. Their scaling dimensions $\Delta_{\phi/2}$ and $\Delta_\phi$ govern the tunneling conductance in the weak backscattering limit, where charge tunnels across the fractionally filled region. In the strong backscattering limit, only electrons tunnel across vacuum, and the conductance scaling is governed by $\Delta_e$. 

For the states with chiral bosonic sectors, i.e., for $p>0$, the scaling dimensions are universal and do not depend on the density-density interactions. For $p<0$, we assume that random tunneling at a disordered interface drives the system to the fixed point where charge and neutral modes decouple~\cite{Kane_Impurity_1995} and compute the scaling dimensions in Appendix~\ref{app.KFP_FP}. In both $p>0$ and $p<0$ cases, the Laughlin quasiparticle excites only the decoupled charge mode, and, thus, its scaling dimension is universally given by $\Delta_\text{Laughlin} = \frac{\nu}{2}$. Only for the $\nu=\frac{3}{10}$ state, the Laughlin quasiparticle is more relevant than the flux quasiparticle.

\subsection{Upstream noise}\label{sec.experimental.upstream}
Upstream noise measurements provide an additional path for distinguishing between different pairing channels. In such experiments, a contact is placed upstream of the voltage source. An injected current creates a hot spot on the source contact's backside, which excites any modes that are present there. If there are upstream neutral modes, they carry this heat to the upstream contact, where it is measured as charge noise. Observing such noise thus indicates the presence of upstream modes; conversely, there is no noise in the case of chiral edges.  

Ref.~\onlinecite{Yutushui_Identifying_2023} derived the noise signatures of half-filled states interfaced with Jain states and showed that any non-Abelian pairing channel can be identified by measuring upstream noise on such interfaces. In particular, the interface of the $\nu=\frac{1}{2}$ paired state with $\ell$ Majoranas and the Jain state at $\nu_q=\frac{q}{2q+1}$ reduces to a single charge mode and $\ell-2q$ Majoranas. 

Since flux attachment does not affect topological stability, the findings of 
Ref.~\onlinecite{Yutushui_Identifying_2023} directly generalize to NG paired states interfaced with NG Jain states. In Tab.~\ref{tab.NoNoise_condition}, we list the filling factors of the corresponding reference states and the noise signatures for $\nu=\frac{3}{8}$ and $\frac{3}{10}$ states.

\begin{table}[htbp]
  \centering\renewcommand{\arraystretch}{1.5}
  \caption{ The noise signatures of interfaces between $\nu_\text{NG}=\frac{\nu^*}{2p\nu^*+1}$ and $\nu_\text{Ref}(q) = \frac{\nu^*_q}{2p\nu^*_q+1}$, where $\nu^*=n+\frac{1}{2}$ with $\ell$-wave pairing of CFs in the half-filled state, and $\nu^*_q = n+\frac{q}{2q+1}$.}
\begin{center}
\setlength\tabcolsep{5pt}
  \begin{tabularx}{\linewidth}{c c | c | c c c c c }
    \hline \hline
    $\frac{3}{8}$ & $\frac{3}{10}$ & &  \multicolumn{5}{c}{Pairing channel}
    \\
   \multicolumn{2}{c|}{$\nu_\text{Ref}(q)$} & $q$ & $\ell=-5$ & $\ell=-3$ & $\ell=-1$ & $\ell=1$ & $\ell=3$ 
    \\\hline
   $\frac{5}{13}$ & $\frac{5}{17}$ & $-2$ & Chiral & Noise & Noise & Noise & Noise 
    \\
   $\frac{2}{5}$ & $\frac{2}{7}$ & $-1$ &Chiral & Chiral & Noise & Noise & Noise 
    \\
    $\frac{1}{3}$ & $\frac{1}{3}$ & 0 & Noise & Noise & Noise & Chiral & Chiral 
    \\
    $\frac{4}{11}$ & $\frac{4}{13}$ & 1 & Noise & Noise & Noise & Noise & Chiral 
    \\ \hline \hline
  \end{tabularx}
  \end{center}

  \label{tab.NoNoise_condition}
\end{table}

\subsection{Coherent conductance}\label{sec.experimental.conductance}
\begin{figure}[t]
 \centering
\includegraphics[width=0.49\linewidth]{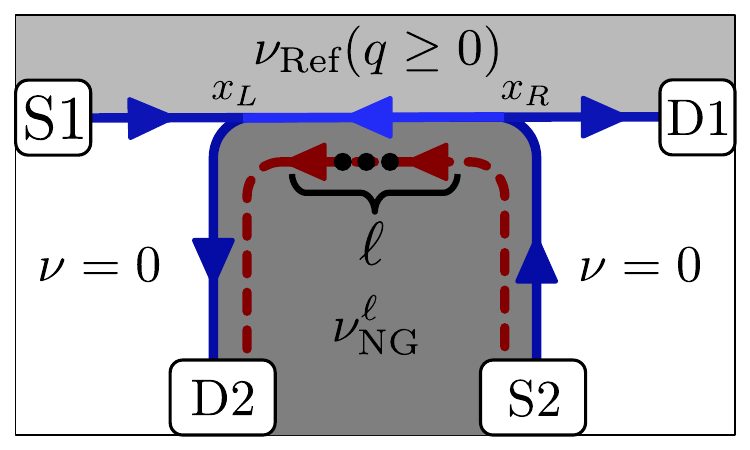} 
 \caption{The $\pi$-junction setup for NG even-denominator states. The blue lines indicate the net charge flow, while red dashed lines denote Majorana modes. 
 }
 \label{fig.Pi_junction} 
 \end{figure}

Refs.~\onlinecite{Yutushui_Identifying_2022,Yutushui_Universal_2025} introduced a $\pi$-geometry setup, Fig.~\ref{fig.Pi_junction}, that identifies the pairing channels of half-filled states via coherent charge conductance. When the transport across the junction $x_L-x_R$ is coherent, i.e., the interface length is shorter than the decoherence length, $|x_L-x_R|\gg L_\text{decoh}$, the (partial or complete) equilibration can be achieved if the counter-propagating modes at the interface localize.  In the window of length scales $\xi_\text{loc}\gg |x_L-x_R|\gg L_\text{decoh}$, the coherent charge conductance value depends on the number of localized Majorana modes and thus reveals the pairing channel.

It is straightforward to generalize this setup to identify pairing channels of any NG even-denominator states; see Fig.~\ref{fig.Pi_junction}. As in the discussion of upstream noise, we interface the NG paired state with a NG Jain state with the same $n$ and $p$. Repeating the analysis of Ref.~\onlinecite{Yutushui_Universal_2025}, we find that for $q\geq0$ and $p>0$, the conductance between S2 and D1 is given by
\begin{align}
    G_\text{S2$\to$D1} = \nu_{m(\ell)}(n,p),
\end{align}
where $m(\ell) = \min(\lfloor{\frac{\ell+1}{2}}\rfloor,q)$ is the number of neutral currents backscattered by Majorana fermions, and $\nu_{m(\ell)}(n,p)$ is given by Eq.~\eqref{eq.filling.ng} with $\nu_0^*=\frac{m(\ell)}{2m(\ell)+1}$. For convenience, we list the coherent conductance values for $\nu=\frac{3}{4}$, $\frac{3}{8}$, and $\frac{3}{10}$ in Appendix~\ref{app.coherent_conductnace}.

\section{Numerics at next-generation even denominator fillings}\label{sec.numerics}
In this section, we introduce trial wave functions for the NG paired states and numerically study them using the methods introduced in Ref.~\onlinecite{Yutushui_phase_2025}. We construct wave functions of NG paired states and BS states at $\nu=\frac{3}{8}$ and $\frac{3}{10}$ for three pairing channels: Moore-Read, anti-Pfaffian, and $f$-wave. We compare these states with the NG composite Fermi liquid and charge density wave states represented by Wigner crystals and bubble states at the same filling factor~\cite{Goerbig_Competition_2004,Fogler_Laughlin_wigner_1997}.

\subsection{First-generation wave functions}
Given a wave function of a CF state $\Psi_\text{CF}$ in the spherical geometry, one obtains the electron wave function via 
\begin{align}
    \Psi_\text{e} = P_\text{LLL}\Psi_\text{CF} \prod_{i<j}\omega_{ij}^{2p},
\end{align}
where the relative position $\omega_{ij}=u_iv_j-u_jv_i$ is defined in terms of Haldane coordinates on a sphere, $(u,v)=(\cos\frac{\theta}{2}e^{i\frac{\varphi}{2}},\sin\frac{\theta}{2}e^{-i\frac{\varphi}{2}})$, and $P_\text{LLL}$ denotes projection into the lowest Landau level. In the case of the Jain states at $\nu=\frac{q}{2pq+1}$, the CF wave function is a single Slater determinant $\Psi_\text{CF} = \chi_q$ of the first $q$ fully filled Landau levels. (For $q<0$, we take the complex conjugate wave function $\Psi_\text{CF} = \chi^*_{|q|}$.)

For the case of a non-Abelian superfluid, the wave function is 
\begin{align}\label{eq.wf_paired}
     \Psi^\text{Paired}_\text{CF} = \text{Pf}\left[\frac{1}{\omega_{ij}}\left(\frac{\omega_{ij}^*}{\omega_{ij}}\right)^r\right],
\end{align}
where $r$ is an integer encoding the pairing $\ell=2r+1$ of the superfluid. For $\ell=1$, the resulting wave function $\Psi_e$ is the celebrated Moore-Read state~\cite{Moore_nonabelions_1991}. For $\ell=-3$, the wave function represents the anti-Pfaffian topological order~\cite{Yutushui_Large_scale_2020,Henderson_Energy_2023}. The $f$-wave pairing corresponds to $\ell=3$~\cite{Wen_Non-Abelian_1991}.

\begin{figure}
    \centering
    \includegraphics[width=0.95\linewidth]{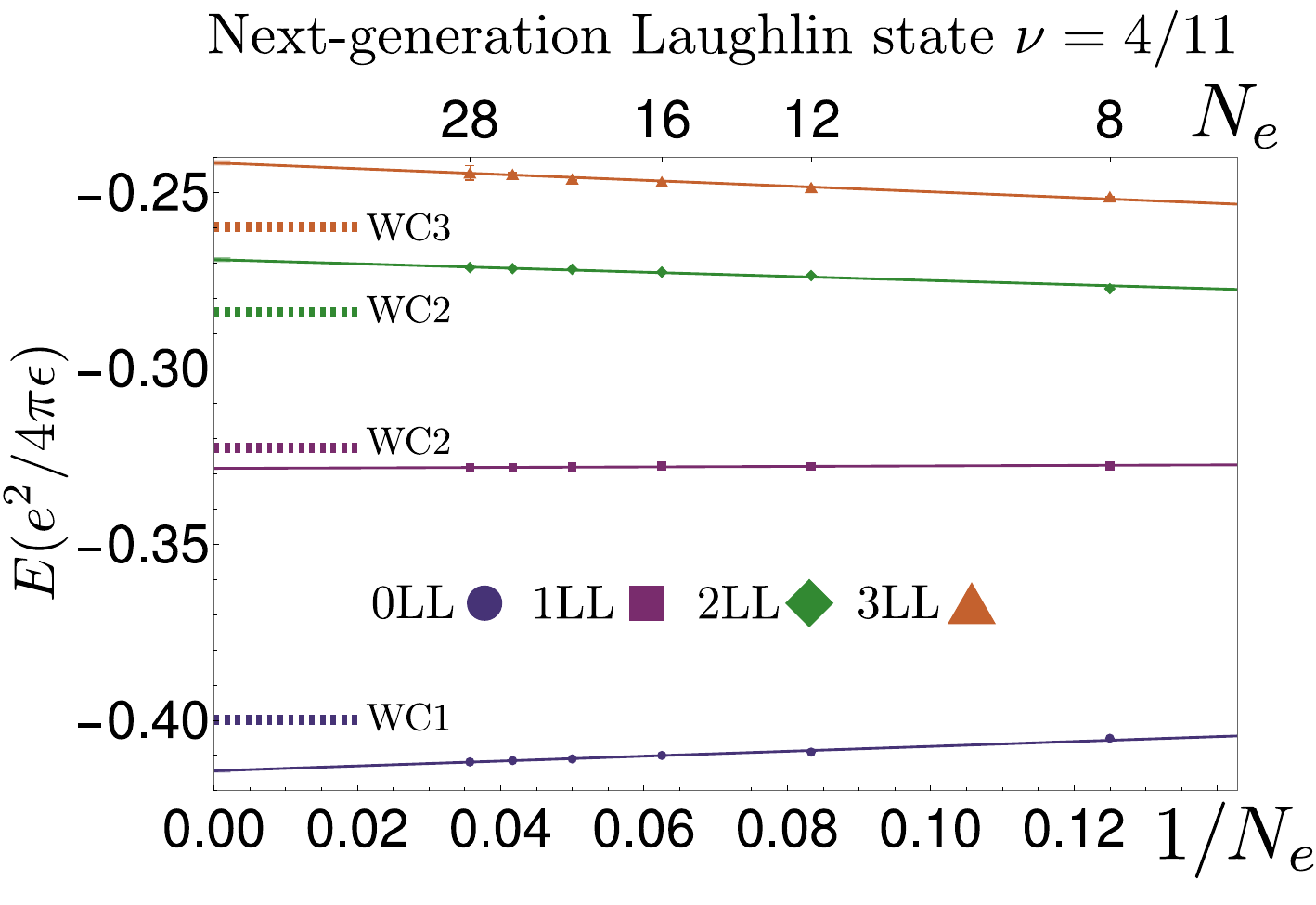}
    \caption{The Coulomb energy of the NG Laughlin state at $\nu=\frac{4}{11}$ in the four lowest LLs. The energies of the optimal triangular Wigner crystals are indicated in matching colors. WC$M$ denotes a crystal of $M$-electron bubbles, which we find to have the lowest energy at $\nu=\frac{4}{11}$ in the $(M-1)$th LL. }
    \label{fig.NG411_energy}
\end{figure}

\subsection{Next-generation wave functions}

Following Refs.~\onlinecite{Mukherjee_Possible_2012,Mukherjee_incompressible_2015}, we construct the $N_e$ electron wave function $\Psi_\text{NG}$ describing CFs at $\nu^*=n+\nu_0^*$ in three steps. First, we build a second quantized representation of $\Psi_{\nu_0^*}$ at filling $\nu_0^*$ and with shift $S_0^*$. The required number of particles $N_{e,0}$ in $\Psi_{\nu_0^*}$ is given by 
\begin{align}
   N_{e,0}=\frac{\nu_0^*(N_e + n(n+S_0^*))}{n + \nu_0^*}.
\end{align}
The $N_{e,0}$-particle Hilbert space is spanned by Slater determinants ${\cal S}^0_I = \det [Y^{q+2n}_{0,m_{I,a}}(\vec r_b)]$ of monopole harmonics \cite{Wu_dirac_1976,Wu_properties_1977}, with $\vect m_{I}\in\{m_{I,1},\ldots, m_{I,N_{e,0}}\}$. We expand $\Psi_{\nu_0^*}$ in this basis according to
\begin{align}
    \Psi_{\nu_0^*} = \sum_{I} C_{I} {\cal S}^0_I~,\qquad C_I = \langle \Psi_{\nu_0^*}|{\cal S}^0_I\rangle.\label{eqn.ng1}
\end{align}

To construct the state of $N_e$ particle at $\nu^*=n+\nu_0^*$, we promote the lowest Landau level Slater determinant ${\cal S}^0_I$ to the $n$th Landau level by replacing  $Y^{q+2n}_{0,m_{I,a}} \rightarrow Y^{q}_{n,m_{I,a}}$ and including all orbitals corresponding to the $n$ filled levels. The resulting wave function at $\nu^*$ is
\begin{align}
    \Psi_{\nu^*} = \sum_{I} C_{I} {\cal S}^n_I,
\end{align}
with the same coefficients as in Eq.~\eqref{eqn.ng1}. Finally, we attach $2p$ fluxes and project the wave function to the lowest Landau level as
\begin{align}\label{eq.NG_WF}
    \Psi_{\nu} = P_\text{LLL} \Psi_{\nu^*} \prod_{i<j}\omega_{i,j}^{2p} \approx \sum_{I} C_{I} {\cal S}^{n,p}_I,
\end{align}
where ${\cal S}^{n,p}_I$ is obtained from ${\cal S}^{n}_I$ by replacing $Y^{q}_{k,m}\to {\cal Y}^{q}_{k,m}$ with CF orbitals defined as 
\begin{align}\label{eq.ng_wf}
 {\cal Y}^{q}_{k,m} = P_\text{LLL} Y^q_{k,m}(r_i)\prod_{i<j}\omega_{i,j}^p~.
\end{align}

The coefficients $C_I$ can be obtained analytically via Jack polynomials in certain cases, e.g., for Laughlin or Moore-Read states. Alternatively, they can be computed via Monte-Carlo sampling for known real-space wave functions, or by numerically diagonalizing a chosen Hamiltonian. We employ all three methods for the different states.

\subsection{Bonderson-Slingerland wave functions}
The BS states at $\nu_\text{BS}=\frac{m}{(2q-1)m+1}$ are described by the wave functions~\cite{Bonderson_hierarchy_2008}
\begin{align}
    \Psi_{\text{BS}-\nu_\text{BS}} = P_\text{LLL}\left[
  \Psi^\text{Paired}_\text{CF}\chi_{m}\chi_1^{2q-1}
    \right],
\end{align}
where $\Psi^\text{Paired}_\text{CF}$ is given by Eq.~\eqref{eq.wf_paired}. We note that the topological order of BS states with $m=2$ can differ from the expected Ising order~\cite{Yutushui_Non_Abelian_2025}, but this complication does not arise at NG even-denominator fillings where $m\geq 3$ is odd.

Specifically, for $\nu=\frac{3}{8}$, we take $m=-3$ and $q=2$, which results in the wave function 
\begin{align}
    \Psi_{\text{BS}-\frac{3}{8}} = P_\text{LLL}\left[
  \Psi^\text{Paired}_\text{CF}\chi_{-3}\chi_1^{3}
    \right],
\end{align}
with $S=\ell_\text{BS}$. We project this wave function by separately acting with $P_\text{LLL}$ on the Pfaffian and $\chi_{-3}$ parts. Explicitly, the projected wave function is given by
\begin{align}
    \Psi_{\text{BS}-\frac{3}{8}} = \Psi_{\frac{1}{2}}\Psi_{\frac{3}{5}}/\chi_1,
\end{align}
where $\Psi_{\frac{1}{2}}=P_\text{LLL}\text{Pf}[\ldots]\chi_1^2$ is the wave function of the half-filled paired state, and $\Psi_{\frac{3}{5}}=P_\text{LLL}\chi_{-3}\chi_1^2$ is the Jain-$\frac{3}{5}$ wave function.

For the $\nu=\frac{3}{10}$ case, we take $m=3$ and $q=2$, and obtain the wave function 
\begin{align}
    \Psi_{\text{BS}-\frac{3}{10}} = P_\text{LLL}\left[
    \Psi^\text{Paired}_\text{CF} \chi_{3}\chi_1^{3}
    \right],
\end{align}
with the shift $S = 6+\ell_\text{BS}$. The projection is performed as
\begin{align}
    \Psi_{\text{BS}-\frac{3}{10}} = \Psi_{\frac{1}{2}}\Psi_{\frac{3}{7}}/\chi_1,
\end{align}
where $\Psi_{\frac{3}{7}}=P_\text{LLL}\chi_{3}\chi_1^2$ is the Jain $\nu=\frac{3}{7}$ wave function.

\begin{figure}[t]
    \centering
    \includegraphics[width=1\linewidth]{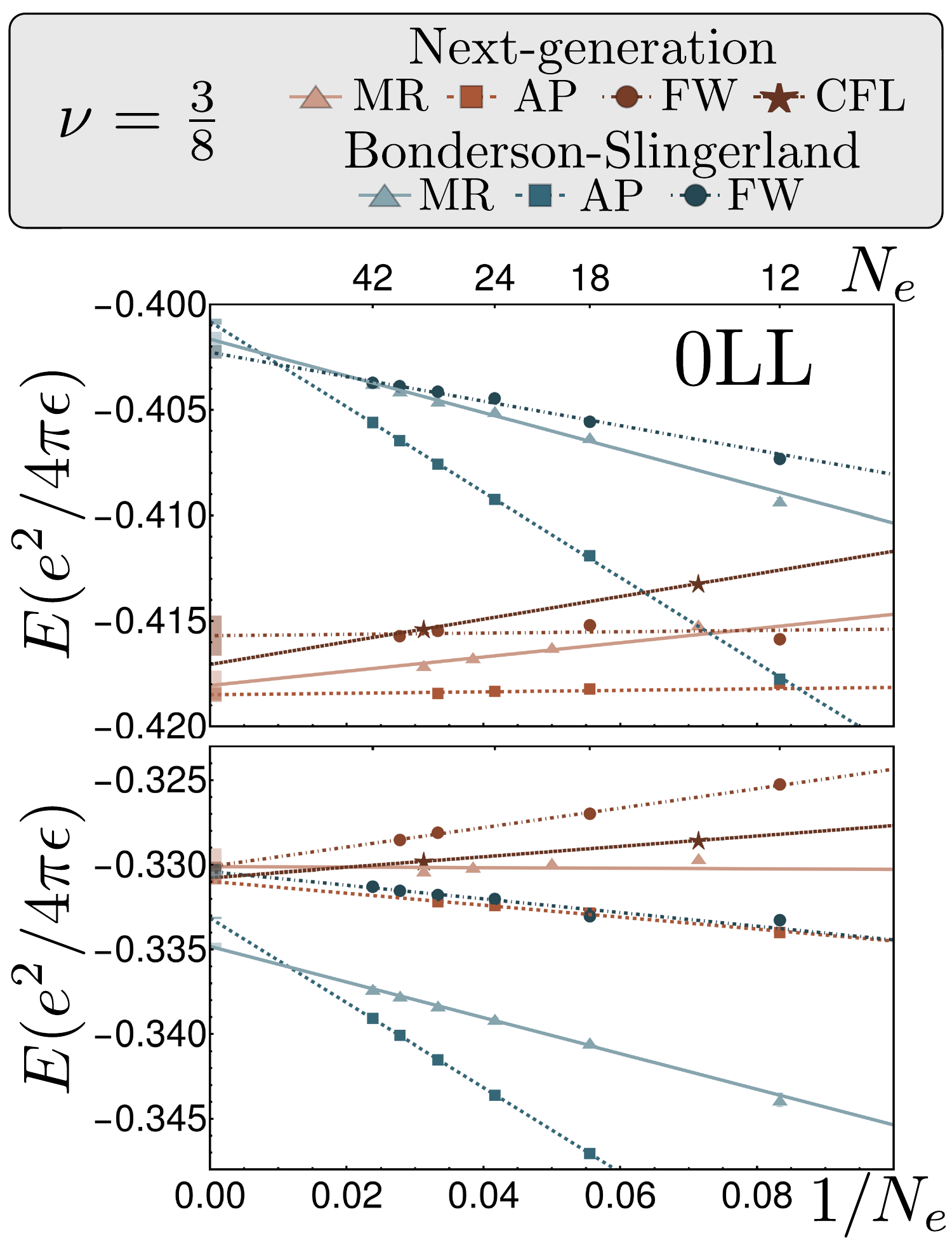}
    \caption{Coulomb energies of different trial states at $\nu=\frac{3}{8}$ in the 0LL (top) and 1LL (bottom). Charge density wave states have higher energy than paired states in the 0LL and 1LL phases. In the 2LL (not shown), Wigner crystals with $M=3$ electron bubbles are favored.}
    \label{fig.Coul_38}
\end{figure}

\begin{figure}[t]
    \centering
    \includegraphics[width=1\linewidth]{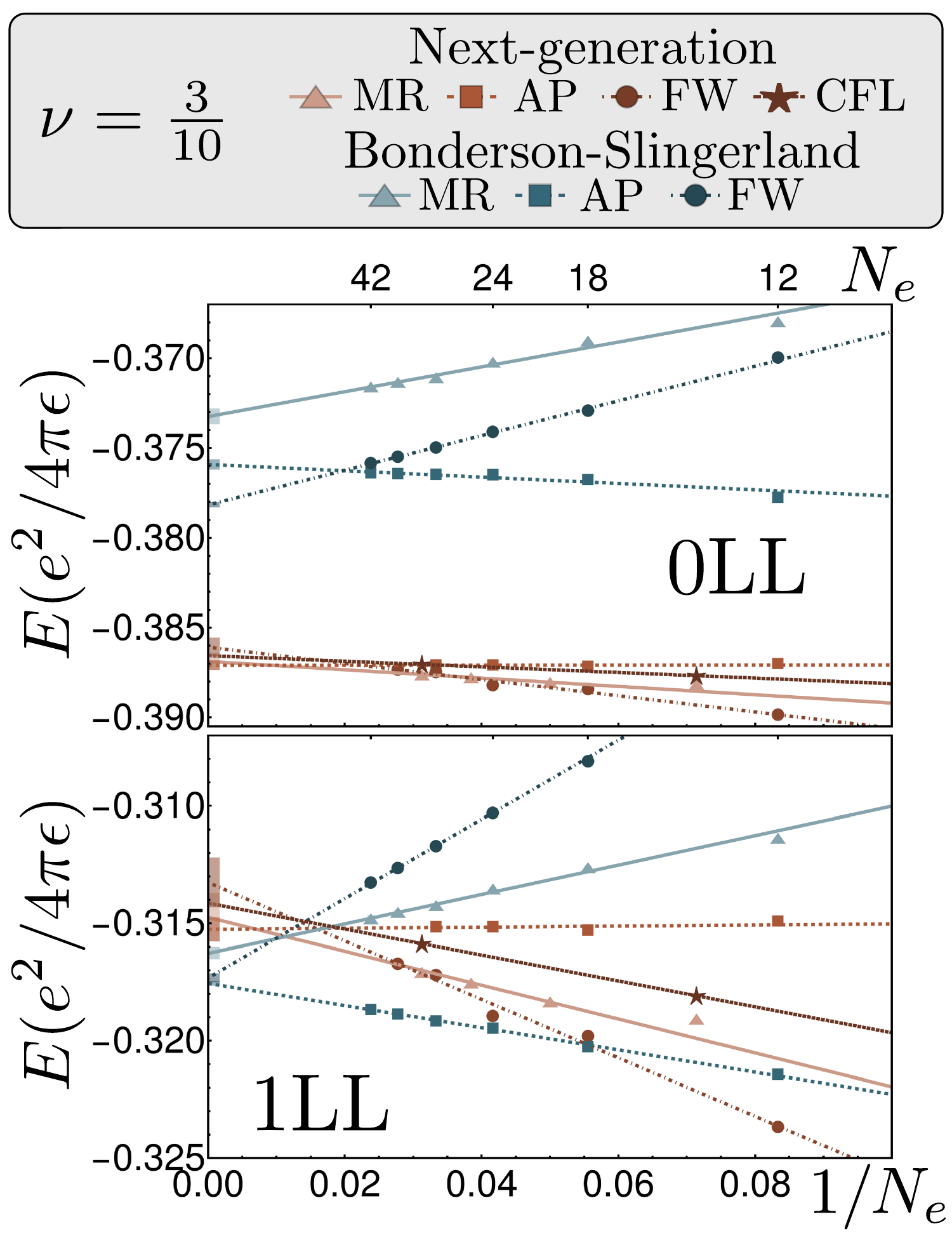}
  \caption{Coulomb energies of different trial states at $\nu=\frac{3}{10}$ in the 0LL (top) and 1LL (bottom). The behavior of charge density wave states is qualitatively the same as in the $\nu=\frac{3}{8}$ case.}
    \label{fig.Coul_310}
\end{figure}

\subsection{Wigner crystals}

At any fractional filling, charge density-wave states compete against FQH liquids. We thus compare their energies to those of the NG and BS states. Specifically, we analyze triangular Wigner crystals formed from $M$-electron bubbles. The Coulomb energies of these states are readily computed with the Hartree-Fock approximation following Ref.~\onlinecite{Goerbig_Competition_2004}.

\subsection{Benchmark: NG Laughlin state at $\nu=\frac{4}{11}$}
As a test case, we study thermodynamic energies of the NG state at $\nu=\frac{4}{11}$ obtained from the Laughlin state in the partially filled first Landau level, i.e., $\nu^*=1+\frac{1}{3}$. The corresponding wave function in the form of Eq.~\eqref{eq.NG_WF} occurs at the shift $S=4$. We numerically compute the static structure factor for up to $N_e=28$ electrons, where $N_{e,0}=8$ CFs form a Laughlin state with just 8512 basis states. Using the technique introduced in Ref.~\onlinecite{Yutushui_phase_2025}, we compute the Coulomb energies in different Landau levels and extrapolate to the thermodynamic limit in Fig.~\ref{fig.NG411_energy}. The thermodynamic energy in the 0LL agrees with previously reported results in Ref.~\onlinecite{Mukherjee_Enigmatic_2014}.

We compare the extrapolated energies of the NG Laughlin state in different Landau levels with the energies of triangular Wigner crystals formed out of $M$ electron bubbles~\cite{Goerbig_Competition_2004}. The energy of the Wigner crystal in a given Landau level is indicated and denoted by WC$M$, where $M$ corresponds to the number of electrons in the bubble that minimizes the energy. We find that the NG Laughlin state is favored in the lowest and first excited Landau levels, while Wigner crystals have lower energies in higher Landau levels. 

\subsection{Next-generation paired states }
For the NG paired states, we obtain the coefficients $C_I$ appearing in Eqs.~\eqref{eqn.ng1} using Jack polynomials for Moore-Read and anti-Pfaffian pairings~\footnote{We generate the exact coefficients using the DiagHam library~\cite{DiagHam}.} and performing Monte Carlo simulations for the $f$-wave case. For the composite Fermi liquid (CFL), we use the Coulomb ground state of the 0LL. In all cases, we then sample the resulting NG states, Eq.~\eqref{eq.NG_WF}, via Monte Carlo techniques to compute the static structure factors for different system sizes. 

\subsubsection{Coulomb interactions in GaAs}
The Coulomb energies in the 0LL and 1LL for the $\nu=\frac{3}{8}$ and $\nu=\frac{3}{10}$ states are shown in Fig.~\ref{fig.Coul_38} and Fig.~\ref{fig.Coul_310}, respectively. The thermodynamic energies of all four NG states are very close and do not permit a clear prediction of which state is most likely to be realized in experiments. In particular, the difference between the NG Moore-Read and anti-Pfaffian energies is below the numerical uncertainty, reflecting an emergent approximate particle-hole symmetry~\cite{Balram_PH_CF_2017}. In contrast, there is a clear distinction between NG states, which are most suitable in the 0LL, and BS states, which are favored in the 1LL; see Appendix~\ref{app.Hartree} for details and supplementary data. In the 2LL and higher Landau levels, we find that Wigner crystals have the lowest energies.

\subsubsection{Bilayer graphene phase diagram}
Next, we investigate the possibility of these states in BLG. The zeroth Landau level of BLG consists of eight energy levels. Four of these levels are composed purely of the 0LL non-relativistic orbitals residing on one sub-lattice site. The other levels contain a mixture of 0LL and 1LL orbitals, whose weights are sensitive to the strength of the magnetic field. The first-excited BLG-LL is composed of 0LL, 1LL, and 2LL, with the mixture strongly affected by the magnetic fields and the displacement fields; see Ref.~\onlinecite{Kumar_Orbitally_2025}. 
 
\begin{figure}[t]
    \centering
    \includegraphics[width=0.95\linewidth]{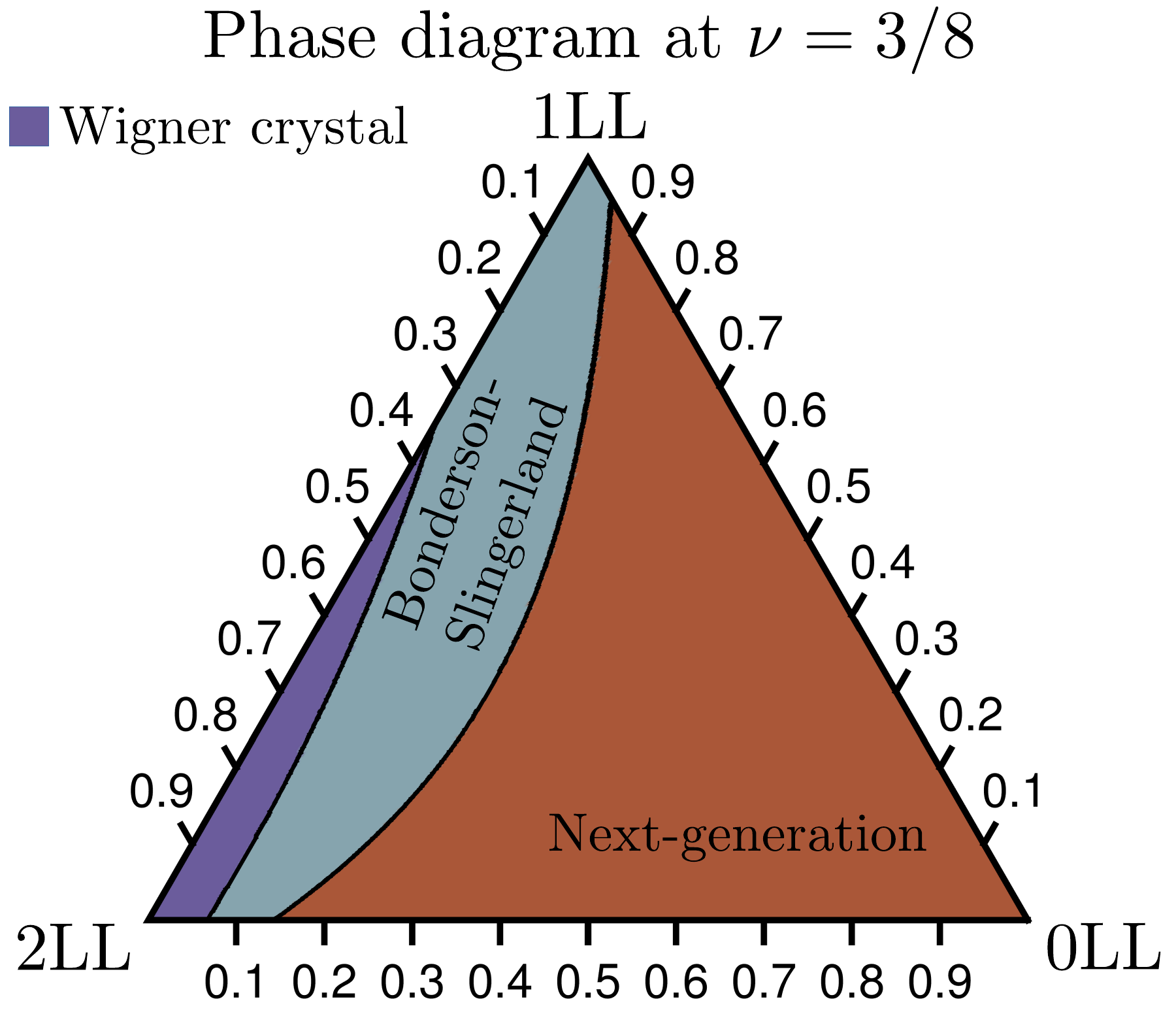}
    \caption{The BLG phase diagram at $\nu=\frac{3}{8}$ based on the energy of trial states in the Landau level composed of 0LL, 1LL, and 2LL relevant to the first two levels of BLG.}
    \label{fig.ternary_phase_38}
\end{figure}

\begin{figure}[t]
    \centering
    \includegraphics[width=0.95\linewidth]{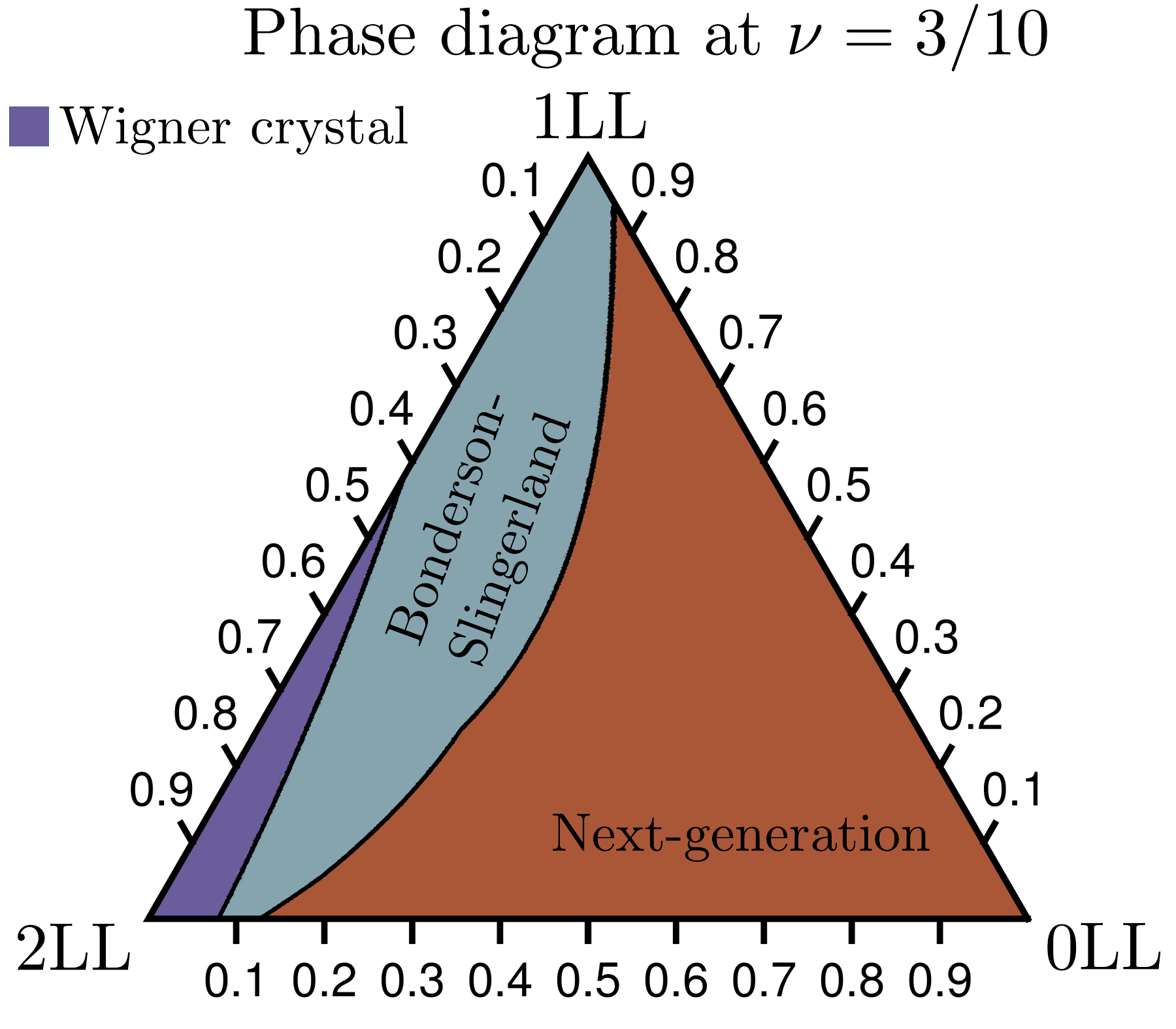}
    \caption{The BLG phase diagram at $\nu=\frac{3}{10}$ based on the energy of trial states.}
    \label{fig.ternary_phase_310}
\end{figure}

All three levels can be captured by tuning the relative weights $|C_n|^2$ of the 0LL, 1LL, and 2LL orbitals; see Fig.~\ref{fig.ternary_phase_38}. The weights are normalized; hence, a point on a ternary plot $(|C_0|^2,|C_1|^2,|C_2|^2)$ specifies any allowed composition. We compute the thermodynamic energies of all the trial states described above and identify the phase with the smallest energy. Each point in the ternary plot is colored according to the phase with the smallest energy. Within numerical accuracy, we cannot reliably distinguish between different pairing channels and associate one color to the type of the phase: the NG, the BS, and the Wigner crystal states. 

We find that the $\nu=\frac{3}{8}$ and $\frac{3}{10}$ fillings have qualitatively similar behavior. In the 0LL levels, NG states are favored. The other four levels in BLG's zeroth LL correspond to points on the 0LL-1LL line. Experimentally expressible magnetic fields typically range between $B=10T$ and $B=30T$, corresponding to $|C_1|^2$ between $0.93$ and $0.82$, close to the boundary between NG and BS states; see Appendix~\ref{app.bilayer}. The Wigner-crystal phase is favored near the 2LL. However, for typical displacement and magnetic fields, the orbital composition does not reach the charge-density-wave region and crosses the boundary between NG and BS states for a relatively low magnetic field $B=10-12T$. We note that we are using unscreened Coulomb interactions. The microscopic correction and Landau level mixing are expected to change the form of the phase diagram.

\section{Discussion}

We have developed a comprehensive theory of NG even-denominator quantum Hall states, encompassing bulk topological order, experimental signatures of their edge states, and a numerical study of trial wave functions. Our work highlights many similarities between these states and the well-studied half-filled ones, but also reveals some qualitative differences. In particular, we point out a competition between BS and NG states at $\nu=\frac{3}{8}$ and $\frac{3}{10}$ that is absent at half-filling, where these orders coincide.

To describe the bulk topological properties, we developed a $K$-matrix formulation of NG paired states and related their quasiparticles to those of BS states. We showed that for each NG state at $\nu=\frac{3}{8}$ and $\frac{3}{10}$, there is a BS state whose quasiparticles have the same charge and statistics. Still, they describe distinct phases in a rotation-invariant system, where they can be distinguished by the quasiparticle spins and the shift quantum number. A similar relationship is known between certain polarized and unpolarized states, but here it occurs between two single-component orders. 

For a specific pairing channel, $\ell_\text{NG}=-3$, we found that NG and BS states exhibit the same shift and describe identical topological orders~\footnote{States with equal $\nu,\kappa_{xy},S$ can be distinct topologically, e.g., by having a different ground state degeneracy on a torus~\cite{balram_Zn_2020}, but that is not the case here.}. Nevertheless, their trial wavefunctions are microscopically very different. We speculate that the two wave functions may differ by topologically trivial modes at intermediate length scales that could be revealed in their entanglement spectrum. Unfortunately, the system sizes available to us are insufficient for testing this hypothesis.

In the process of deriving the bulk properties of the NG states, we also revealed several general properties of topological phases obtained via flux attachment. Specifically, we derived the transformation of the thermal Hall conductance and showed that the topological stability of their edge states remains unaffected by flux attachment.

We have also systematically analyzed several experimental signatures of the NG paired states. In particular, we computed the scaling dimensions of tunneling operators for different pairing channels and proposed electric conductance, thermal transport, and noise measurements that can identify the topological order.  

Finally, we have numerically computed the energies of NG and BS trial wavefunctions for Coulomb interactions in semiconductors and BLG. We find that NG states are systematically favored in the zeroth Landau level, while BS have lower energies in the first excited Landau level. In higher Landau levels, Wigner crystals are favored over either one.

\begin{acknowledgments}
 This work was supported by the Israel Science Foundation (ISF) under grant 3281/25 and by the
Minerva Foundation with funding from the Federal German
Ministry for Education and Research.
\end{acknowledgments}

\appendix

\section{Proof of equivalence}\label{app.equiv}
In this Appendix, we prove the equivalence of topological orders obtained by: (i) attaching two fluxes to an Abelian state at $\nu_0^*$ and then performing particle-hole conjugation, and (ii) placing a hole state at $\nu_0^*$ on top of one filled Landau level of holes and then attaching two fluxes. 
\subsubsection{The shift quantum number}
First, we verify that the filling factors and the shifts obtained by the two prescriptions match. Suppose the CF state at $\nu_0^*$ has shift $S_0^*$. In the prescription (i), we first attach two fluxes that change the quantum numbers according to
\begin{align}
    (\nu_0^*,S_0^*)\to\left(\frac{\nu_0^*}{2\nu_0^*+1},S_0^*+2\right).
\end{align}
Recall that particle-hole conjugate transforms the filling and shift as $(\nu,S)\to(\bar{\nu},\bar{S})=\left(1-\nu,\frac{1-S\nu}{1-\nu_0^*}\right)$. Hence, we find the shift in the prescription (i)
\begin{align}
   \left(\nu_1,S_1\right) =  \left(\frac{1+\nu_0^* }{   2\nu_0^*+1},\frac{1-\nu_0^*S_0^*}{1+\nu_0^*}\right).
\end{align}

In prescription (ii), we first place the $\nu_0^*$ state in the first excited Landau level. The number of particles fully filling the lowest Landau level is $N_{e,\text{filled}}=N_\phi^*+1$. In the partially filled first excited Landau level, there are two additional states. Hence, the relation between the number of fluxes $N^*_\phi$ and the number of particles in the first Landau level is $N^*_\phi+2=N_{e,0}/\nu_0^* -S_0^*$. The total number $N_e=N_{e,0}+N_{e,\text{filled}}$ of particles includes $N_{e,\text{filled}}$ particles fully filling the lowest Landau level. Expressing the number of fluxes $N_\phi^*$ in terms of the total number of particles $N_e$, we find $N_\phi^* = (1+\nu_0^*)^{-1}N_e - S^*$ with $S^*=\frac{1+(S_0^*+2)\nu_0^*}{1+\nu_0^*}$. Finally, attaching two fluxes to the hole-like state at $\nu^*=-(1+\nu_0^*)$ results in the filling factor and the shift 
\begin{align}
   \left(\nu_2,S_2\right) =\left(\frac{1+\nu_0^*}{2(1+\nu_0^*)-1},2-S^*\right).
\end{align}
that coincides with the $(\nu_1,S_1)$ from the prescription (i).

\subsubsection{$K$-matrix}
We now confirm the equivalence of topological orders explicitly on the level of $K$-matrices. For a CF topological order at $\nu_0^*$ parametrized by a $K_0^*$-matrix and charge vector $\vect{t}_0$, the prescription (i) results in an electron $K$-matrix
\begin{align}
    K_1 = \begin{pmatrix}
        1 & 0 \\ 0 & -K_0^*-
        2\vect{t}_0\vect{t}_0^T        
    \end{pmatrix}
\end{align}
and $\vect{t}_1 = (1,\vect{t}_0)$. Similarly, we obtain an electron $K$-matrix from the prescription (ii)
\begin{align}
    K_2 = 2\vect{t}_2\vect{t}_2^T - \begin{pmatrix}
        1 & 0 \\ 0 & K_0^*  
    \end{pmatrix} = \begin{pmatrix}
        1 & 2\vect{t}_0^T \\ 2\vect{t}_0 & 2\vect{t}_0\vect{t}_0^T-K_0^* 
    \end{pmatrix} 
\end{align}
and the same charge vector $\vect{t}_2 = (1,\vect{t}_0)$. The $W\in \text{SL}(d+1, \mathbb{Z})$ transformation 
\begin{align}
    W = \begin{pmatrix}
        1 & 2\vect{t}_0^T \\ 0 & -\hat{I}  
    \end{pmatrix}
\end{align}
where $\hat{I}$ is a $\text{dim}(K_0^*)$-dimensional identity matrix, relates the two theories 
\begin{align}
    WK_2W^T=K_1,\qquad W^T\vect{t}_2=\vect{t}_1.
\end{align}
Consequently, the two prescriptions yield identical Abelian topological orders for any $K_0^*$ and $\vect{t}_0$.

\section{Details of Hartree-Fock calculation and supplemental data}\label{app.Hartree}
In the main text, we compare the energies of paired and gapless CFL states with those of Wigner crystal states for different interactions. Here, we outline the Hartree-Fock calculation and provide additional data.

\begin{figure}[t]
    \centering
    \includegraphics[width=0.95\linewidth]{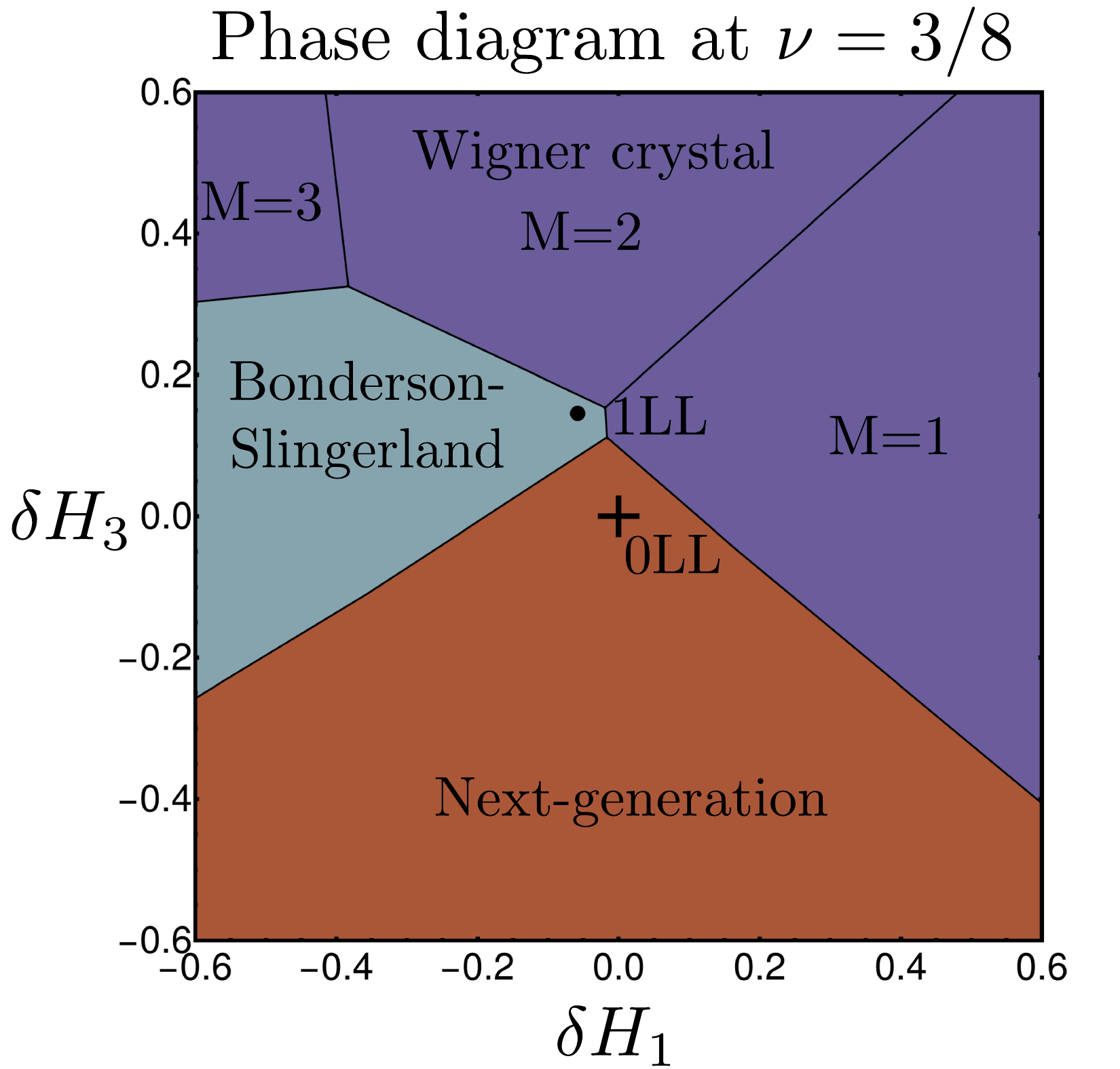}
    \caption{The phase diagram at $\nu=\frac{3}{8}$ based on the energy of trial states. The pure 0LL Coulomb interactions (denoted by `$+$') are perturbed by the $\delta H_1$ and $\delta H_3$ Haldane pseudopotentials. The values corresponding to 1LL are indicated (with other $\delta H_{M\geq5}$ the same as in the 0LL).}
    \label{fig.phase_38}
\end{figure}

\subsection{Coulomb interactions for Hartree-Fock calculations}
The $n$th Landau level orbitals result in the form factors 
\begin{align}
    F_n(q)  = L_n\left(\frac{q^2}{2}\right)e^{-\frac{q^2}{4}},
\end{align}
where $ L_n(x)$ are Laguerre polynomials~\cite{MacDonald_Introduction_1994}. The Hartree and Fock parts of the effective potential in momentum space are 
\begin{align}
    &V_\text{H}(q) = V(q) |F_n(q)|^2,\\
    &V_\text{F}(q) = \frac{1}{2\pi}\int d^2p V(p) |F_n(p)|^2 e^{-i(p_xq_y-p_yq_x)}.
\end{align}
Here, $V(q)=\frac{2\pi}{|q|}$ is the Coulomb interacting potential in momentum space. Following Ref.~\onlinecite{Fogler_Laughlin_wigner_1997,Goerbig_Competition_2004}, we compute the Hartree-Fock energy of the triangular Wigner crystal formed of the $M$ electron bubbles. 

\begin{figure}[t]
	\centering
	\includegraphics[width=0.95\linewidth]{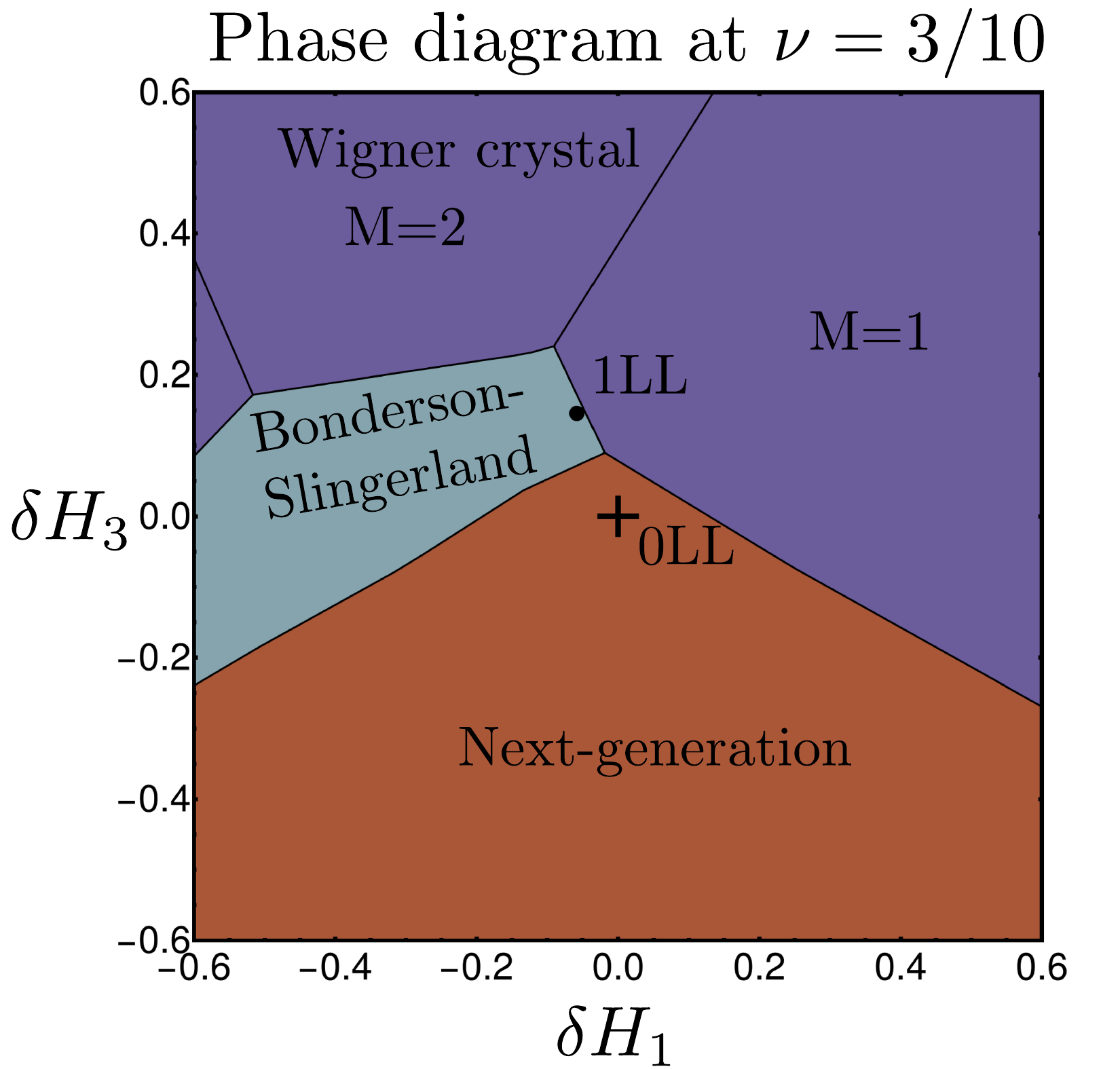}
	\caption{The phase diagram at $\nu=\frac{3}{10}$ based on the energy of trial states as a function of the first and the third pseudopotentials.}
	\label{fig.phase_310}
\end{figure}

\subsection{Pseudopotential interactions for Hartree-Fock}
We now define the interaction potentials corresponding to an individual pseudopotential. We recall that the Haldane pseudopotentials for an effective potential $V_\text{H}(q)$ are
\begin{align}
    H_m = \int \frac{d^2q}{2\pi}V_\text{H}(q) L_m(q^2) e^{-\frac{q^2}{2}}.
\end{align}
Using the orthogonality of the  Laguerre polynomials, we define a modification 
\begin{align}
    \delta V_m(q) =2 L_m(q^2)e^{-\frac{q^2}{2}}
\end{align}
to the effective potentials $V_\text{H}(q)$ that result in the deviation $\delta H_m = 1$ for a given $m$. The corresponding Fock potentials are  $V_\text{F}^m(q) = (-1)^m \delta V_m(q)$.

\begin{figure}
    \centering
    \includegraphics[width=1\linewidth]{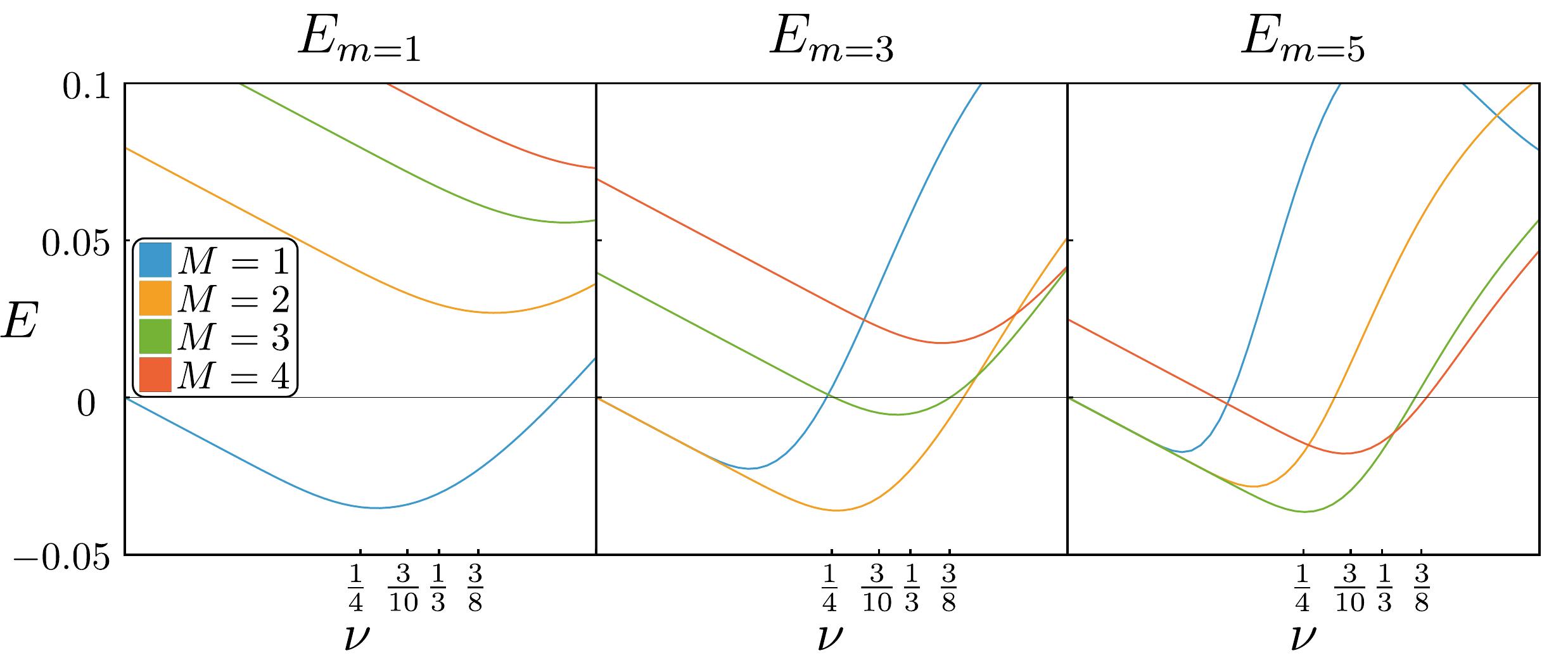}
    \caption{Hartree-Fock energies of Wigner crystal states with $M$ electron bubbles for interactions consisting of a single pseudopotential $H_m=1$.}
    \label{fig.E_hald_nu}
\end{figure}

The energies per particle of the Wigner crystal with $M$ electron bubbles for $V_\text{H}(q) =  \delta V_m(q)$ for $m=1,3,5$ are plotted in Fig.~\ref{fig.E_hald_nu} for the filling factor range $\nu\in[0,\frac{1}{2}]$. We note that the Hartree-Fock energies vanish for $2M-1\leq m$ as $\nu$ approach zero.

\subsection{Haldane pseudopotential phase diagram}
The two-body interactions are conveniently parameterized by Haldane pseudopotentials. In the context of the FQH effect, only a few of the lowest pseudopotentials are often important and determine the phase of a system. In particular, pseudopotentials suitable for the hole bands of GaAs, where the next-generation states are observed, are calculated in Refs.~\onlinecite{Simion_Magnetic_2014,Simion_Non_Abelian_2017}. We, therefore, study perturbations of the Coulomb interactions in 0LL by the two most relevant Haldane pseudopotentials, $\delta H_1$ and $\delta H_3$. We compute the energies of various trial states as a function of the $\delta H_m$
\begin{align}
    E(\delta H_1,\delta H_3) = E^\text{0LL} + \sum_{m=1,3} \delta H_m E_{m},
\end{align}
and identify the lowest energy phase. The points in the $\delta H_1-\delta H_3$ plane are colored according to the phase with the minimal energy. 

Due to the lowest Landau level projection scheme, $\delta H_1$ does not contribute to the energies of the NG states for $\nu=\frac{3}{10}$. However, the Wigner crystal and BS states are susceptible to changes in all odd pseudopotentials, resulting in a non-trivial competition; see Fig.~\ref{fig.phase_310}.

We find that for pure 0LL Coulomb interactions ($\delta H_M=0$), the NG paired states are favored at both fillings. Decreasing $\delta H_3$ and increasing $\delta H_1$ drives the system into the BS states with either Moore-Read ($\ell=1$) or anti-Pfaffian ($\ell=-3$) pairings. We also find that $M=1$ Wigner crystals can be favored by a slight increase of $\delta H_1$ and $\delta H_3$. The screening of Coulomb interaction or Landau level mixing in electron-doped GaAs samples can modify the 0LL interaction and, presumably, favor the Wigner crystals or CFLs, explaining the absence of these plateaus in the present experimental data.

\subsection{Bilayer graphene phase diagram}\label{app.bilayer}
\begin{figure}[t]
    \centering
    \includegraphics[width=0.95\linewidth]{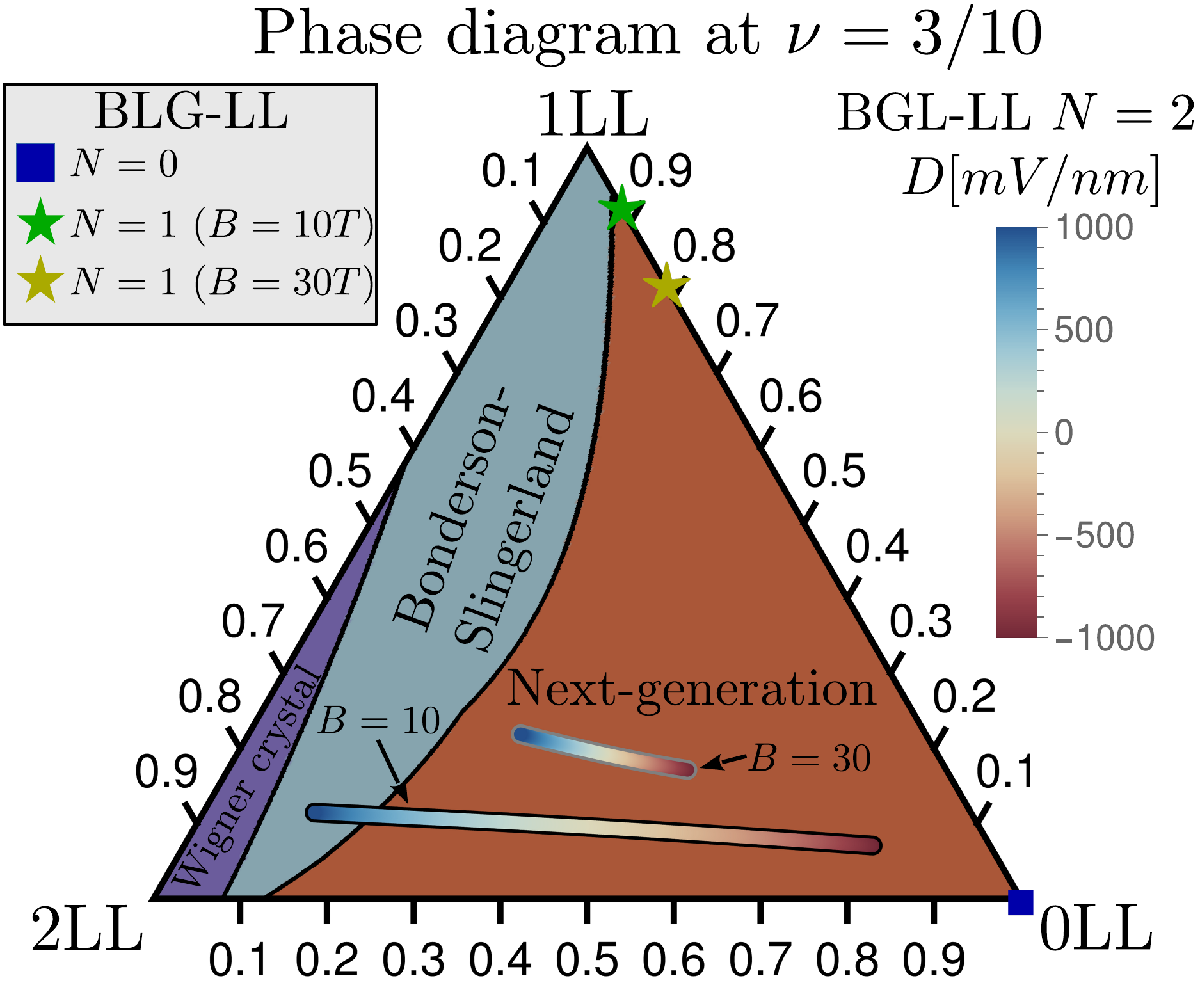}
    \caption{The BLG phase diagram at $\nu=\frac{3}{10}$ based on the energy of trial states. The points in the phase diagram corresponding to the lowest ($N=0,1$) BLG-LL at $B=10T$ and $B=30T$ are indicated by the square and stars. In the first excited LL ($N=2$), the composition changes with the displacement field as indicated by the color bar.}
    \label{fig.ternary_phase_310_BLG}
\end{figure}

Here, we augment Fig.~\ref{fig.ternary_phase_310} of the main text with information about the orbital compositions of the first three Landau levels of BLG for $B=10T$ and $30T$ in Fig.~\ref{fig.ternary_phase_310_BLG}. The values of $(|C_0|^2,|C_1|^2,|C_2|^2)$ for the corresponding Landau level with the microscopic parameter are obtained following Ref.~\onlinecite{Kumar_Orbitally_2025}. 

\section{Bonderson-Slingerland states}\label{app.BS}
The edge theory of the BS state at $\nu_\text{BS}=\frac{\pm m}{\pm (2q-1) m +1}$ is described by 
\begin{align}
{\cal L}_\text{BS} = {\cal L}_{K_\text{BS}} + {\cal L}_{\ell_\text{BS}},
\end{align}
where $K_\text{BS} = \pm \mathbb{I}_m + q \vect{t}\vect{t}^T$ and $\vect{t}=(1,1,\ldots)$. The corresponding thermal conductance is $\kappa_{xy} = m + \ell_\text{BS}$ for $m>0$ and $\kappa_{xy} =2- m + \ell_\text{BS}$ for $m<0$. The wave function for $\ell_\text{BS}=2r+1$ is given 
\begin{align}\label{sec.app.BSWF}
    \Psi_\text{BS} = P_\text{LLL}\left[\frac{1}{\omega_{ij}}\left(\frac{\omega^*_{ij}}{\omega_{ij}}\right)^{r}\right]\phi_m\phi_1^{2q-1}
\end{align}
occurs at $S^\text{BS}_\ell=m+2q-1+\ell_\text{BS}$.

\subsection{Equivalence of the BS and NG topological orders}\label{app.BSequalnextgen}
The NG state with $2p$ fluxes and $\nu^*=\pm(1+\frac{1}{2})$, and the BS states with the bosonic part forming a Jain state with $\nu^*=\mp3$ and $q=2p\pm 1$ fluxes, occur at the same filling factor
\begin{align}\label{eq.equivalence_BS}
    \nu_\text{NG}=\frac{\pm(1+\frac{1}{2})}{\pm 2p (1+\frac{1}{2})+1} = \nu_\text{BS}=\frac{\mp 3}{\mp (2p\pm1) 3+1}.
\end{align}
We now show that, when the pairing channels obey $\ell_\text{NG}=-6\pm \ell_\text{BS}$, the two states represent the same intrinsic topological order, in the sense of having the same quasiparticle charges and statistics. 

We first demonstrate the equivalence between the NG and BS topological orders for $p=0$ by identifying the $W\in \text{SL}(n,\mathbb{Z})$ transformation that relates the two states. The equivalence for $p\neq 0$ follows, since flux attachment commutes with the $W\in \text{SL}(n,\mathbb{Z})$ transformation. For $p=0$, the filling factors are $\nu=\pm\frac{3}{2}$. We prove the equivalence for $\nu=\frac{3}{2}$. The $\nu=-\frac{3}{2}$ case follows trivially from $K\to-K$ and $\ell_\text{BS}\to -\ell_\text{BS}$.  

We start with a BS state with a pair of Majorana modes, $\ell_\text{BS}=2$, which we absorb into the $K$ matrix, i.e., $K^*_\text{BS} = \text{diag}(-\mathbb{I}_3, 1)$ and $\vect{t}_\text{BS}=(1,1,1,0)$. Upon attaching $q=1$ fluxes, we obtain
\begin{align}\label{sec.app.KBS}
    K_\text{BS} =K^*_\text{BS} +q\vect{t}_\text{BS}\vect{t}_\text{BS}^T =\left(\begin{array}{cccc}
 0 & 1 & 1 & 0 \\
 1 & 0 & 1 & 0 \\
 1 & 1 & 0 &  0\\
 0& 0 & 0 & 1 \\
\end{array}\right).
\end{align}
We then apply the transformation 
\begin{align}
    W=
    \left(\begin{array}{cccc}
 1 & 0 & 0 & 1 \\
 1 & 0 & 1 & 0 \\
 -1 & 1 & -1 &  -1\\
 2& -1 & 1 & 1 \\
\end{array}\right)
    \in \text{SL}(4,\mathbb{Z}),
\end{align}
such that $K_\text{NG} = W^TK_\text{BS}W = \text{diag}(2,1,-1,-1)$ and $\vect{t}_\text{NG}=W^T\vect{t}_\text{BS}=(1,1,0,0)$. The latter $K$-matrix represents the $\nu^*_\text{NG}= \frac{3}{2}$ NG state with $\ell_\text{NG}=- 4$ bosonized Majoranas. We have thus established the equivalence between
\begin{align}
     {\cal L}_\text{NG} + {\cal L}_{\mp4} \equiv {\cal L}_{\text{BS}} + {\cal L}_{\pm 2} ,
\end{align}
where ${\cal L}_{\text{BS}}$ and ${\cal L}_\text{NG}$ contain three and two bosonic modes, respectively, with Majorana modes factored out, and the sign corresponds to the sign in Eq.~\eqref{eq.equivalence_BS}. Adding $\delta \ell$ Majorana on both the right and left-hand sides of the equation does not affect this equivalence, establishing it for general $\ell$.

\subsection{Inequivalence of the BS and NG states}\label{app.BSinequalnextgen}
Despite hosting quasiparticles with the same charge and statistics, the NG and BS states with the same filling and thermal conductance, in general, represent distinct universality classes in clean systems. The shift corresponding to these states is different in most cases. For odd numbers of Majoranas, Eq.~\eqref{eqn.nextgenshift} and Eq.~\eqref{sec.app.BSWF} imply that the two states exhibit
\begin{align}
&\kappa_\text{NG}= 1 \pm (1+\frac{\ell_\text{NG}}{2}) ,\quad
&&\kappa_\text{BS} =1 \mp 2+\frac{\ell_\text{BS}}{2},
\\
& S_\text{NG}= 2p \pm (2+\frac{\ell_\text{NG}}{3}) ,\quad
    &&S_\text{BS} = 2p\mp 2+ \ell_\text{BS}.
\end{align}
Only for $\ell_\text{NG}=-3 = \mp \ell_\text{BS}$ are the two states genuinely the same, in the sense of having the same topological order and shift.

\subsection{Microscopic comparison of NG and BS wavefunctions}\label{app.BSinequalmicroscopics}

Finally, we comment on the significantly different microscopic behaviors of the BS and NG wave functions, which exhibit the same filling factor and shift. For concreteness, we focus on the $\nu=\frac{3}{8}$ filling. At this filling, the NG state with anti-Pfaffian pairing, $\ell_\text{NG} = -3$, exhibits the same topological order as the BS with $\ell_\text{BS}=3$, with the shift $S=3$. 

First, we compute the overlaps between the trial wave functions in Tab.~\ref{tab.overlaps}. The overlaps decay quite rapidly, even though they remain significant for larger Hilbert spaces. At the same time, the overlaps of the NG paired and CFL states are large and comparable to the overlap of the underlying half-filled states. We thus speculate that the CF orbitals ${\cal S}^{n,p}_I$ in Eq.~\eqref{eq.NG_WF} remain largely orthogonal as their predecessors, the fermion orbitals ${\cal S}^{n}_I$.

Next, we compare the density correlation functions for $N_e=30$ electrons. The density correlation functions are significantly different. The  BS state has a `shoulder' near a small $R$ that was argued to be a characteristic of paired states~\cite{Balram_parton_2018}. Such a shoulder is typically present for the exact ground states of the first excited Landau level. This behavior aligns with the BS states being favorable in the 1LL. By contrast, neither NG CFLs nor paired states exhibit such a shoulder, consistent with them being favored in the 0LL.

%The substantial difference between wave functions describing the same topological order is unusual for FQH states. We speculate that a possible source for this difference may be 
% \david{I killed most of what came below (there are also comments as part of it if you care). Actually, I wanted to speculate on non-topo modes here, but it seems that you don't actually need to use any localization and t-instability, so there is really no basis for saying that } 

% Next, we compare the density correlation functions for $N_e=30$ electrons. The density correlation functions are significantly different. The  BS state has a `shoulder' near a small $R$ that was argued to be a characteristic of paired states~\cite{Balram_parton_2018}. Such a shoulder is typically present for the exact ground states of the first excited Landau level. This behavior aligns with the BS states being the lowest energy states in the 1LL among the states tested. Similarly to the overlaps, the density correlation functions of the NG paired and CFL states are more similar, nevertheless visually distinct \david{sound like an oxymoron. They are the same but different. So what do you want to say?}. Neither of them has a shoulder, which is consistent with being favored in the 0LL. 

\begin{figure}[t]
    \centering
    \includegraphics[width=0.95\linewidth]{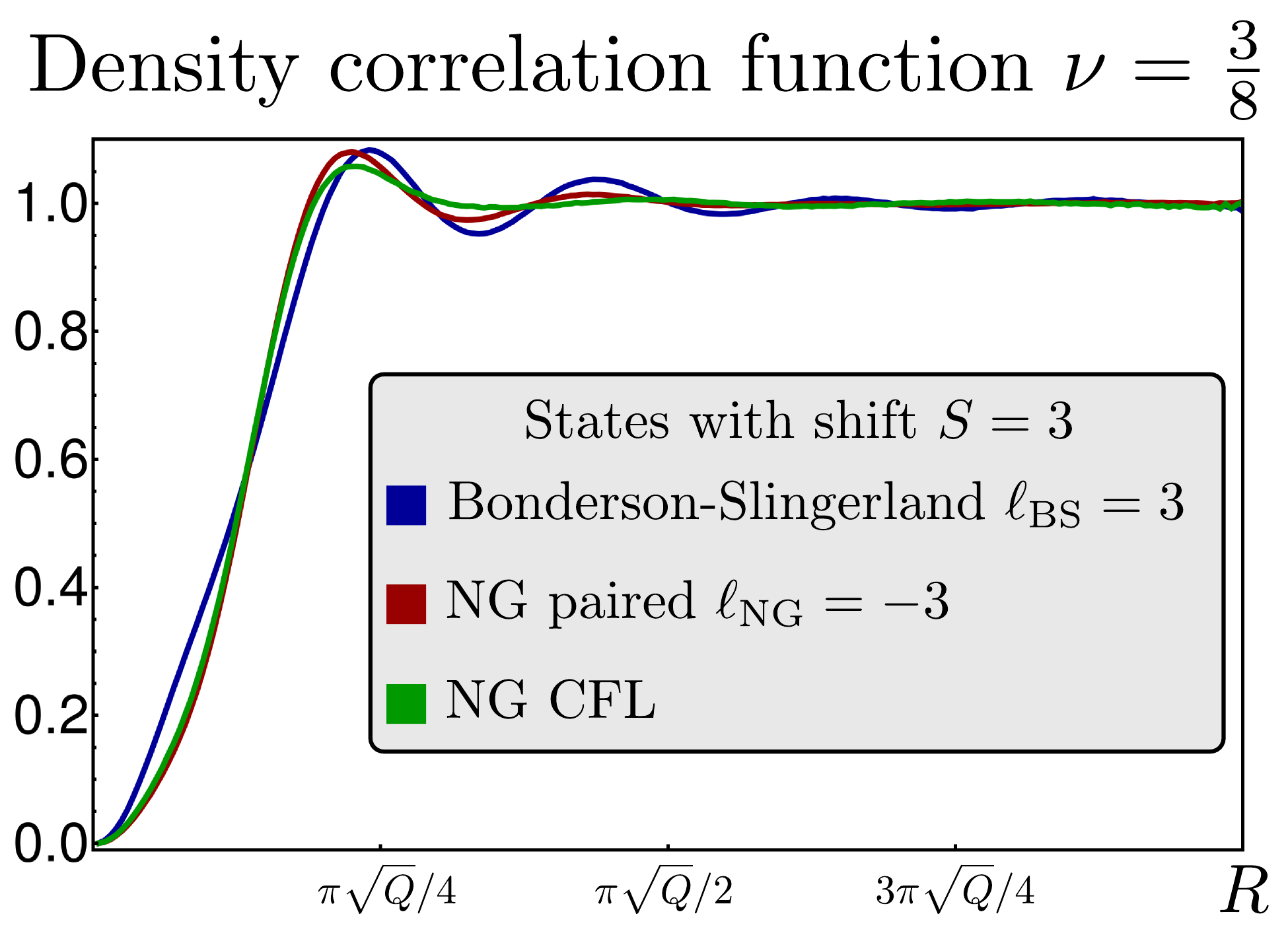}
    \caption{Density-density correlation function of the $\ell_\text{NG}=-3$ NG paired state, composite Fermi liquid, and $\ell_\text{BS}=3$ BS state that all occur at the same shift for $\nu=\frac{3}{8}$. The particle number is $N_e=30$. The BS state is visually distinct from the NG states and has a shoulder for a small $R$ characteristic of 1LL states.}
    \label{fig:placeholder}
\end{figure}

\begin{table}[b]
    \centering\renewcommand{\arraystretch}{1.3}
    \caption{Overlaps of the NG and BS states at $\nu=\frac{3}{8}$ with $\ell_\text{NG}=-3$ and $\ell_\text{BS}=3$ pairing, respectively. The overlaps are computed with Monte-Carlo simulations, and only significant digits are shown. We specify the dimension of the $L_z=0$ and $L^2=0$ Hilbert spaces. For reference, the squared overlaps between the first-generation paired and CFL states at $\nu_0^*$ used to construct the NG states are $\approx 92\%$ in the first column and $\approx 66\%$ in the last.} 
    \begin{tabular}{c|cccc}
    \hline\hline
       $(N_e,N_\phi)$ & $(12,29)$ & $(18,45)$ & $(24,61)$ & $(30,77)$ \\\hline
        Dim $L_z=0$ & $1.4\times 10^{6}$& $2.5\times 10^{10}$ & $5.5\times 10^{14}$ & $1.4\times 10^{19}$\\
        Dim $L^2=0$ & 1,330 & $6.1\times 10^6$& $5.6\times 10^{10}$  & $7.2\times 10^{14}$
        \\\hline
        $|\left\langle\text{NG}|\text{BS}\right\rangle|^2$ & $0.539$ & $0.30$ & $0.18$ & $0.10$\\
        $|\left\langle\text{NG}|\text{NG-CFL}\right\rangle|^2$ & $0.912$ & ----- & ----- & $0.62$ \\
        \hline\hline
    \end{tabular}
    \label{tab.overlaps}
\end{table}

\section{Abelian states}\label{app.Abelian}
Here, we briefly outline the construction of the NG and BS states for Abelian states, which can be fully captured within the $K$ matrix formalism.

\subsection{Abelian first-generation states}
For Abelian states with an even number $\ell=2r$ of Majoranas, we use the $K$-matrix from Ref.~\onlinecite{Yutushui_daughters_2024}, i.e.
\begin{align}\label{eq.paired}
    K^*_\text{paired}=\text{sgn}(r)\;\text{diag}(K_\text{SF},\mathbb{I}_{|r|-1}),
\end{align}
where $K_\text{SF}=\mathbb{I}_2-\sigma_x$ with $\sigma_x$ Pauli matrix. The $|r|+1$ dimension charge vector is $\vect{t}_\text{paired}=(1,\ldots,1)$. The last $|r|-1$ components of the $K$-matrix, $\mathbb{I}_{|r|-1}$, represent an IQH state, and thus, have the shift vector $\vect{s}_{r-1} =(\frac{1}{2},\frac{3}{2},\ldots , |r|-\frac{3}{2})$. We assume that the superfluid state is formed on top of the $|r|-1$ filled levels, and the angular momenta of the paired fermions are increased by this amount. Consequently, we take the shift vector to be
\begin{align}
 \vect{s}^*_\text{paired} = \text{sgn}(r)\left(|r|-\frac{1}{2},|r|-\frac{1}{2},\vect{s}_r\right).
\end{align}
The shift vector after flux attachment is $\vect{s}_\text{paired} = \vect{s}^*_\text{paired} + p$, yielding the shift $S_\text{paired} = 2p + \text{sgn}(r)(2|r|-1)$. The shifts of the paired states with odd $\ell$ and even $\ell+\text{sgn}(\ell)$ Majoranas are the same.

\subsection{Abelian next-generation states}
The next-generation states before flux attachment are given by $K_\text{NG}^* = \text{diag}(K^*_\text{paired},\mathbb{I}_n)$ with $\vect{t}_\text{NG}=(1,\ldots,1)$. The paired state is formed in the $n$th Landau level, and its shift vector is offset by $n$, i.e., $\vect{s}_\text{paired}=\vect{s}_\text{paired}^*+ n$. Taking into account the last $n$ components corresponding to the IQH state, we get the shift vector $\vect{s}^*_\text{NG}= (\vect{s}^*_\text{paired}+n,\vect{s}_n)$. Finally, after the flux attachment, the $K$-matrix and the shift vector transform according to Eq.~\eqref{eq.flux_attachment} and Eq.~\eqref{eq.cfshift}.

For instance, we consider $\nu_\text{NG}=\frac{3}{8}$ ($n=1$) with $\ell_\text{NG} = -4$ Majoranas. The paired $K$-matrix is $3$-dimensional and given by Eq.~\eqref{eq.paired}. Placing it on top of the single filled level and attaching $2p=2$ fluxes, we obtain
\begin{align}\label{eq.K_ng-4}
       K_\text{NG}=
    \left(\begin{array}{cccc}
 3 & 5 & 4 & 2 \\
 5 & 3 & 4 & 2 \\
 4 & 4 & 3 & 2\\
 2& 2 & 2 & 3 \\
\end{array}\right),\quad
\vect{s}_\text{NG}=\left(\frac{1}{2},\frac{1}{2},\frac{3}{2},\frac{3}{2}\right),
\end{align}
and $\vect{t}_\text{NG}=(1,1,1,1)$, yielding $S_\text{NG}= \frac{7}{3}$. 

% \david{So you are saying this prescription gives you a fractional shift, and you also get that for BS, even though all non-Abelian BS states have integer shifts? This sounds a bit worrying}

\subsection{Abelian Bonderson-Slingerland states}
To construct the $K$-matrix of Abelian BS states, we use the perspective of flux condensation. Starting with $K_\text{paired}$, we condense flux quasiparticles into a bosonic Jain state given by $K_\text{Jain} = \text{sgn}(m)\mathbb{I}_{|m|}+q\vect{t}\vect{t}^T$ and $s_\text{m}=\text{sgn}(m)\left(\frac{1}{2},\ldots,\frac{2|m|-1}{2}\right)+\frac{q}{2}$. The resulting $K$-matrix is 
\begin{align}
    &K_\text{BS}= \left(\begin{array}{cccc}
 K_\text{paired} & \vect{1}_{|r|+1,m} \\
 \vect{1}_{m,|r|+1} & K_\text{Jain} 
\end{array}\right),\\
&\vect{s}_\text{BS} = \left(\vect{s}_\text{paired}, s_\text{m}\right),
\end{align}
and $\vect{t}_\text{BS} = (1,\ldots,1,0,\ldots,0)$ with $|r|+1$ ones and $|m|$ zeros. Specifically, for $\ell_\text{BS}=2$, $m=-2$ and $q=-1$, we get
\begin{align}\label{eq.app.bs}
       K_\text{BS}=
    \left(\begin{array}{cccc}
 3 & 1 & 1 & 1 \\
 1 & 3 & 1 & 1 \\
 1 & 1 & -2 & -1\\
 1 & 1 & -1& -2 \\
\end{array}\right),\;\;
\vect{s}_\text{BS}=\left(\frac{3}{2},\frac{3}{2},-1,-2\right),
\end{align}
and $\vect{t}_\text{BS}=(1,1,0,0)$. This topological order is equivalent to the one described by Eq.~\eqref{eq.K_ng-4}, as can be seen by $W^TK_\text{NG}W =K_\text{BS}$, $W^T\vect{t}_\text{NG} =\vect{t}_\text{BS}$ with 
\begin{align}
    W=
    \left(\begin{array}{cccc}
 0 & -1 & 1 & 1 \\
 0 & 1 & -1 & 0 \\
 1 & -1 & 1 &  -1\\
 0 & 2 & -1 & 0 \\
\end{array}\right)
    \in \text{SL}(4,\mathbb{Z}).
\end{align}
However, the shift of the BS states is $S_\text{BS}=1$, and the two states, Eq.~\eqref{eq.app.bs} and Eq.~\eqref{eq.K_ng-4}, are different in the presence of rotational symmetry.

\section{Kane-Fisher-Polchinski fixed point for the next-generation states}\label{app.KFP_FP}
For chiral states, the scaling dimensions of tunneling operators are universal and do not depend on non-topological data, such as density-density interactions parametrized by the velocity matrix in Eq.~\eqref{eq.LK}. However, even chiral paired states with $\ell$ become non-chiral when promoted to the next generation for $\nu^*<0$. Nevertheless, in the presence of ubiquitous disorder, the edge can flow to a fixed point where the charge and neutral modes decouple \cite{Moore_Classification_1997,Moore_Critical_2002}. If all neutral modes have the same chirality, as in the case of $\ell\geq0$, the scaling dimensions of tunneling operators become universal. For $\ell=-1$, the one Majorana mode counter-propagating to the rest of the neutral sector \textit{cannot} couple relevantly or marginally, and thus, does not change the universal result. Hence, our analysis holds for the pairing channels with $\ell\geq-1$ irrespective of the sign of $\nu^*$. For $\ell\leq-2$, the scaling dimensions are generally non-universal despite the charge-neutral decoupling. 

\subsection{The $\nu_\text{NG}=\frac{3}{4}$ KFP fixed point}
At the edge of the $\nu_\text{NG}=\frac{3}{4}$ state, disordered tunneling between the two counter-propagating modes in the basis where $K=\begin{pmatrix}
        0 & 2 \\ 
        2 & 1
    \end{pmatrix}$ is 
\begin{align}
    {\cal L}_\text{dis} = \xi(x)e^{i(2\phi_1-\phi_2)} +  \text{H.c.}
\end{align}
To analyze the problem in the presence of tunneling, we first transform to the charge-neutral basis via  
\begin{align}
     \begin{pmatrix}
       \phi_c \\ 
        \phi_n
    \end{pmatrix} =\begin{pmatrix}
        1 & 1 \\ 
        2 & -1
    \end{pmatrix}\begin{pmatrix}
       \phi_1 \\ 
        \phi_2
    \end{pmatrix}. 
\end{align}
The tunneling in this basis is described by
\begin{align}
    {\cal L}_\text{dis} = \xi(x)e^{i\phi_n} +  \text{H.c.}
\end{align}
Absorbing the disordered tunneling, with a disorder-dependent gauge transformation, reveals that the density-density interactions of the form 
\begin{align}
    {\cal L}=2h_{nc}\partial_x\phi_c\partial_x\phi_n
\end{align}
are irrelevant. Hence, the IR-fixed point Lagrangian has $h_{nc}=0$, and the scaling dimension of quasiparticle $e^{i\vect{m}\cdotnone\vect{\phi}}$ can be readily inferred as
\begin{align}
    \Delta_{\vect{m}}= \frac{1}{2}\;\vect{m}^T\cdotnone\hat{\Delta}\cdotnone \vect{m}= \frac{1}{2}\vect{m}^T\cdotnone\begin{pmatrix}
        \frac{5}{12}&-\frac{1}{6}\\-\frac{1}{6}& \frac{2}{3}
    \end{pmatrix}\cdotnone\vect{m},
\end{align}
where $\hat{\Delta} = W\cdotnone |K_\text{cn}^{-1}| \cdotnone W^T$ involves an absolute value of the diagonal matrix $K_\text{cn} =W^T\cdotnone K \cdotnone W= \diag(\frac{4}{3},-\frac{1}{3})$. Notice that for the chiral edge, the eigenvalues of $K$ are positive and $\Delta$ reduces to $K^{-1}$. We thus read off the scaling dimension of the electron operators 
\begin{align}
    \psi_{e1} = e^{i(2\phi_1+\phi_2)},\qquad \psi_{e2}=\gamma_k e^{2i\phi_2},
\end{align}
to be $\Delta_{e1} = \frac{5}{6}$ and  $\Delta_{e2} = \frac{11}{6}$. For the flux and half-flux operators 
\begin{align}
    \psi_\phi = e^{i\phi_2},\qquad\psi_{\phi/2} = e^{i\phi_2/2}\prod_{k=1}^{|\ell|}\sigma_k,
\end{align}
we find $\Delta_\phi =\frac{5}{24}$ and $\Delta_\phi =\frac{6|\ell|+5}{96}$.

\subsection{The $\nu_\text{NG}=\frac{3}{10}$ KFP fixed point}
In the basis where $K=\begin{pmatrix}
    2&4\\4&3
\end{pmatrix}$, the disordered tunneling is 
\begin{align}
    {\cal L}_\text{dis} = \xi(x)e^{i(2\phi_1-\phi_2)} +  \text{H.c.}
\end{align}
The scaling dimension of $e^{i\vect{m}^T\cdotnone\vect{\phi}}$ is given by
\renewcommand*{\arraystretch}{1.2}
\begin{align}
    \Delta_{\vect{m}} = \frac{1}{2}\vect{m}^T\begin{pmatrix}
        \frac{11}{30}&-\frac{4}{15}\\-\frac{4}{15}& \frac{7}{15}
    \end{pmatrix}\vect{m}.
\end{align} 
\renewcommand*{\arraystretch}{1}
We find the scaling dimension of electron operators 
\begin{align}
    \psi_{e1} = e^{i(4\phi_1 + 3\phi_2)},\qquad \psi_{e2}=\gamma_k e^{i(2\phi_1+4\phi_2)}
\end{align}
to be $\Delta_{e1} = \frac{11}{6}$ and  $\Delta_{e2} = \frac{17}{6}$.
The flux and half-flux quasiparticles 
\begin{align}
    \psi_\phi = e^{i\phi_1},\qquad\psi_{\phi/2} = e^{i\phi_1/2}\prod_{k=1}^{|\ell|}\sigma_k,
\end{align}
have scaling dimension $\Delta_\phi =\frac{11}{60}$ and $\Delta_\phi =\frac{15|\ell|+11}{240}$, respectively.

\subsection{The $\nu_\text{NG}=\frac{5}{8}$ KFP fixed point}
For $\nu_\text{NG}=\frac{5}{8}$ the $K$-matrix is
\begin{align}
    K=\begin{pmatrix}
    0&2&2\\2&1&2\\2&2&1
\end{pmatrix}.
\end{align}
The scaling dimension of $e^{i\vect{m}^T\cdotnone\vect{\phi}}$ is\renewcommand*{\arraystretch}{1.2}
\begin{align}
    \Delta_{\vect{m}} = \frac{1}{2}\vect{m}^T\begin{pmatrix}
        \frac{17}{30}&-\frac{3}{20}&-\frac{3}{20}\\
        -\frac{3}{20}& \frac{7}{10 }& -\frac{3}{10}\\
       -\frac{3}{20} &-\frac{3}{10}&\frac{7}{10 }
    \end{pmatrix}\vect{m}.
\end{align} 
\renewcommand*{\arraystretch}{1}
The electron operators 
\begin{align}
    \psi_{e1} = e^{i(2\phi_1 + 2\phi_2+\phi_3)},\qquad \psi_{e2}=\gamma_k e^{i(2\phi_2+2\phi_3)}
\end{align}
have scaling dimensions $\Delta_{e1} = \frac{11}{10}$ and  $\Delta_{e2} = \frac{21}{10}$. The flux and half-flux quasiparticles 
\begin{align}
    \psi_\phi = e^{i\phi_1},\qquad\psi_{\phi/2} = e^{i\phi_1/2}\prod_{k=1}^{|\ell|}\sigma_k,
\end{align}
have scaling dimension $\Delta_\phi =\frac{17}{80}$ and $\Delta_\phi =\frac{20|\ell|+17}{320}$, respectively.

\section{Next-generation state coherent conductances}\label{app.coherent_conductnace}
Here we tabulate the values of the coherent charge conductance in the $\pi$-geometry for the $\nu_\text{NG}=\frac{3}{4}$, $\nu_\text{NG}=\frac{3}{8}$, and $\nu_\text{NG}=\frac{3}{10}$ in Tab.~\ref{tab.condunctance_3_4}, Tab.~\ref{tab.condunctance_3_8}, and Tab.~\ref{tab.condunctance_3_10}, respectively. For the case when the NG Jain states are chiral, $q\geq-1$, the conductance values are readily obtained using the methods outlined in Refs.~\onlinecite {Yutushui_Localization_2024,Yutushui_Universal_2025}.

% --- TABLE 1 ---
\begin{table}[t]
	\centering
	\setlength{\tabcolsep}{2pt} % Reduced padding to prevent overflow
	\renewcommand{\arraystretch}{1.3} 
	\caption{The coherent charge conductance between S2 and D1 in the $\pi$-junction in Fig.~\ref{fig.Pi_junction} for $\nu_\text{NG}(1,1)=\frac{3}{8}$.}
	
	% w{c}{1cm} = Fixed width (1cm) + Centered text + Safe for RevTeX
	\begin{tabular}{c c !{\vrule width .9pt} w{c}{1cm} | w{c}{1cm} | w{c}{1cm} | w{c}{1cm} | w{c}{1cm} | w{c}{1cm} }
		\hline \hline
		$q$ & $\ell=$ & $-5,-6$ & $-3,-4$ & $-1,-2$ &$0$& $1,2$ & $3,4$ \\ 
		\hline
		$-1$ & $\nu_q=\frac{2}{5}$  & $\frac{3}{8}$ & $\frac{3}{8}$ & $\frac{3}{8}$ & $\frac{1}{3}$ & $\frac{1}{3}$ & $\frac{1}{3}$ \\
		$0$  & $\nu_q=\frac{1}{3}$  & $\frac{1}{3}$ & $\frac{1}{3}$ & $\frac{1}{3}$ & $\frac{1}{3}$ & $\frac{1}{3}$ & $\frac{1}{3}$ \\
		$1$  & $\nu_q=\frac{4}{11}$ & $\frac{1}{3}$ & $\frac{1}{3}$ & $\frac{1}{3}$ & $\frac{1}{3}$ & $\frac{4}{11}$ & $\frac{4}{11}$ \\
		$2$  & $\nu_q=\frac{7}{19}$ & $\frac{1}{3}$ & $\frac{1}{3}$ & $\frac{1}{3}$ & $\frac{1}{3}$ & $\frac{4}{11}$ & $\frac{7}{19}$ \\
		\hline\hline
	\end{tabular}
	\label{tab.condunctance_3_8}
\end{table}

% --- TABLE 2 ---
\begin{table}[t]
	\centering
	\setlength{\tabcolsep}{2pt}
	\renewcommand{\arraystretch}{1.3}
	\caption{The coherent charge conductance between S2 and D1 in the $\pi$-junction in Fig.~\ref{fig.Pi_junction} for $\nu_\text{NG}(1,-1)=\frac{3}{4}$.}
	\begin{tabular}{c c !{\vrule width .9pt} w{c}{1cm} | w{c}{1cm} | w{c}{1cm} | w{c}{1cm} | w{c}{1cm} | w{c}{1cm} }
		\hline\hline 
		$q$ & $\ell=$ & $-5,-6$ & $-3,-4$ & $-1,-2$ &$0$& $1,2$ & $3,4$ \\ 
		\hline
		$-1$ & $\nu_q=\frac{2}{3}$  & $\frac{2}{3}$   & $\frac{2}{3}$   & $\frac{2}{3}$   & $\frac{6}{11}$  & $\frac{6}{11}$  & $\frac{6}{11}$ \\
		$0$  & $\nu_q=1$            & $\frac{3}{4}$   & $\frac{3}{4}$   & $\frac{3}{4}$   & $\frac{3}{4}$   & $\frac{3}{4}$   & $\frac{3}{4}$ \\
		$1$  & $\nu_q=\frac{4}{5}$  & $\frac{12}{19}$ & $\frac{12}{19}$ & $\frac{12}{19}$ & $\frac{12}{19}$ & $\frac{3}{4}$   & $\frac{3}{4}$ \\
		$2$  & $\nu_q=\frac{7}{9}$  & $\frac{21}{34}$ & $\frac{21}{34}$ & $\frac{21}{34}$ & $\frac{21}{34}$ & $\frac{84}{115}$& $\frac{3}{4}$ \\
		\hline\hline
	\end{tabular}
	\label{tab.condunctance_3_4}
\end{table}

% --- TABLE 3 ---
\begin{table}[t]
	\centering
	\setlength{\tabcolsep}{2pt}
	\renewcommand{\arraystretch}{1.3}
	\caption{The coherent charge conductance between S2 and D1 in the $\pi$-junction in Fig.~\ref{fig.Pi_junction} for $\nu_\text{NG}(1,-2)=\frac{3}{10}$.}
	\begin{tabular}{c c !{\vrule width .9pt} w{c}{1cm} | w{c}{1cm} | w{c}{1cm} | w{c}{1cm} | w{c}{1cm} | w{c}{1cm} }
		\hline\hline 
		$q$ & $\ell=$ & $-5,-6$ & $-3,-4$ & $-1,-2$ &$0$& $1,2$ & $3,4$ \\ 
		\hline
		$-1$ & $\nu_q=\frac{2}{7}$   & $\frac{2}{7}$   & $\frac{2}{7}$   & $\frac{2}{7}$   & $\frac{6}{23}$  & $\frac{6}{23}$  & $\frac{6}{23}$ \\
		$0$  & $\nu_q=\frac{1}{3}$   & $\frac{3}{10}$  & $\frac{3}{10}$  & $\frac{3}{10}$  & $\frac{3}{10}$  & $\frac{3}{10}$  & $\frac{3}{10}$ \\
		$1$  & $\nu_q=\frac{4}{13}$  & $\frac{12}{43}$ & $\frac{12}{43}$ & $\frac{12}{43}$ & $\frac{12}{43}$ & $\frac{3}{10}$  & $\frac{3}{10}$ \\
		$2$  & $\nu_q=\frac{7}{23}$  & $\frac{21}{76}$ & $\frac{21}{76}$ & $\frac{21}{76}$ & $\frac{21}{76}$ & $\frac{84}{283}$& $\frac{3}{10}$ \\
		\hline\hline
	\end{tabular}
	\label{tab.condunctance_3_10}
\end{table}

\bibliography{ref}

\begin{thebibliography}{120}
\providecommand{\natexlab}[1]{#1}
\providecommand{\url}[1]{\texttt{#1}}
\expandafter\ifx\csname urlstyle\endcsname\relax
  \providecommand{\doi}[1]{doi: #1}\else
  \providecommand{\doi}{doi: \begingroup \urlstyle{rm}\Url}\fi

\bibitem[Wen(1990)]{Wen_topological_1990}
X.~G. Wen.
\newblock Topological orders in rigid states.
\newblock \emph{International Journal of Modern Physics B}, 04\penalty0
  (02):\penalty0 239--271, 1990.
\newblock \doi{10.1142/S0217979290000139}.
\newblock URL \url{https://doi.org/10.1142/S0217979290000139}.

\bibitem[Jain(2007)]{Jain_composite_2007}
J.~K. Jain.
\newblock \emph{Composite {Fermions}}.
\newblock Cambridge University Press, 2007.
\newblock \doi{10.1017/CBO9780511607561}.

\bibitem[Halperin and Jain(2020)]{Halperin_FQH_2020}
Bertrand~I Halperin and Jainendra~K Jain.
\newblock \emph{Fractional Quantum {Hall} Effects}.
\newblock WORLD SCIENTIFIC, 2020.
\newblock \doi{10.1142/11751}.
\newblock URL \url{https://www.worldscientific.com/doi/abs/10.1142/11751}.

\bibitem[Willett et~al.(1987)Willett, Eisenstein, St\"ormer, Tsui, Gossard, and
  English]{Willett_observation_1987}
R.~Willett, J.~P. Eisenstein, H.~L. St\"ormer, D.~C. Tsui, A.~C. Gossard, and
  J.~H. English.
\newblock Observation of an even-denominator quantum number in the fractional
  quantum {Hall} effect.
\newblock \emph{Phys. Rev. Lett.}, 59\penalty0 (15):\penalty0 1776, October
  1987.
\newblock \doi{10.1103/PhysRevLett.59.1776}.
\newblock URL \url{https://link.aps.org/doi/10.1103/PhysRevLett.59.1776}.

\bibitem[Haldane(1983)]{Haldane_fqh_1983}
F.~D.~M. Haldane.
\newblock Fractional quantization of the {Hall} effect: {A} hierarchy of
  incompressible quantum fluid states.
\newblock \emph{Phys. Rev. Lett.}, 51\penalty0 (7):\penalty0 605, 1983.
\newblock URL
  \url{https://journals.aps.org/prl/abstract/10.1103/PhysRevLett.51.605}.

\bibitem[Moore and Read(1991)]{Moore_nonabelions_1991}
G.~Moore and N.~Read.
\newblock Nonabelions in the fractional quantum {Hall} effect.
\newblock \emph{Nucl. Phys. B}, 360\penalty0 (2-3):\penalty0 362, 1991.
\newblock URL
  \url{http://www.sciencedirect.com/science/article/pii/055032139190407O}.

\bibitem[Greiter et~al.(1991)Greiter, Wen, and
  Wilczek]{Greiter_half_filled_1991}
M.~Greiter, X.~G. Wen, and F.~Wilczek.
\newblock Paired {Hall} state at half filling.
\newblock \emph{Phys. Rev. Lett.}, 66:\penalty0 3205, Jun 1991.
\newblock \doi{10.1103/PhysRevLett.66.3205}.
\newblock URL \url{https://link.aps.org/doi/10.1103/PhysRevLett.66.3205}.

\bibitem[Read and Green(2000)]{Read_paired_2000}
N.~Read and D.~Green.
\newblock Paired states of fermions in two dimensions with breaking of parity
  and time-reversal symmetries and the fractional quantum {Hall} effect.
\newblock \emph{Phys. Rev. B}, 61\penalty0 (15):\penalty0 10267, 2000.
\newblock URL
  \url{https://journals.aps.org/prb/abstract/10.1103/PhysRevB.61.10267}.

\bibitem[Suen et~al.(1992{\natexlab{a}})Suen, Engel, Santos, Shayegan, and
  Tsui]{Suen_Observation_1992}
Y.~W. Suen, L.~W. Engel, M.~B. Santos, M.~Shayegan, and D.~C. Tsui.
\newblock Observation of a \ensuremath{\nu}=1/2 fractional quantum {Hall} state
  in a double-layer electron system.
\newblock \emph{Phys. Rev. Lett.}, 68:\penalty0 1379--1382, Mar
  1992{\natexlab{a}}.
\newblock \doi{10.1103/PhysRevLett.68.1379}.
\newblock URL \url{https://link.aps.org/doi/10.1103/PhysRevLett.68.1379}.

\bibitem[Suen et~al.(1992{\natexlab{b}})Suen, Santos, and
  Shayegan]{Suen_Correlated_1992}
Y.~W. Suen, M.~B. Santos, and M.~Shayegan.
\newblock Correlated states of an electron system in a wide quantum well.
\newblock \emph{Phys. Rev. Lett.}, 69:\penalty0 3551--3554, Dec
  1992{\natexlab{b}}.
\newblock \doi{10.1103/PhysRevLett.69.3551}.
\newblock URL \url{https://link.aps.org/doi/10.1103/PhysRevLett.69.3551}.

\bibitem[Suen et~al.(1994)Suen, Manoharan, Ying, Santos, and
  Shayegan]{Suen_Origin_1994}
Y.~W. Suen, H.~C. Manoharan, X.~Ying, M.~B. Santos, and M.~Shayegan.
\newblock Origin of the \ensuremath{\nu}=1/2 fractional quantum {Hall} state in
  wide single quantum wells.
\newblock \emph{Phys. Rev. Lett.}, 72:\penalty0 3405--3408, May 1994.
\newblock \doi{10.1103/PhysRevLett.72.3405}.
\newblock URL \url{https://link.aps.org/doi/10.1103/PhysRevLett.72.3405}.

\bibitem[Shabani et~al.(2009{\natexlab{a}})Shabani, Gokmen, and
  Shayegan]{Shabani_Correlated_2009}
J.~Shabani, T.~Gokmen, and M.~Shayegan.
\newblock Correlated states of electrons in wide quantum wells at low fillings:
  The role of charge distribution symmetry.
\newblock \emph{Phys. Rev. Lett.}, 103:\penalty0 046805, Jul
  2009{\natexlab{a}}.
\newblock \doi{10.1103/PhysRevLett.103.046805}.
\newblock URL \url{https://link.aps.org/doi/10.1103/PhysRevLett.103.046805}.

\bibitem[Shabani et~al.(2009{\natexlab{b}})Shabani, Gokmen, Chiu, and
  Shayegan]{shabani_evidence_2009}
J.~Shabani, T.~Gokmen, Y.~T. Chiu, and M.~Shayegan.
\newblock Evidence for developing fractional quantum {Hall} states at even
  denominator $1/2$ and $1/4$ fillings in asymmetric wide quantum wells.
\newblock \emph{Physical Review Letters}, 103:\penalty0 256802, Dec
  2009{\natexlab{b}}.
\newblock \doi{10.1103/PhysRevLett.103.256802}.
\newblock URL \url{https://link.aps.org/doi/10.1103/PhysRevLett.103.256802}.

\bibitem[Shabani et~al.(2013)Shabani, Liu, Shayegan, Pfeiffer, West, and
  Baldwin]{Shabani_Phase_2013}
J.~Shabani, Yang Liu, M.~Shayegan, L.~N. Pfeiffer, K.~W. West, and K.~W.
  Baldwin.
\newblock Phase diagrams for the stability of the
  $\ensuremath{\nu}=\frac{1}{2}$ fractional quantum {Hall} effect in electron
  systems confined to symmetric, wide {GaAs} quantum wells.
\newblock \emph{Phys. Rev. B}, 88:\penalty0 245413, Dec 2013.
\newblock \doi{10.1103/PhysRevB.88.245413}.
\newblock URL \url{https://link.aps.org/doi/10.1103/PhysRevB.88.245413}.

\bibitem[Hasdemir et~al.(2015)Hasdemir, Liu, Deng, Shayegan, Pfeiffer, West,
  Baldwin, and Winkler]{Hasdemir_tilted_2015}
S.~Hasdemir, Yang Liu, H.~Deng, M.~Shayegan, L.~N. Pfeiffer, K.~W. West, K.~W.
  Baldwin, and R.~Winkler.
\newblock $\ensuremath{\nu}=1/2$ fractional quantum {Hall} effect in tilted
  magnetic fields.
\newblock \emph{Phys. Rev. B}, 91:\penalty0 045113, Jan 2015.
\newblock \doi{10.1103/PhysRevB.91.045113}.
\newblock URL \url{https://link.aps.org/doi/10.1103/PhysRevB.91.045113}.

\bibitem[Dorozhkin et~al.(2023)Dorozhkin, Kapustin, Fedorov, Umansky, and
  Smet]{Dorozhkin_Unconventional_2023}
S.~I. Dorozhkin, A.~A. Kapustin, I.~B. Fedorov, V.~Umansky, and J.~H. Smet.
\newblock Unconventional fractional quantum {Hall} states in a wide quantum
  well.
\newblock \emph{JETP Letters}, 117\penalty0 (1):\penalty0 68--74, Jan 2023.
\newblock ISSN 1090-6487.
\newblock \doi{10.1134/S0021364022602974}.
\newblock URL \url{https://doi.org/10.1134/S0021364022602974}.

\bibitem[Liu et~al.(2014)Liu, Graninger, Hasdemir, Shayegan, Pfeiffer, West,
  Baldwin, and Winkler]{Liu_hole_2014}
Yang Liu, A.~L. Graninger, S.~Hasdemir, M.~Shayegan, L.~N. Pfeiffer, K.~W.
  West, K.~W. Baldwin, and R.~Winkler.
\newblock Fractional quantum hall effect at $\ensuremath{\nu}=1/2$ in hole
  systems confined to gaas quantum wells.
\newblock \emph{Phys. Rev. Lett.}, 112:\penalty0 046804, Jan 2014.
\newblock \doi{10.1103/PhysRevLett.112.046804}.
\newblock URL \url{https://link.aps.org/doi/10.1103/PhysRevLett.112.046804}.

\bibitem[Zibrov et~al.(2018)Zibrov, Spanton, Zhou, Kometter, Taniguchi,
  Watanabe, and Young]{Zibrov_Even_Denominator_2018}
A.~A. Zibrov, E.~M. Spanton, H.~Zhou, C.~Kometter, T.~Taniguchi, K.~Watanabe,
  and A.~F. Young.
\newblock Even-denominator fractional quantum {Hall} states at an isospin
  transition in monolayer graphene.
\newblock \emph{Nature Physics}, 14\penalty0 (9):\penalty0 930--935, 2018.
\newblock ISSN 1745-2481.
\newblock \doi{10.1038/s41567-018-0190-0}.
\newblock URL \url{https://doi.org/10.1038/s41567-018-0190-0}.

\bibitem[Kim et~al.(2019)Kim, Balram, Taniguchi, Watanabe, Jain, and
  Smet]{Kim_Even_Denominator_f_wave_2019}
Y.~Kim, A.~C. Balram, T.~Taniguchi, K.~Watanabe, J.~K. Jain, and J.~H. Smet.
\newblock Even-denominator fractional quantum {Hall} states in higher {Landau}
  levels of graphene.
\newblock \emph{Nature Physics}, 15\penalty0 (2):\penalty0 154--158, Feb 2019.
\newblock ISSN 1745-2481.
\newblock \doi{10.1038/s41567-018-0355-x}.
\newblock URL \url{https://doi.org/10.1038/s41567-018-0355-x}.

\bibitem[Ki et~al.(2014)Ki, Fal’ko, Abanin, and
  Morpurgo]{Ki_bilyaer_graphene_2014}
D.~K. Ki, V.~I. Fal’ko, D.~A. Abanin, and A.~F. Morpurgo.
\newblock Observation of even denominator fractional quantum {Hall} effect in
  suspended bilayer graphene.
\newblock \emph{Nano Letters}, 14\penalty0 (4):\penalty0 2135--2139, 2014.
\newblock \doi{10.1021/nl5003922}.
\newblock URL \url{https://doi.org/10.1021/nl5003922}.

\bibitem[Kim et~al.(2015)Kim, Lee, Jung, Skákalová, Taniguchi, Watanabe, Kim,
  and Smet]{Kim_bilayer_graphene_2015}
Y.~Kim, D.~S. Lee, S.~Jung, V.~Skákalová, T.~Taniguchi, K.~Watanabe, J.~S.
  Kim, and J.~H. Smet.
\newblock Fractional quantum {Hall} states in bilayer graphene probed by
  transconductance fluctuations.
\newblock \emph{Nano Letters}, 15\penalty0 (11):\penalty0 7445--7451, 2015.
\newblock \doi{10.1021/acs.nanolett.5b02876}.
\newblock URL \url{https://doi.org/10.1021/acs.nanolett.5b02876}.

\bibitem[Li et~al.(2017)Li, Tan, Chen, Zeng, Taniguchi, Watanabe, Hone, and
  Dean]{Li_bilayer_graphene_2017}
J.~I.~A. Li, C.~Tan, S.~Chen, Y.~Zeng, T.~Taniguchi, K.~Watanabe, J.~Hone, and
  C.~R. Dean.
\newblock Even-denominator fractional quantum {Hall} states in bilayer
  graphene.
\newblock \emph{Science}, 358\penalty0 (6363):\penalty0 648--652, 2017.
\newblock \doi{10.1126/science.aao2521}.
\newblock URL \url{https://www.science.org/doi/abs/10.1126/science.aao2521}.

\bibitem[Zibrov et~al.(2017)Zibrov, Kometter, Zhou, Spanton, Taniguchi,
  Watanabe, Zaletel, and Young]{Zibrov_Tunable_bilayer_graphene_2017}
A.~A. Zibrov, C.~Kometter, H.~Zhou, E.~M. Spanton, T.~Taniguchi, K.~Watanabe,
  M.~P. Zaletel, and A.~F. Young.
\newblock Tunable interacting composite fermion phases in a half-filled
  bilayer-graphene {Landau} level.
\newblock \emph{Nature}, 549\penalty0 (7672):\penalty0 360--364, 2017.
\newblock ISSN 1476-4687.
\newblock \doi{10.1038/nature23893}.
\newblock URL \url{https://doi.org/10.1038/nature23893}.

\bibitem[Huang et~al.(2022)Huang, Fu, Hickey, Alem, Lin, Watanabe, Taniguchi,
  and Zhu]{Huang_Valley_bilayer_graphene_2022}
K.~Huang, H.~Fu, D.~R. Hickey, N.~Alem, X.~Lin, K.~Watanabe, T.~Taniguchi, and
  J.~Zhu.
\newblock Valley isospin controlled fractional quantum {Hall} states in bilayer
  graphene.
\newblock \emph{Phys. Rev. X}, 12:\penalty0 031019, Jul 2022.
\newblock \doi{10.1103/PhysRevX.12.031019}.
\newblock URL \url{https://link.aps.org/doi/10.1103/PhysRevX.12.031019}.

\bibitem[{Kumar} et~al.(2024){Kumar}, {Haug}, {Kim}, {Yutushui}, {Khudiakov},
  {Bhardwaj}, {Ilin}, {Watanabe}, {Taniguchi}, {Mross}, and
  {Ronen}]{Kumar_Quarter_2024}
R.~{Kumar}, A.~{Haug}, J.~{Kim}, M.~{Yutushui}, K.~{Khudiakov}, V.~{Bhardwaj},
  A.~{Ilin}, K.~{Watanabe}, T.~{Taniguchi}, D.~F. {Mross}, and Y.~{Ronen}.
\newblock {Quarter- and half-filled quantum {Hall} states and their competing
  interactions in bilayer graphene}.
\newblock \emph{arXiv e-prints}, art. arXiv:2405.19405, May 2024.
\newblock \doi{10.48550/arXiv.2405.19405}.

\bibitem[Assouline et~al.(2024)Assouline, Wang, Zhou, Cohen, Yang, Zhang,
  Taniguchi, Watanabe, Mong, Zaletel, and
  Young]{Assouline_Energy_Gap_bilayer_graphene_2024}
A.~Assouline, T.~Wang, H.~Zhou, L.~A. Cohen, F.~Yang, R.~Zhang, T.~Taniguchi,
  K.~Watanabe, R.~S.~K. Mong, M.~P. Zaletel, and A.~F. Young.
\newblock Energy gap of the even-denominator fractional quantum {Hall} state in
  bilayer graphene.
\newblock \emph{Phys. Rev. Lett.}, 132:\penalty0 046603, Jan 2024.
\newblock \doi{10.1103/PhysRevLett.132.046603}.
\newblock URL \url{https://link.aps.org/doi/10.1103/PhysRevLett.132.046603}.

\bibitem[Haug et~al.(2025)Haug, Kumar, Firon, Yutushui, Watanabe, Taniguchi,
  Mross, and Ronen]{Haug_Interaction_2025}
Andr{\'e} Haug, Ravi Kumar, Tomer Firon, Misha Yutushui, Kenji Watanabe,
  Takashi Taniguchi, David~F Mross, and Yuval Ronen.
\newblock Interaction-driven quantum phase transitions between topological and
  crystalline orders of electrons.
\newblock \emph{arXiv preprint arXiv:2504.18626}, 2025.

\bibitem[Singh et~al.(2025)Singh, Wang, Gupta, Baldwin, Pfeiffer, and
  Shayegan]{Singh_fractional_2025}
Siddharth~Kumar Singh, Chengyu Wang, Adbhut Gupta, Kirk~W Baldwin, Loren~N
  Pfeiffer, and Mansour Shayegan.
\newblock Fractional quantum hall state at $\nu= 1/2$ with energy gap up to 6
  k, and possible transition from one-to two-component state.
\newblock \emph{arXiv preprint arXiv:2510.03983}, 2025.

\bibitem[{Kumar} et~al.(2025){Kumar}, {Firon}, {Haug}, {Yutushui}, {Ner Gaon},
  {Watanabe}, {Taniguchi}, {Mross}, and {Ronen}]{Kumar_Orbitally_2025}
Ravi {Kumar}, Tomer {Firon}, Andr{\'e} {Haug}, Misha {Yutushui}, Alon {Ner
  Gaon}, Kenji {Watanabe}, Takashi {Taniguchi}, David~F. {Mross}, and Yuval
  {Ronen}.
\newblock {Orbitally tuned composite-fermion metal-to-superfluid transitions}.
\newblock \emph{arXiv e-prints}, art. arXiv:2512.21383, December 2025.
\newblock \doi{10.48550/arXiv.2512.21383}.

\bibitem[Chen et~al.(2024)Chen, Huang, Li, Tong, Kuang, Xi, Watanabe,
  Taniguchi, Liu, Zhu, Lu, Zhang, Wu, and Wang]{chen_tunable_2023}
Y.~Chen, Y.~Huang, Q.~Li, B.~Tong, G.~Kuang, C.~Xi, K.~Watanabe, T.~Taniguchi,
  G.~Liu, Z.~Zhu, L.~Lu, F.-C. Zhang, Y.-H. Wu, and L.~Wang.
\newblock Tunable even- and odd-denominator fractional quantum {Hall} states in
  trilayer graphene.
\newblock \emph{Nature Communications}, 15\penalty0 (1):\penalty0 6236, Jul
  2024.
\newblock ISSN 2041-1723.
\newblock \doi{10.1038/s41467-024-50589-2}.
\newblock URL \url{https://doi.org/10.1038/s41467-024-50589-2}.

\bibitem[Chanda et~al.(2025)Chanda, Kaur, Singh, Watanabe, Taniguchi, Jain,
  Khanna, Balram, and Bid]{chanda_Even_TLG_2025}
Tanima Chanda, Simrandeep Kaur, Harsimran Singh, Kenji Watanabe, Takashi
  Taniguchi, Manish Jain, Udit Khanna, Ajit~C Balram, and Aveek Bid.
\newblock Even denominator fractional quantum {Hall} states in the zeroth
  {Landau} level of monolayer-like band of aba trilayer graphene.
\newblock \emph{arXiv preprint arXiv:2502.06245}, 2025.

\bibitem[Falson et~al.(2015)Falson, Maryenko, Friess, Zhang, Kozuka, Tsukazaki,
  Smet, and Kawasaki]{Falson_Zno_2015}
J.~Falson, D.~Maryenko, B.~Friess, D.~Zhang, Y.~Kozuka, A.~Tsukazaki, J.~H.
  Smet, and M.~Kawasaki.
\newblock Even-denominator fractional quantum {Hall} physics in {ZnO}.
\newblock \emph{Nature Physics}, 11\penalty0 (4):\penalty0 347--351, Apr 2015.
\newblock ISSN 1745-2481.
\newblock \doi{10.1038/nphys3259}.
\newblock URL \url{https://doi.org/10.1038/nphys3259}.

\bibitem[Falson et~al.(2018)Falson, Tabrea, Zhang, Sodemann, Kozuka, Tsukazaki,
  Kawasaki, v.~Klitzing, and Smet]{Falson_Zno_2018}
J.~Falson, D.~Tabrea, D.~Zhang, I.~Sodemann, Y.~Kozuka, A.~Tsukazaki,
  M.~Kawasaki, K.~v.~Klitzing, and J.~H. Smet.
\newblock A cascade of phase transitions in an orbitally mixed half-filled
  {Landau} level.
\newblock \emph{Science Advances}, 4\penalty0 (9):\penalty0 eaat8742, 2018.
\newblock \doi{10.1126/sciadv.aat8742}.
\newblock URL \url{https://www.science.org/doi/abs/10.1126/sciadv.aat8742}.

\bibitem[Falson(2019)]{Falson_Phase_2019}
Joseph Falson.
\newblock Phase transitions at $\ensuremath{\nu}=5/2$ in {ZnO}-based
  heterostructures.
\newblock \emph{Physica E: Low-dimensional Systems and Nanostructures},
  110:\penalty0 49--51, 2019.
\newblock ISSN 1386-9477.
\newblock \doi{https://doi.org/10.1016/j.physe.2019.01.027}.
\newblock URL
  \url{https://www.sciencedirect.com/science/article/pii/S1386947718318587}.

\bibitem[Shi et~al.(2020)Shi, Shih, Gustafsson, Rhodes, Kim, Watanabe,
  Taniguchi, Papi{\'{c}}, Hone, and Dean]{Shi_even_wse2_2020}
Qianhui Shi, En-Min Shih, Martin~V. Gustafsson, Daniel~A. Rhodes, Bumho Kim,
  Kenji Watanabe, Takashi Taniguchi, Zlatko Papi{\'{c}}, James Hone, and
  Cory~R. Dean.
\newblock Odd- and even-denominator fractional quantum {Hall} states in
  monolayer {WSe}$_2$.
\newblock \emph{Nature Nanotechnology}, 15\penalty0 (7):\penalty0 569--573, Jul
  2020.
\newblock ISSN 1748-3395.
\newblock \doi{10.1038/s41565-020-0685-6}.
\newblock URL \url{https://doi.org/10.1038/s41565-020-0685-6}.

\bibitem[Stern(2010)]{Stern_non_Abelian_2010}
Ady Stern.
\newblock Non-abelian states of matter.
\newblock \emph{Nature}, 464\penalty0 (7286):\penalty0 187--193, Mar 2010.
\newblock ISSN 1476-4687.
\newblock \doi{10.1038/nature08915}.
\newblock URL \url{https://doi.org/10.1038/nature08915}.

\bibitem[Banerjee et~al.(2018)Banerjee, Heiblum, Umansky, Feldman, Oreg, and
  Stern]{Banerjee_Observation_2018}
M.~Banerjee, M.~Heiblum, V.~Umansky, D.~E. Feldman, Y.~Oreg, and A.~Stern.
\newblock Observation of half-integer thermal {Hall} conductance.
\newblock \emph{Nature}, 559\penalty0 (7713):\penalty0 205, July 2018.
\newblock ISSN 1476-4687.
\newblock \doi{10.1038/s41586-018-0184-1}.
\newblock URL \url{https://www.nature.com/articles/s41586-018-0184-1}.

\bibitem[Dutta et~al.(2022)Dutta, Umansky, Banerjee, and
  Heiblum]{Dutta_Isolated_2022}
B.~Dutta, V.~Umansky, M.~Banerjee, and M.~Heiblum.
\newblock Isolated ballistic non-{Abelian} interface channel.
\newblock \emph{Science}, 377\penalty0 (6611):\penalty0 1198--1201, 2022.
\newblock \doi{10.1126/science.abm6571}.
\newblock URL \url{https://www.science.org/doi/abs/10.1126/science.abm6571}.

\bibitem[Paul et~al.(2024)Paul, Tiwari, Melcer, Umansky, and
  Heiblum]{Paul_Thermal_2024}
Arup~Kumar Paul, Priya Tiwari, Ron Melcer, Vladimir Umansky, and Moty Heiblum.
\newblock Topological thermal {Hall} conductance of even-denominator fractional
  states.
\newblock \emph{Phys. Rev. Lett.}, 133:\penalty0 076601, Aug 2024.
\newblock \doi{10.1103/PhysRevLett.133.076601}.
\newblock URL \url{https://link.aps.org/doi/10.1103/PhysRevLett.133.076601}.

\bibitem[Son(2015)]{Son_is_2015}
D.~T. Son.
\newblock Is the composite {Fermion} a {Dirac} particle?
\newblock \emph{Phys. Rev. X}, 5:\penalty0 031027, 2015.
\newblock ISSN 2160-3308.
\newblock \doi{10.1103/PhysRevX.5.031027}.
\newblock URL \url{https://link.aps.org/doi/10.1103/PhysRevX.5.031027}.

\bibitem[{Mishmash} et~al.(2018){Mishmash}, {Mross}, {Alicea}, and
  {Motrunich}]{Mishmash_numerical_2018}
R.~V. {Mishmash}, D.~F. {Mross}, J.~{Alicea}, and O.~I. {Motrunich}.
\newblock {Numerical exploration of trial wave functions for the
  particle-hole-symmetric Pfaffian}.
\newblock \emph{Phys. Rev. B}, 98\penalty0 (8):\penalty0 081107, Aug 2018.
\newblock \doi{10.1103/PhysRevB.98.081107}.

\bibitem[Yutushui and Mross(2020)]{Yutushui_Large_scale_2020}
M.~Yutushui and D.~F. Mross.
\newblock Large-scale simulations of particle-hole-symmetric {Pfaffian} trial
  wave functions.
\newblock \emph{Phys. Rev. B}, 102:\penalty0 195153, Nov 2020.
\newblock \doi{10.1103/PhysRevB.102.195153}.
\newblock URL \url{https://link.aps.org/doi/10.1103/PhysRevB.102.195153}.

\bibitem[W\'ojs and Quinn(2005)]{Wojs_3body_2005}
A.~W\'ojs and J.~J. Quinn.
\newblock Three-body correlations and finite-size effects in {Moore-Read}
  states on a sphere.
\newblock \emph{Physical Review B}, 71:\penalty0 045324, Jan 2005.
\newblock \doi{10.1103/PhysRevB.71.045324}.
\newblock URL \url{https://link.aps.org/doi/10.1103/PhysRevB.71.045324}.

\bibitem[Wang et~al.(2009)Wang, Sheng, and Haldane]{wang_particle_hole_2009}
Hao Wang, D.~N. Sheng, and F.~D.~M. Haldane.
\newblock Particle-hole symmetry breaking and the $\nu=\frac{5}{2}$ fractional
  quantum {Hall} effect.
\newblock \emph{Phys. Rev. B}, 80\penalty0 (24):\penalty0 241311, December
  2009.
\newblock \doi{10.1103/PhysRevB.80.241311}.
\newblock URL \url{https://link.aps.org/doi/10.1103/PhysRevB.80.241311}.

\bibitem[Rezayi and Simon(2011)]{Rezayi_breaking_2011}
E.~H. Rezayi and S.~H. Simon.
\newblock Breaking of particle-hole symmetry by {Landau} level mixing in the
  $\nu = 5/2$ quantized {Hall} state.
\newblock \emph{Phys. Rev. Lett.}, 106:\penalty0 116801, Mar 2011.
\newblock \doi{10.1103/PhysRevLett.106.116801}.
\newblock URL \url{https://link.aps.org/doi/10.1103/PhysRevLett.106.116801}.

\bibitem[Zaletel et~al.(2015)Zaletel, Mong, Pollmann, and
  Rezayi]{zaletel_infinite_2015}
M.~P. Zaletel, R.~S.~K. Mong, F.~Pollmann, and E.~H. Rezayi.
\newblock Infinite density matrix renormalization group for multicomponent
  quantum {Hall} systems.
\newblock \emph{Phys. Rev. B}, 91\penalty0 (4):\penalty0 045115, January 2015.
\newblock \doi{10.1103/PhysRevB.91.045115}.
\newblock URL \url{https://link.aps.org/doi/10.1103/PhysRevB.91.045115}.

\bibitem[Tylan-Tyler and Lyanda-Geller(2015)]{Tylan_Phase_2015}
Anthony Tylan-Tyler and Yuli Lyanda-Geller.
\newblock Phase diagram and edge states of the $\ensuremath{\nu}=5/2$
  fractional quantum {Hall} state with {Landau} level mixing and finite well
  thickness.
\newblock \emph{Phys. Rev. B}, 91:\penalty0 205404, May 2015.
\newblock \doi{10.1103/PhysRevB.91.205404}.
\newblock URL \url{https://link.aps.org/doi/10.1103/PhysRevB.91.205404}.

\bibitem[Rezayi(2017)]{Rezayi_Landau_2017}
E.~H. Rezayi.
\newblock Landau level mixing and the ground state of the $\nu=5/2$ quantum
  {Hall} effect.
\newblock \emph{Phys. Rev. Lett.}, 119\penalty0 (2):\penalty0 026801, July
  2017.
\newblock \doi{10.1103/PhysRevLett.119.026801}.
\newblock URL \url{https://link.aps.org/doi/10.1103/PhysRevLett.119.026801}.

\bibitem[Antoni\ifmmode~\acute{c}\else \'{c}\fi{}
  et~al.(2018)Antoni\ifmmode~\acute{c}\else \'{c}\fi{}, Vu\ifmmode
  \check{c}\else \v{c}\fi{}i\ifmmode \check{c}\else
  \v{c}\fi{}evi\ifmmode~\acute{c}\else \'{c}\fi{}, and
  Milovanovi\ifmmode~\acute{c}\else \'{c}\fi{}]{Antoni_Paired_2018}
L.~Antoni\ifmmode~\acute{c}\else \'{c}\fi{}, J.~Vu\ifmmode \check{c}\else
  \v{c}\fi{}i\ifmmode \check{c}\else \v{c}\fi{}evi\ifmmode~\acute{c}\else
  \'{c}\fi{}, and M.~V. Milovanovi\ifmmode~\acute{c}\else \'{c}\fi{}.
\newblock Paired states at 5/2: Particle-hole {Pfaffian} and particle-hole
  symmetry breaking.
\newblock \emph{Phys. Rev. B}, 98:\penalty0 115107, Sep 2018.
\newblock \doi{10.1103/PhysRevB.98.115107}.
\newblock URL \url{https://link.aps.org/doi/10.1103/PhysRevB.98.115107}.

\bibitem[Zhao et~al.(2023)Zhao, Balram, and Jain]{Zhao_CF_pairing_LLM_2023}
T.~Zhao, A.~C. Balram, and J.~K. Jain.
\newblock Composite fermion pairing induced by {Landau} level mixing.
\newblock \emph{Physical Review Letters}, 130:\penalty0 186302, May 2023.
\newblock \doi{10.1103/PhysRevLett.130.186302}.
\newblock URL \url{https://link.aps.org/doi/10.1103/PhysRevLett.130.186302}.

\bibitem[Simon(2018)]{Simon_equilibration_2018}
S.~H. Simon.
\newblock Interpretation of thermal conductance of the $\nu=$~5/2 edge.
\newblock \emph{Phys. Rev. B}, 97:\penalty0 121406(R), Mar 2018.
\newblock \doi{10.1103/PhysRevB.97.121406}.
\newblock URL \url{https://link.aps.org/doi/10.1103/PhysRevB.97.121406}.

\bibitem[Ma and Feldman(2019{\natexlab{a}})]{Feldman_equilibration_2019}
K.~K.~W. Ma and D.~E. Feldman.
\newblock Partial equilibration of integer and fractional edge channels in the
  thermal quantum {Hall} effect.
\newblock \emph{Phys. Rev. B}, 99:\penalty0 085309, Feb 2019{\natexlab{a}}.
\newblock \doi{10.1103/PhysRevB.99.085309}.
\newblock URL \url{https://link.aps.org/doi/10.1103/PhysRevB.99.085309}.

\bibitem[Simon and Rosenow(2020)]{Simon_equilibration_2020}
S.~H. Simon and B.~Rosenow.
\newblock Partial equilibration of the anti-{Pfaffian} edge due to {Majorana}
  disorder.
\newblock \emph{Phys. Rev. Lett.}, 124:\penalty0 126801, Mar 2020.
\newblock \doi{10.1103/PhysRevLett.124.126801}.
\newblock URL \url{https://link.aps.org/doi/10.1103/PhysRevLett.124.126801}.

\bibitem[Asasi and Mulligan(2020)]{Asasi_equilibration_2020}
H.~Asasi and M.~Mulligan.
\newblock Partial equilibration of anti-{Pfaffian} edge modes at
  $\ensuremath{\nu}=5/2$.
\newblock \emph{Phys. Rev. B}, 102:\penalty0 205104, Nov 2020.
\newblock \doi{10.1103/PhysRevB.102.205104}.
\newblock URL \url{https://link.aps.org/doi/10.1103/PhysRevB.102.205104}.

\bibitem[Lotri{\v{c}} et~al.(2025)Lotri{\v{c}}, Wang, Zaletel, Simon, and
  Parameswaran]{Lotric_Majorana_2025}
Tev{\v{z}} Lotri{\v{c}}, Taige Wang, Michael~P Zaletel, Steven~H Simon, and
  SA~Parameswaran.
\newblock Majorana edge reconstruction and the $\backslash nu= 5/2$ non-abelian
  thermal hall puzzle.
\newblock \emph{arXiv preprint arXiv:2507.07161}, 2025.

\bibitem[Mross et~al.(2018)Mross, Oreg, Stern, Margalit, and
  Heiblum]{Mross_theory_2018}
D.~F. Mross, Y.~Oreg, A.~Stern, G.~Margalit, and M.~Heiblum.
\newblock Theory of disorder-induced half-integer thermal {Hall} conductance.
\newblock \emph{Phys. Rev. Lett.}, 121:\penalty0 026801, Jul 2018.
\newblock \doi{10.1103/PhysRevLett.121.026801}.
\newblock URL \url{https://link.aps.org/doi/10.1103/PhysRevLett.121.026801}.

\bibitem[Wang et~al.(2018)Wang, Vishwanath, and
  Halperin]{Wang_topological_2018}
C.~Wang, A.~Vishwanath, and B.~I. Halperin.
\newblock Topological order from disorder and the quantized {Hall} thermal
  metal: Possible applications to the $\nu=5/2$ state.
\newblock \emph{Phys. Rev. B}, 98:\penalty0 045112, Jul 2018.
\newblock \doi{10.1103/PhysRevB.98.045112}.
\newblock URL \url{https://link.aps.org/doi/10.1103/PhysRevB.98.045112}.

\bibitem[Lian and Wang(2018)]{Lian_theory_2018}
B.~Lian and J.~Wang.
\newblock Theory of the disordered $\nu=$~5/2 quantum thermal {Hall} state:
  Emergent symmetry and phase diagram.
\newblock \emph{Phys. Rev. B}, 97:\penalty0 165124, Apr 2018.
\newblock \doi{10.1103/PhysRevB.97.165124}.
\newblock URL \url{https://link.aps.org/doi/10.1103/PhysRevB.97.165124}.

\bibitem[Zhu et~al.(2020)Zhu, Sheng, and Yang]{zhu2020}
W.~Zhu, D.~N. Sheng, and Kun Yang.
\newblock Topological interface between pfaffian and anti-pfaffian order in
  $\ensuremath{\nu}=5/2$ quantum hall effect.
\newblock \emph{Phys. Rev. Lett.}, 125:\penalty0 146802, Sep 2020.
\newblock \doi{10.1103/PhysRevLett.125.146802}.
\newblock URL \url{https://link.aps.org/doi/10.1103/PhysRevLett.125.146802}.

\bibitem[Fulga et~al.(2020)Fulga, Oreg, Mirlin, Stern, and Mross]{fulga2020}
I.~C. Fulga, Yuval Oreg, Alexander~D. Mirlin, Ady Stern, and David~F. Mross.
\newblock Temperature enhancement of thermal hall conductance quantization.
\newblock \emph{Phys. Rev. Lett.}, 125:\penalty0 236802, Dec 2020.
\newblock \doi{10.1103/PhysRevLett.125.236802}.
\newblock URL \url{https://link.aps.org/doi/10.1103/PhysRevLett.125.236802}.

\bibitem[Sp\aa{}nsl\"att et~al.(2019)Sp\aa{}nsl\"att, Park, Gefen, and
  Mirlin]{Spanslatt_Noise_2019}
C.~Sp\aa{}nsl\"att, J.~Park, Y.~Gefen, and A.~D. Mirlin.
\newblock Topological classification of shot noise on fractional quantum {Hall}
  edges.
\newblock \emph{Phys. Rev. Lett.}, 123:\penalty0 137701, Sep 2019.
\newblock \doi{10.1103/PhysRevLett.123.137701}.
\newblock URL \url{https://link.aps.org/doi/10.1103/PhysRevLett.123.137701}.

\bibitem[Park et~al.(2020)Park, Sp\aa{}nsl\"att, Gefen, and
  Mirlin]{Park_noise_2020}
J.~Park, C.~Sp\aa{}nsl\"att, Y.~Gefen, and A.~D. Mirlin.
\newblock Noise on the non-{Abelian} $\ensuremath{\nu}=5/2$ fractional quantum
  {Hall} edge.
\newblock \emph{Phys. Rev. Lett.}, 125:\penalty0 157702, Oct 2020.
\newblock \doi{10.1103/PhysRevLett.125.157702}.
\newblock URL \url{https://link.aps.org/doi/10.1103/PhysRevLett.125.157702}.

\bibitem[Manna et~al.(2024)Manna, Das, Goldstein, and
  Gefen]{Manna_Full_Classification_2022}
S.~Manna, A.~Das, M.~Goldstein, and Y.~Gefen.
\newblock Full classification of transport on an equilibrated $5/2$ edge via
  shot noise.
\newblock \emph{Phys. Rev. Lett.}, 132:\penalty0 136502, Mar 2024.
\newblock \doi{10.1103/PhysRevLett.132.136502}.
\newblock URL \url{https://link.aps.org/doi/10.1103/PhysRevLett.132.136502}.

\bibitem[Yutushui and Mross(2023)]{Yutushui_Identifying_2023}
Misha Yutushui and David~F. Mross.
\newblock Identifying {non-Abelian} anyons with upstream noise.
\newblock \emph{Phys. Rev. B}, 108:\penalty0 L241102, Dec 2023.
\newblock \doi{10.1103/PhysRevB.108.L241102}.
\newblock URL \url{https://link.aps.org/doi/10.1103/PhysRevB.108.L241102}.

\bibitem[Manna and Das(2023)]{Manna_Experimentally_2023}
S.~Manna and A.~Das.
\newblock Experimentally motivated order of length scales affect shot noise,
  2023.

\bibitem[Ma and Feldman(2019{\natexlab{b}})]{Ken_sixteenfold_2019}
K.~K.~W. Ma and D.~E. Feldman.
\newblock The sixteenfold way and the quantum {Hall} effect at half-integer
  filling factors.
\newblock \emph{Phys. Rev. B}, 100:\penalty0 035302, Jul 2019{\natexlab{b}}.
\newblock \doi{10.1103/PhysRevB.100.035302}.
\newblock URL \url{https://link.aps.org/doi/10.1103/PhysRevB.100.035302}.

\bibitem[Yutushui et~al.(2022)Yutushui, Stern, and
  Mross]{Yutushui_Identifying_2022}
Misha Yutushui, Ady Stern, and David~F. Mross.
\newblock Identifying the $\ensuremath{\nu}=\frac{5}{2}$ topological order
  through charge transport measurements.
\newblock \emph{Phys. Rev. Lett.}, 128:\penalty0 016401, Jan 2022.
\newblock \doi{10.1103/PhysRevLett.128.016401}.
\newblock URL \url{https://link.aps.org/doi/10.1103/PhysRevLett.128.016401}.

\bibitem[Yutushui et~al.(2025{\natexlab{a}})Yutushui, Stern, and
  Mross]{Yutushui_Universal_2025}
Misha Yutushui, Ady Stern, and David~F. Mross.
\newblock Universal charge conductance at abelian--non-abelian quantum hall
  interfaces.
\newblock \emph{Phys. Rev. Lett.}, 134:\penalty0 206301, May
  2025{\natexlab{a}}.
\newblock \doi{10.1103/PhysRevLett.134.206301}.
\newblock URL \url{https://link.aps.org/doi/10.1103/PhysRevLett.134.206301}.

\bibitem[Lee et~al.(2001)Lee, Scarola, and Jain]{Lee_Stripe_2001}
Seung-Yeop Lee, Vito~W. Scarola, and J.~K. Jain.
\newblock Stripe formation in the fractional quantum hall regime.
\newblock \emph{Phys. Rev. Lett.}, 87:\penalty0 256803, Nov 2001.
\newblock \doi{10.1103/PhysRevLett.87.256803}.
\newblock URL \url{https://link.aps.org/doi/10.1103/PhysRevLett.87.256803}.

\bibitem[Scarola et~al.(2002)Scarola, Jain, and Rezayi]{Scarola_Possible_2002}
V.~W. Scarola, J.~K. Jain, and E.~H. Rezayi.
\newblock Possible pairing-induced even-denominator fractional quantum hall
  effect in the lowest landau level.
\newblock \emph{Phys. Rev. Lett.}, 88:\penalty0 216804, May 2002.
\newblock \doi{10.1103/PhysRevLett.88.216804}.
\newblock URL \url{https://link.aps.org/doi/10.1103/PhysRevLett.88.216804}.

\bibitem[W\'ojs et~al.(2004)W\'ojs, Yi, and Quinn]{Wojs_Fractional_2004}
Arkadiusz W\'ojs, Kyung-Soo Yi, and John~J. Quinn.
\newblock Fractional quantum hall states of clustered composite fermions.
\newblock \emph{Phys. Rev. B}, 69:\penalty0 205322, May 2004.
\newblock \doi{10.1103/PhysRevB.69.205322}.
\newblock URL \url{https://link.aps.org/doi/10.1103/PhysRevB.69.205322}.

\bibitem[Mukherjee et~al.(2012)Mukherjee, Mandal, W\'ojs, and
  Jain]{Mukherjee_Possible_2012}
Sutirtha Mukherjee, Sudhansu~S. Mandal, Arkadiusz W\'ojs, and Jainendra~K.
  Jain.
\newblock Possible anti-pfaffian pairing of composite fermions at
  $\ensuremath{\nu}=3/8$.
\newblock \emph{Phys. Rev. Lett.}, 109:\penalty0 256801, Dec 2012.
\newblock \doi{10.1103/PhysRevLett.109.256801}.
\newblock URL \url{https://link.aps.org/doi/10.1103/PhysRevLett.109.256801}.

\bibitem[{Mukherjee} and {Mandal}(2015)]{Mukherjee_incompressible_2015}
S.~{Mukherjee} and S.~S. {Mandal}.
\newblock {Incompressible states of the interacting composite fermions in
  negative effective magnetic fields at $\nu~=$~4/13,~5/17, and 3/10}.
\newblock \emph{Phys. Rev. B}, 92\penalty0 (23):\penalty0 235302, Dec 2015.
\newblock \doi{10.1103/PhysRevB.92.235302}.

\bibitem[Bonderson and Slingerland(2008)]{Bonderson_hierarchy_2008}
P.~Bonderson and J.~K. Slingerland.
\newblock Fractional quantum {Hall} hierarchy and the second {Landau} level.
\newblock \emph{Phys. Rev. B}, 78:\penalty0 125323, Sep 2008.
\newblock \doi{10.1103/PhysRevB.78.125323}.
\newblock URL \url{https://link.aps.org/doi/10.1103/PhysRevB.78.125323}.

\bibitem[Wen and Zee(1992)]{Wen_shift_1992}
X.~G. Wen and A.~Zee.
\newblock Shift and spin vector: {New} topological quantum numbers for the
  {Hall} fluids.
\newblock \emph{Phys. Rev. Lett.}, 69\penalty0 (6):\penalty0 953, August 1992.
\newblock \doi{10.1103/PhysRevLett.69.953}.
\newblock URL \url{https://link.aps.org/doi/10.1103/PhysRevLett.69.953}.

\bibitem[Lee et~al.(2007)Lee, Ryu, Nayak, and Fisher]{Lee_particle_hole_2007}
S.~S. Lee, S.~Ryu, C.~Nayak, and M.~P.~A. Fisher.
\newblock Particle-hole symmetry and the \ensuremath{\nu}=~5/2 quantum {Hall}
  state.
\newblock \emph{Phys. Rev. Lett.}, 99\penalty0 (23):\penalty0 236807, December
  2007.
\newblock ISSN 0031-9007, 1079-7114.
\newblock \doi{10.1103/PhysRevLett.99.236807}.
\newblock URL \url{https://link.aps.org/doi/10.1103/PhysRevLett.99.236807}.

\bibitem[Levin et~al.(2007)Levin, Halperin, and
  Rosenow]{Levin_particle_hole_2007}
M.~Levin, B.~I. Halperin, and B.~Rosenow.
\newblock Particle-hole symmetry and the {Pfaffian} state.
\newblock \emph{Phys. Rev. Lett.}, 99\penalty0 (23):\penalty0 236806, December
  2007.
\newblock ISSN 0031-9007, 1079-7114.
\newblock \doi{10.1103/PhysRevLett.99.236806}.
\newblock URL \url{https://link.aps.org/doi/10.1103/PhysRevLett.99.236806}.

\bibitem[Wen(1991)]{Wen_Non-Abelian_1991}
X.~G. Wen.
\newblock Non-{Abelian} statistics in the fractional quantum {Hall} states.
\newblock \emph{Phys. Rev. Lett.}, 66:\penalty0 802--805, Feb 1991.
\newblock \doi{10.1103/PhysRevLett.66.802}.
\newblock URL \url{https://link.aps.org/doi/10.1103/PhysRevLett.66.802}.

\bibitem[Moore and Seiberg(1989)]{moore_classical_1989}
Gregory Moore and Nathan Seiberg.
\newblock Classical and quantum conformal field theory.
\newblock \emph{Communications in Mathematical Physics}, 123\penalty0
  (2):\penalty0 177--254, 1989.

\bibitem[Moore and Seiberg(1990)]{Moore_Lectures_1990}
Gregory Moore and Nathan Seiberg.
\newblock \emph{Lectures on RCFT}, pages 263--361.
\newblock Springer US, Boston, MA, 1990.
\newblock ISBN 978-1-4615-3802-8.
\newblock \doi{10.1007/978-1-4615-3802-8_8}.
\newblock URL \url{https://doi.org/10.1007/978-1-4615-3802-8_8}.

\bibitem[Pan et~al.(2003)Pan, Stormer, Tsui, Pfeiffer, Baldwin, and
  West]{Pan_Fractional_2003}
W.~Pan, H.~L. Stormer, D.~C. Tsui, L.~N. Pfeiffer, K.~W. Baldwin, and K.~W.
  West.
\newblock Fractional quantum {Hall} effect of composite fermions.
\newblock \emph{Phys. Rev. Lett.}, 90:\penalty0 016801, Jan 2003.
\newblock \doi{10.1103/PhysRevLett.90.016801}.
\newblock URL \url{https://link.aps.org/doi/10.1103/PhysRevLett.90.016801}.

\bibitem[Pan et~al.(2015)Pan, Baldwin, West, Pfeiffer, and
  Tsui]{Pan_Fractional_2015}
W.~Pan, K.~W. Baldwin, K.~W. West, L.~N. Pfeiffer, and D.~C. Tsui.
\newblock Fractional quantum {Hall} effect at {Landau} level filling
  $\ensuremath{\nu}=4/11$.
\newblock \emph{Phys. Rev. B}, 91:\penalty0 041301, Jan 2015.
\newblock \doi{10.1103/PhysRevB.91.041301}.
\newblock URL \url{https://link.aps.org/doi/10.1103/PhysRevB.91.041301}.

\bibitem[Samkharadze et~al.(2015)Samkharadze, Arnold, Pfeiffer, West, and
  Cs\'athy]{Samkharadze_Observation_411_2015}
N.~Samkharadze, I.~Arnold, L.~N. Pfeiffer, K.~W. West, and G.~A. Cs\'athy.
\newblock Observation of incompressibility at $\ensuremath{\nu}=4/11$ and
  $\ensuremath{\nu}=5/13$.
\newblock \emph{Phys. Rev. B}, 91:\penalty0 081109, Feb 2015.
\newblock \doi{10.1103/PhysRevB.91.081109}.
\newblock URL \url{https://link.aps.org/doi/10.1103/PhysRevB.91.081109}.

\bibitem[Note1()]{Note1}
Note1.
\newblock States at $\nu =\protect \frac {1}{4}$ can be described by
  conventional four-flux CFs, and $\nu =\protect \frac {3}{4}$ as their hole
  conjugates~\cite {Huang_non_Abelian_2024,Yutushui_phase_2025}. At the end of
  Sec.~\ref {sec.equivalence}, we prove the equivalence of topological orders
  obtained by different CF prescriptions.

\bibitem[Wang et~al.(2022)Wang, Gupta, Singh, Chung, Pfeiffer, West, Baldwin,
  Winkler, and Shayegan]{Wang_even_3_4_2022}
Chengyu Wang, A.~Gupta, S.~K. Singh, Y.~J. Chung, L.~N. Pfeiffer, K.~W. West,
  K.~W. Baldwin, R.~Winkler, and M.~Shayegan.
\newblock Even-denominator fractional quantum {Hall} state at filling factor
  $\ensuremath{\nu}=3/4$.
\newblock \emph{Phys. Rev. Lett.}, 129:\penalty0 156801, Oct 2022.
\newblock \doi{10.1103/PhysRevLett.129.156801}.
\newblock URL \url{https://link.aps.org/doi/10.1103/PhysRevLett.129.156801}.

\bibitem[Wang et~al.(2023)Wang, Gupta, Madathil, Singh, Chung, Pfeiffer,
  Baldwin, and Shayegan]{Wang_Next_generation_2023}
Chengyu Wang, Adbhut Gupta, Pranav~T. Madathil, Siddharth~K. Singh, Yoon~Jang
  Chung, Loren~N. Pfeiffer, Kirk~W. Baldwin, and Mansour Shayegan.
\newblock Next-generation even-denominator fractional quantum {Hall} states of
  interacting composite fermions.
\newblock \emph{Proceedings of the National Academy of Sciences}, 120\penalty0
  (52):\penalty0 e2314212120, 2023.
\newblock \doi{10.1073/pnas.2314212120}.
\newblock URL \url{https://www.pnas.org/doi/abs/10.1073/pnas.2314212120}.

\bibitem[Huang and Wu(2024)]{Huang_non_Abelian_2024}
Kai-Wen Huang and Ying-Hai Wu.
\newblock Non-abelian fractional quantum {Hall} states at filling factor 3/4.
\newblock \emph{arXiv preprint arXiv:2408.16275}, 2024.

\bibitem[Yutushui and Mross(2025)]{Yutushui_phase_2025}
M.~Yutushui and D.~F. Mross.
\newblock Phase diagram of compressible and paired states in the quarter-filled
  {Landau} level.
\newblock \emph{Phys. Rev. B}, 111:\penalty0 035106, Jan 2025.
\newblock \doi{10.1103/PhysRevB.111.035106}.
\newblock URL \url{https://link.aps.org/doi/10.1103/PhysRevB.111.035106}.

\bibitem[Bonderson et~al.(2012)Bonderson, Feiguin, M\"oller, and
  Slingerland]{Bonderson_Competing_2012}
Parsa Bonderson, Adrian~E. Feiguin, Gunnar M\"oller, and J.~K. Slingerland.
\newblock Competing topological orders in the $\ensuremath{\nu}=12/5$ quantum
  {Hall} state.
\newblock \emph{Phys. Rev. Lett.}, 108:\penalty0 036806, Jan 2012.
\newblock \doi{10.1103/PhysRevLett.108.036806}.
\newblock URL \url{https://link.aps.org/doi/10.1103/PhysRevLett.108.036806}.

\bibitem[Kane and Fisher(1995)]{Kane_Impurity_1995}
C.~L. Kane and M.~P.~A. Fisher.
\newblock Impurity scattering and transport of fractional quantum {Hall} edge
  states.
\newblock \emph{Phys. Rev. B}, 51\penalty0 (19):\penalty0 13449--13466, 1995.
\newblock \doi{10.1103/PhysRevB.51.13449}.
\newblock URL
  \url{https://www.scopus.com/inward/record.uri?eid=2-s2.0-0001191058&doi=10.1103%2fPhysRevB.51.13449&partnerID=40&md5=2afe4b51a63a7a267088f7ba0e155d33}.

\bibitem[Moore and Wen(1998)]{Moore_Classification_1997}
J.~E. Moore and X.~G. Wen.
\newblock Classification of disordered phases of quantum {Hall} edge states.
\newblock \emph{Phys. Rev. B}, 57:\penalty0 10138--10156, Apr 1998.
\newblock \doi{10.1103/PhysRevB.57.10138}.
\newblock URL \url{https://link.aps.org/doi/10.1103/PhysRevB.57.10138}.

\bibitem[Moore and Wen(2002)]{Moore_Critical_2002}
J.~E. Moore and X.~G. Wen.
\newblock Critical points in edge tunneling between generic fractional quantum
  {Hall} states.
\newblock \emph{Phys. Rev. B}, 66:\penalty0 115305, Sep 2002.
\newblock \doi{10.1103/PhysRevB.66.115305}.
\newblock URL \url{https://link.aps.org/doi/10.1103/PhysRevB.66.115305}.

\bibitem[Haldane(1995)]{Haldane_Stability_1995}
F.~D.~M. Haldane.
\newblock Stability of chiral {Luttinger} liquids and {Abelian} quantum {Hall}
  states.
\newblock \emph{Phys. Rev. Lett.}, 74:\penalty0 2090--2093, Mar 1995.
\newblock \doi{10.1103/PhysRevLett.74.2090}.
\newblock URL \url{https://link.aps.org/doi/10.1103/PhysRevLett.74.2090}.

\bibitem[Note2()]{Note2}
Note2.
\newblock Notice that off-diagonal terms of the $K$-matrix are zero, reflecting
  the fact that CFs of state A do not feel fluxes on CFs of state B and vice
  versa.

\bibitem[Note3()]{Note3}
Note3.
\newblock The $\ell =-1$ case is equivalent to the present case with the
  extended $K$-matrix $\protect \tilde {K}= \protect \text {daig}(K,-1)$ and
  charge vector $(\protect \bm {t},0)$.

\bibitem[Note4()]{Note4}
Note4.
\newblock The topological stability for non-Abelian edges has been analyzed for
  several examples~\cite {Lee_particle_hole_2007,Levin_particle_hole_2007,
  Bishara_PH_Read_Rezayi_2008,Yutushui_Identifying_2023}, but we are not aware
  of a generalization of Eq.~\protect \eqref {eqn.haldantstability} for the
  non-Abelian case.

\bibitem[Chang(2003)]{Chang_Chiral_LL_2003}
A.~M. Chang.
\newblock Chiral {Luttinger} liquids at the fractional quantum {Hall} edge.
\newblock \emph{Rev. Mod. Phys.}, 75:\penalty0 1449--1505, Nov 2003.
\newblock \doi{10.1103/RevModPhys.75.1449}.
\newblock URL \url{https://link.aps.org/doi/10.1103/RevModPhys.75.1449}.

\bibitem[Lin et~al.(2012)Lin, Dillard, Kastner, Pfeiffer, and
  West]{Lin_Measurements_2012}
X.~Lin, C.~Dillard, M.~A. Kastner, L.~N. Pfeiffer, and K.~W. West.
\newblock Measurements of quasiparticle tunneling in the $\nu=5/2$ fractional
  quantum {Hall} state.
\newblock \emph{Phys. Rev. B}, 85:\penalty0 165321, Apr 2012.
\newblock \doi{10.1103/PhysRevB.85.165321}.
\newblock URL \url{https://link.aps.org/doi/10.1103/PhysRevB.85.165321}.

\bibitem[Baer et~al.(2014)Baer, R\"ossler, Ihn, Ensslin, Reichl, and
  Wegscheider]{Baer_experimental_2014}
S.~Baer, C.~R\"ossler, T.~Ihn, K.~Ensslin, C.~Reichl, and W.~Wegscheider.
\newblock Experimental probe of topological orders and edge excitations in the
  second {Landau} level.
\newblock \emph{Phys. Rev. B}, 90:\penalty0 075403, Aug 2014.
\newblock \doi{10.1103/PhysRevB.90.075403}.
\newblock URL \url{https://link.aps.org/doi/10.1103/PhysRevB.90.075403}.

\bibitem[Fu et~al.(2016)Fu, Wang, Shan, Xiong, Pfeiffer, West, Kastner, and
  Lin]{Fu_Competing_2016}
Hailong Fu, Pengjie Wang, Pujia Shan, Lin Xiong, Loren~N. Pfeiffer, Ken West,
  Marc~A. Kastner, and Xi~Lin.
\newblock Competing $\nu$ = 5/2 fractional quantum {Hall} states in confined
  geometry.
\newblock \emph{Proceedings of the National Academy of Sciences}, 113\penalty0
  (44):\penalty0 12386--12390, 2016.
\newblock \doi{10.1073/pnas.1614543113}.
\newblock URL \url{https://www.pnas.org/doi/abs/10.1073/pnas.1614543113}.

\bibitem[Radu et~al.(2008)Radu, Miller, Marcus, Kastner, Pfeiffer, and
  West]{Radu_quasi_particle_2008}
Iuliana~P. Radu, J.~B. Miller, C.~M. Marcus, M.~A. Kastner, L.~N. Pfeiffer, and
  K.~W. West.
\newblock Quasi-particle properties from tunneling in the
  $\ensuremath{\nu}=5/2$ fractional quantum {Hall} state.
\newblock \emph{Science}, 320\penalty0 (5878):\penalty0 899, 2008.
\newblock ISSN 0036-8075.
\newblock \doi{10.1126/science.1157560}.

\bibitem[Goerbig et~al.(2004)Goerbig, Lederer, and
  Smith]{Goerbig_Competition_2004}
M.~O. Goerbig, P.~Lederer, and C.~Morais Smith.
\newblock {Competition between quantum-liquid and electron-solid phases in
  intermediate {Landau} levels}.
\newblock \emph{Phys. Rev. B}, 69:\penalty0 115327, 2004.

\bibitem[Fogler and Koulakov(1997)]{Fogler_Laughlin_wigner_1997}
M.~M. Fogler and A.~A. Koulakov.
\newblock {Laughlin liquid to charge-density-wave transition at high {Landau}
  levels}.
\newblock \emph{Phys. Rev. B}, 55:\penalty0 9326--9329, 1997.

\bibitem[Henderson et~al.(2023)Henderson, M\"oller, and
  Simon]{Henderson_Energy_2023}
G.~J. Henderson, G.~M\"oller, and S.~H. Simon.
\newblock Energy minimization of paired composite fermion wave functions in the
  spherical geometry.
\newblock \emph{Phys. Rev. B}, 108:\penalty0 245128, Dec 2023.
\newblock \doi{10.1103/PhysRevB.108.245128}.
\newblock URL \url{https://link.aps.org/doi/10.1103/PhysRevB.108.245128}.

\bibitem[Wu and Yang(1976)]{Wu_dirac_1976}
T.~T. Wu and C.~N. Yang.
\newblock Dirac monopole without strings: {Monopole} harmonics.
\newblock \emph{Nucl. Phys. B}, 107\penalty0 (3):\penalty0 365, 1976.
\newblock ISSN 0550-3213.
\newblock \doi{http://dx.doi.org/10.1016/0550-3213(76)90143-7}.
\newblock URL
  \url{http://www.sciencedirect.com/science/article/pii/0550321376901437}.

\bibitem[Wu and Yang(1977)]{Wu_properties_1977}
T.~T. Wu and C.~N. Yang.
\newblock Some properties of monopole harmonics.
\newblock \emph{Phys. Rev. D}, 16\penalty0 (4):\penalty0 1018, August 1977.
\newblock \doi{10.1103/PhysRevD.16.1018}.
\newblock URL \url{http://link.aps.org/doi/10.1103/PhysRevD.16.1018}.

\bibitem[Yutushui et~al.(2025{\natexlab{b}})Yutushui, Hermanns, and
  Mross]{Yutushui_Non_Abelian_2025}
Misha Yutushui, Maria Hermanns, and David~F. Mross.
\newblock Non-abelian phases from the condensation of abelian anyons.
\newblock \emph{Phys. Rev. Lett.}, 135:\penalty0 056501, Jul
  2025{\natexlab{b}}.
\newblock \doi{10.1103/3yvl-4hws}.
\newblock URL \url{https://link.aps.org/doi/10.1103/3yvl-4hws}.

\bibitem[Mukherjee et~al.(2014)Mukherjee, Mandal, Wu, W\'ojs, and
  Jain]{Mukherjee_Enigmatic_2014}
Sutirtha Mukherjee, Sudhansu~S. Mandal, Ying-Hai Wu, Arkadiusz W\'ojs, and
  Jainendra~K. Jain.
\newblock Enigmatic $4/11$ state: A prototype for unconventional fractional
  quantum hall effect.
\newblock \emph{Phys. Rev. Lett.}, 112:\penalty0 016801, Jan 2014.
\newblock \doi{10.1103/PhysRevLett.112.016801}.
\newblock URL \url{https://link.aps.org/doi/10.1103/PhysRevLett.112.016801}.

\bibitem[Note5()]{Note5}
Note5.
\newblock We generate the exact coefficients using the DiagHam library~\cite
  {DiagHam}.

\bibitem[Balram and Jain(2017)]{Balram_PH_CF_2017}
Ajit~C. Balram and J.~K. Jain.
\newblock Particle-hole symmetry for composite fermions: An emergent symmetry
  in the fractional quantum hall effect.
\newblock \emph{Phys. Rev. B}, 96:\penalty0 245142, Dec 2017.
\newblock \doi{10.1103/PhysRevB.96.245142}.
\newblock URL \url{https://link.aps.org/doi/10.1103/PhysRevB.96.245142}.

\bibitem[Note6()]{Note6}
Note6.
\newblock States with equal $\nu ,\kappa _{xy},S$ can be distinct
  topologically, e.g., by having a different ground state degeneracy on a
  torus~\cite {balram_Zn_2020}, but that is not the case here.

\bibitem[MacDonald(1994)]{MacDonald_Introduction_1994}
A.~H. MacDonald.
\newblock {Introduction to the Physics of the Quantum {Hall} Regime}, 1994.
\newblock URL \url{https://arxiv.org/abs/cond-mat/9410047}.

\bibitem[Simion and Lyanda-Geller(2014)]{Simion_Magnetic_2014}
G.~E. Simion and Y.~B. Lyanda-Geller.
\newblock Magnetic field spectral crossings of luttinger holes in quantum
  wells.
\newblock \emph{Phys. Rev. B}, 90:\penalty0 195410, Nov 2014.
\newblock \doi{10.1103/PhysRevB.90.195410}.
\newblock URL \url{https://link.aps.org/doi/10.1103/PhysRevB.90.195410}.

\bibitem[Simion and Lyanda-Geller(2017)]{Simion_Non_Abelian_2017}
George Simion and Yuli Lyanda-Geller.
\newblock Non-abelian $\ensuremath{\nu}=\frac{1}{2}$ quantum hall state in
  ${\mathrm{\ensuremath{\Gamma}}}_{8}$ valence band hole liquid.
\newblock \emph{Phys. Rev. B}, 95:\penalty0 161111, Apr 2017.
\newblock \doi{10.1103/PhysRevB.95.161111}.
\newblock URL \url{https://link.aps.org/doi/10.1103/PhysRevB.95.161111}.

\bibitem[Balram et~al.(2018)Balram, Barkeshli, and Rudner]{Balram_parton_2018}
A.~C. Balram, M.~Barkeshli, and M.~S. Rudner.
\newblock Parton construction of a wave function in the anti-{Pfaffian} phase.
\newblock \emph{Phys. Rev. B}, 98:\penalty0 035127, Jul 2018.
\newblock \doi{10.1103/PhysRevB.98.035127}.
\newblock URL \url{https://link.aps.org/doi/10.1103/PhysRevB.98.035127}.

\bibitem[Yutushui et~al.(2024{\natexlab{a}})Yutushui, Hermanns, and
  Mross]{Yutushui_daughters_2024}
Misha Yutushui, Maria Hermanns, and David~F. Mross.
\newblock Paired fermions in strong magnetic fields and daughters of
  even-denominator hall plateaus.
\newblock \emph{Phys. Rev. B}, 110:\penalty0 165402, Oct 2024{\natexlab{a}}.
\newblock \doi{10.1103/PhysRevB.110.165402}.
\newblock URL \url{https://link.aps.org/doi/10.1103/PhysRevB.110.165402}.

\bibitem[Yutushui et~al.(2024{\natexlab{b}})Yutushui, Park, and
  Mirlin]{Yutushui_Localization_2024}
Misha Yutushui, Jinhong Park, and Alexander~D. Mirlin.
\newblock Localization and conductance in fractional quantum hall edges.
\newblock \emph{Phys. Rev. B}, 110:\penalty0 035402, Jul 2024{\natexlab{b}}.
\newblock \doi{10.1103/PhysRevB.110.035402}.
\newblock URL \url{https://link.aps.org/doi/10.1103/PhysRevB.110.035402}.

\bibitem[Bishara et~al.(2008)Bishara, Fiete, and
  Nayak]{Bishara_PH_Read_Rezayi_2008}
W.~Bishara, G.~A. Fiete, and C.~Nayak.
\newblock Quantum {Hall} states at $\nu=\frac{2}{k+2}$: Analysis of the
  particle-hole conjugates of the general level-$k$ read-rezayi states.
\newblock \emph{Phys. Rev. B}, 77:\penalty0 241306(R), Jun 2008.
\newblock \doi{10.1103/PhysRevB.77.241306}.
\newblock URL \url{https://link.aps.org/doi/10.1103/PhysRevB.77.241306}.

\bibitem[N.~Regnault and et~al.()]{DiagHam}
Z.~Papic N.~Regnault, G.~Moeller and et~al.
\newblock Diagham.
\newblock \url{https://www.nick-ux.org/diagham/index.php/Main_Page}.

\bibitem[Balram et~al.(2020)Balram, Jain, and Barkeshli]{balram_Zn_2020}
Ajit~C. Balram, J.~K. Jain, and Maissam Barkeshli.
\newblock \ensuremath{{\mathbb{Z}}_{n}} superconductivity of composite bosons
  and the $7/3$ fractional quantum hall effect.
\newblock \emph{Phys. Rev. Res.}, 2:\penalty0 013349, Mar 2020.
\newblock \doi{10.1103/PhysRevResearch.2.013349}.
\newblock URL \url{https://link.aps.org/doi/10.1103/PhysRevResearch.2.013349}.

\end{thebibliography}
\end{document}